\documentclass[onecolumn,showpacs,superscriptaddress,tighten,floatfix,nofootinbib,prd]{revtex4} 
\usepackage{graphicx}
\usepackage{epsf}
\usepackage{epsfig}
\usepackage{amssymb,amsmath}
\usepackage[usenames]{color}
\usepackage{amssymb}
\usepackage{times}
\usepackage{mathrsfs}
\usepackage{hyperref}
\usepackage[normalem]{ulem}
\usepackage{bm}
\hypersetup{
  colorlinks=true,        % false: boxed links; true: colored links
  linkcolor=blue,         % color of internal links
  citecolor=Cyan,         % color of links to bibliography
  urlcolor=NavyBlue
}

\usepackage{float}

\usepackage{amsbsy,amssymb,amsmath,setspace}

\usepackage{datetime}

\usepackage{graphicx,color,epstopdf} % epstopdf is used to convert eps figures 
\usepackage[dvipsnames]{xcolor}
\usepackage{moreverb}
\usepackage{dsfont}
\usepackage{upgreek}
\usepackage{multirow}
\usepackage{framed,mdframed}
\usepackage{tensor}
\usepackage{relsize} % e.g. used for \mathsmaller
\usepackage{enumerate}

\usepackage{wrapfig} % Insert wraped figure 
\usepackage{accents}
\usepackage{microtype} % Slightly tweak font spacing for aesthetics

\usepackage{booktabs} % Horizontal rules in tables
\usepackage{paralist} % Used for the compactitem environment which makes bullet 
\usepackage{pgf}
\usepackage{tikz}\usetikzlibrary{shapes.misc} % shedding of the section titles

\usetikzlibrary{decorations.fractals}
\usetikzlibrary{decorations.pathmorphing,backgrounds}
\usetikzlibrary{decorations.shapes}
\usetikzlibrary{decorations.footprints}
\usetikzlibrary{shapes,arrows,positioning,calc}
\tikzset{paint/.style={ draw=#1!50!black, fill=#1!50 },
    decorate with/.style=
    {decorate,decoration={shape backgrounds,shape=#1,shape size=2mm}}}
\usetikzlibrary{arrows}

\definecolor{shadecolor}{rgb}{0.85,0.85,0.85}

\newcommand{\Lag}{\mathscr{L}}
\newcommand{\Eul}{{\mathscr{E}}}
\newcommand{\spacetime}{\mathcal{V}^4}
\newcommand{\mattertime}{\mathcal{M}^4}
\newcommand{\matter}{\mathcal{M}^3}

\newcommand{\scE}{{\mathscr{E}}}

\newcommand{\calE}{\mathcal{E}}
\newcommand{\FF}{{\boldsymbol{F}}}

\newcommand{\QQ}{{\boldsymbol{Q}}}

\newcommand{\dev}{{\rm dev}}

\newcommand{\reaSet}{{\rm I\!R}}
\newcommand{\halb}{\frac{1}{2}}

\definecolor{qqqqff}{rgb}{0.,0.,1.}
\definecolor{ffqqqq}{rgb}{1.,0.,0.}

\newcommand{\pd}{\partial}
\newcommand{\rmd}{{\rm d}}
\newcommand{\cgrad}[2]{ \xi^{#1}_{\phantom{#1}#2} }	% Config 
\newcommand{\proptime}{\tau}
\newcommand{\uLag}{U}
\newcommand{\dist}[2]{ A^{#1}_{\ \, #2} }	% Distortion field A
\newcommand{\idist}[2]{ A^{#1}_{\ #2} }	% Inverse distortion
\newcommand{\plast}[2]{ P^{#1}_{\ #2} }	% Distortion field A
\newcommand{\F}[2]{ x^{#1}_{\ #2}\, }				% Deformation gradient F
\newcommand{\devG}[1]{ \mathring{G}_{#1} }				% deviatoric part 
\newcommand{\devGMix}[2]{ \mathring{G}^{#1}_{\ #2} }				% deviatoric 
\newcommand{\devsigma}[1]{ \mathring{\sigma}_{#1} }				% deviatoric part of \sigma
\newcommand{\Cauchy}[1]{\sigma_{#1} }				% Cauchy stress 
\newcommand{\Cauchymix}[2]{\sigma^{#1}_{\phantom{#1}#2} }		% Mixed Cauchy 
\newcommand{\EMTD}[2]{\Sigma^{#1}_{\ \, #2} }				% Our energy momentum DENSITY
\newcommand{\EMT}[2]{\mathcal{T}^{#1}_{\ \, #2} }				% Our energy-momentum, pure tensor
\newcommand{\EMTGR}[2]{T^{#1}_{\ \, #2} }				% energy-momentum of GR
\newcommand{\DT}[1]{\overset{\boldsymbol{\cdot}}{#1} }				% Our 

\newcommand{\ED}{E}				% Total Energy which is scalar-DENSITY, i.e. with \sqrt{-g}
\newcommand{\E}{\mathcal{E}}	% Total Energy which is pure scalar, i.e. NO \sqrt{-g}, like in GR
\newcommand{\LED}{\mathcal{L}}	% Eulerian Lagrangain density , i.e. with \sqrt{-g}, like in GR

\newcommand{\myxi}{\xi}

\newcommand{\G}{\mathcal{G}}
\newcommand{\LagrProj}{\mathcal{G}}
\newcommand{\LagrProjEff}{G}
\newcommand{\Geff}[1]{G_{#1}}
\newcommand{\GeffContr}[1]{G^{#1}}
\newcommand{\GeffMix}[2]{G^{#1}_{\ #2}}

\newcommand{\Kab}[1]{{\kappa_{#1}}}
\newcommand{\KabContr}[1]{{\kappa^{#1}}}
\newcommand{\KAB}{\kappa}
\newcommand{\Kmn}{\kappa}
\newcommand{\KMN}[1]{\kappa_{#1}}
\newcommand{\KMNContr}[1]{\kappa^{#1}}
\newcommand{\traceG}{G}
\newcommand{\entropy}{s}
\newcommand{\Entropy}{S}

\newcommand{\EqTitle}[2]{\texorpdfstring{#1}{#2}}
\newcommand{\oo}{\hphantom{\mu}}
\newcommand{\lapse}{\alpha}
\newcommand{\shift}{\beta}

\newcommand{\gammabar}[1]{\gamma^{#1\frac{1}{2}}}

\newcommand{\mbf}[1]{\mathrm{#1}}

\newcommand{\sA}{{\mathsmaller A}}
\newcommand{\sB}{{\mathsmaller B}}
\newcommand{\sC}{{\mathsmaller C}}

\newcommand{\sK}{{\mathsmaller K}}
\newcommand{\sM}{{\mathsmaller M}}
\newcommand{\sN}{{\mathsmaller N}}
\newcommand{\sL}{{\mathsmaller L}}

\newcommand{\sFrm}{{\rm \mathsmaller F}}
\newcommand{\sErm}{{\rm \mathsmaller E}}
\newcommand{\sLrm}{{\rm \mathsmaller L}}
\newcommand{\sRrm}{{\rm \mathsmaller R}}
\newcommand{\sGrm}{{\rm \mathsmaller G}}
\newcommand{\sMrm}{{\rm \mathsmaller M}}

\newcommand{\Lie}[1]{\mathcal{L}_{#1}}

\newcommand{\ER}[1]{ \textcolor{magenta}   {\small\textbf{
ER: #1}} }

\newcommand{\IP}[1]{ \textcolor{Red}   {\small\textbf{
IP: #1}} }
\renewcommand{\v}{\boldsymbol{v}}
\renewcommand{\u}{\boldsymbol{u}}
\newcommand{\x}{\boldsymbol{x}}
\newcommand{\y}{\boldsymbol{y}}
\newcommand{\q}{\boldsymbol{q}}

\newcommand{\ie}{i.e.~}
\newcommand{\eg}{e.g.~}
\begin{document}
%%%%%%%%%%%%%%%%%%%%%%%%%%%%%%%

%----------------------------------------------------------------------------------------
%	TITLE SECTION
%----------------------------------------------------------------------------------------

\allowdisplaybreaks

\title[Unified causal hyperbolic continuum mechanics in general relativity]{
A new causal general relativistic formulation for dissipative continuum fluid and solid mechanics and 
its solution with high-order ADER schemes} 

\date{\today}
\label{firstpage}

\author{Ilya Peshkov\footnote{The work by I.P. has been started while being at 
Institut de Math\'{e}matiques de Toulouse, France}}
\affiliation{
	Laboratory of Applied Mathematics, University of Trento,
	Via Mesiano 77, 38123 Trento, Italy
}

%\affiliation{
%Institut de Math\'{e}matiques de Toulouse,  
%Toulouse, France
%%; Sobolev Institute of 
%%Mathematics, Novosibirsk, Russia, 
%}

\author{Evgeniy Romenski}
\affiliation{
Sobolev 
Institute of Mathematics and Novosibirsk State University, Novosibirsk, Russia
}
\affiliation{
	Laboratory of Applied Mathematics, University of Trento,
	Via Mesiano 77, 38123 Trento, Italy
}

\author{Francesco Fambri\footnote{The work by F.F. has been started while being 
at University of Trento, Italy}}
\affiliation{Max-Planck Institute for Plasma Physics, Boltzmannstr. 2, 85748 
Garching, Germany}
\author{Michael Dumbser}
\affiliation{
Laboratory of Applied Mathematics, University of Trento,
Via Mesiano 77, 38123 Trento, Italy
}

\begin{abstract}
We present a unified causal general relativistic formulation of dissipative and 
non-dissipative continuum mechanics. The presented theory is the first general 
relativistic theory that can deal simultaneously with viscous fluids as well 
as irreversible deformations in solids and hence it also provides a fully 
covariant formulation of the Newtonian continuum mechanics in arbitrary 
curvilinear 
spacetimes. In such a formulation, the matter is considered as a Riemann-Cartan manifold with 
non-vanishing torsion and the main field of the theory being the non-holonomic 
basis tetrad field 
also called four-distortion field.
Thanks to the variational nature of the 
governing equations, the theory is compatible with the variational structure of 
the Einstein field equations. Symmetric hyperbolic equations are the 
only admissible equations in our unified theory and thus, all perturbations propagate at finite 
speeds (even in the diffusive regime) and the Cauchy problem for the governing 
PDEs is locally well-posed for 
arbitrary and regular initial data which is very important for the 
numerical treatment of the presented model. Nevertheless, the numerical 
solution of the discussed hyperbolic equations is a challenging task because of 
the presence of the stiff algebraic source terms of relaxation type and 
non-conservative differential terms. Our 
numerical strategy is thus based on an advanced family of high-accuracy ADER 
Discontinuous Galerkin and Finite Volume methods which provides a very 
efficient framework for general relaxation hyperbolic PDE systems. An extensive 
range of numerical examples is presented demonstrating the applicability of our 
theory to relativistic flows of viscous fluids and deformation of solids in 
Minkowski and curved spacetimes. 
\end{abstract}

%\pacs{
%04.25.D-, % Numerical relativity
%%04.25.dg, % Numerical studies of black holes and black-hole binaries
%}
\maketitle

%%%%%%%%%%%%%%%%%%%%%%%%
\section{Introduction}\label{sec.Intro}

\subsection{Classical fluid dynamics and Eckart-Landau-Lifshitz theory}

Paradoxically, a century after the formulation of the General Relativity (GR) 
theory of gravity by Einstein, there is still no a dissipative continuous 
theory compatible with the \textit{variational nature} of GR. It is 
known though that the Euler equations for perfect fluids, nonlinear 
elasticity theory of perfect elastic solids, and Maxwell equations in vacuum do 
admit a fully covariant and variational formulation compatible with GR, 
e.g. see
\cite{RezzollaZanottiBook,Carter1972,Kijowski1992,Karlovini2003,
Wernig-Pichler2006,Broda2008,Gundlach2012}. What 
makes dissipative systems so special? It is of course the way how the 
dissipation is represented in the classical/modern continuum mechanics. Thus, 
any continuum model relies on the fundamental mass, momentum and energy 
conservation laws. However, in order to be applied to a certain physical 
system, the conservation laws have to be supplemented by \textit{constitutive 
laws} which relate the state of that system to the external stimuli. Thus, the constitutive theory 
of 
the modern dissipative 
continuum mechanics relies on the Classical Irreversible 
Thermodynamics (CIT), e.g. see \cite{deGrootMazur1984,LebonJou2008}, which, in turn,
relies on the famous phenomenological constitutive laws such as 
Newton's viscous law, Fourier's law of heat conduction, and Fick's law of 
diffusion, etc. For example, the entire 
fluid mechanics of viscous fluids is built around Newton's viscous law by using 
it directly as in 
the Navier-Stokes equations or generalizing and/or extending it to more complex 
media (non-linear viscosity approach). The key feature of all such laws is the 
\textit{steady-state} assumption, that is the flow (or a transfer process) 
should be microscopically in a steady-state (time independent) regime, 
i.e. the time is completely removed from the microscopic time evolution\footnote{In other 
words, it is implied that the steady-state is reached at an infinite 
rate.} and the history of the microscopic evolution preceding the 
steady-state state is disregarded completely. It is well known that the 
steady-state assumption provides a good approximation to reality if the 
characteristic length/time scale of the process is sufficiently longer than a 
microscopic characteristic length/time scale, i.e. the steady-state is reached 
significantly faster than the macroscopic characteristic time of the process.
Nevertheless, being acceptable from the engineering standpoint, the 
steady-state-based transport theory has the following conceptual issues that, in particular, 
make it difficult building of a consistent with GR continuous dissipative theory:
\begin{itemize}
\item First of all, we note that the steady-state is actually a deceptive state. 
Indeed, what is macroscopically 
seen as a time-independent process is, in fact, the result of the
competitive \textit{dynamics} between the external energy supply 
and the internal (microscopic) dissipative processes in the system. 

\item Secondly, by ignoring the 
time 
in the constitutive relations, one obtains parabolic conservation laws which is known to violate 
the 
causality principle (superluminal signal speeds). Moreover, in contrast to the 
non-relativistic case, linear parabolic PDEs have ill-posed initial value 
problem~\cite{Hiscock1988,RezzollaZanottiBook,Romatschke2010} in the 
relativistic settings (unbounded growth of short-wavelength perturbations, 
which necessarily leads to grid-dependent numerical results that do not 
converge as the spatial resolution is refined) which 
makes them practically unusable for the numerical simulations. Note that even in the 
non-relativistic framework, a nonlinear viscosity parabolic model might be 
ill-posed~\cite{Schaeffer1987,Barker2015}.

\item Thirdly, a variational 
formulation 
for the parabolic dissipative theory is not known and most likely does not exist. Hence, any 
possible coupling of the Navier-Stokes stress with the matter energy-momentum coming 
from the Einstein field equations destroys the Euler-Lagrange structure of the latter.
\end{itemize}

%In particular, disregarding the time in the classical constitutive relations is 
%known to result in that the mathematical model, parabolic partial differential 
%equations (PDEs), admit infinite velocity for perturbation propagation which, 
%of course, violates the \textit{causality} principle. 

For completeness, we recall that formally the Navier-Stokes-Fourier equations can be written in the 
relativistic settings and there are two versions of such equations due to 
Eckart~\cite{Eckart1940} and Landau and 
Lifshitz~\cite{LandauLifshitzHydro} which differ by the definition of the 
4-velocity~\cite{RezzollaZanottiBook}. Both formulations suffer from the above issues. This is a 
good illustration of the 
non-universality of the phenomenological constitutive theory and that
a ``\textit{good approximation}'' of specific experimental data (e.g. stress strain-rate relation) 
does not necessarily results in
physical consistency and mathematical regularity of the governing equations.
Nevertheless, attempts to fix the stability issues of the Eckart and Landau-Lifshitz theories have 
continued. 
Thus, thanks to the separate treatment of the momentum density and energy current density, V\'an 
and 
Bir\'o 
\cite{Van2012} were able to build a stable modification of the Eckart theory. Also, recent results 
by Freist\"uhler and Temple 
\cite{Freistuhler2017} suggest that the 
Eckart-Landau-Lifshitz second-order 
equations can be still modified in such a way that the resulting equations can be obtained as a 
uniform limit of \textit{second-order} symmetric hyperbolic equations in the sense of 
\cite{Hughes1977} and thus can provide a causal formulation for dissipative fluids. 
However, our main counterargument for using phenomenological dissipative theories for modeling 
general relativistic flows is that they do not admit a variational formulation and therefore, the 
dissipative stress has to be added to the canonical matter energy-momentum tensor in an \textit{ad 
hoc} manner.

\subsection{M\"{u}ller-Israel-Stewart theory}

The well-known fix allowing (to some degree) to avoid acausal and unstable 
behavior 
of the relativistic parabolic dissipative theory, as well as of  the 
non-relativistic Navier-Stokes equations is their 
``\textit{hyperbolization}'' via the Maxwell-Cattaneo 
procedure~\cite{Maxwell1867,Cattaneo1948} when the 
original second-order parabolic PDEs are transformed into a new \textit{extended} 
first-order hyperbolic system in which the stress tensor (or heat flux, mass flux) is promoted to 
the 
independent state variable governed by its own evolution equation of 
relaxation type. This naive hyperbolization then had become more mature after 
works 
by M\"{u}ller~\cite{Mueller1966PhD,Muller1967}, Israel \cite{Israel1976} and 
Stewart~\cite{Stewart1977}, known now as the M\"{u}ller-Israel-Stewart theory or 
Extended Irreversible Thermodynamics (EIT), which, in turn, later transformed 
into an alternative \textit{divergence formulation} and, which has also benefited from a close 
connection with the kinetic theory of gases, particularly through the 
\textit{moment method} of Grad~\cite{Grad1949,Torrilhon2016}, and is called 
the \textit{divergence-type formulation} of Extended Irreversible 
Thermodynamics, or as Rational Extended Thermodynamics 
(RET) by Mu\"uller and Ruggeri and others 
~\cite{MullerRuggeri1986,MullerRuggeri1998,Ruggeri2015,Geroch1990,Reula2018}. 
In the non-relativistic settings, there is also a very similar formulation known also as Extended 
Irreversible Thermodynamics~\cite{EIT2010,LebonJou2008}. 
One of the 
central ideas in such theories is the hierarchical structure of the equations when the flux of the 
$ k $-th evolution equation enters as the density field in the $ k+1 $ equation, and so on. This 
results 
in the ever-increasing rank of the state variables by one. For a more 
comprehensive 
reviews of the existing relativistic theories of dissipative fluids, 
we refer the reader to~\cite{RezzollaZanottiBook,Romatschke2010,Andersson2007a}. Currently, the 
M\"{u}ller-Israel-Stewart theory presents the state of the art of the relativistic dissipative 
fluid dynamics and is used in numerical simulation of Heavy Ion Collisions to study the properties 
of the quark-gluon plasma \cite{Bouras2009,Bouras2010,Takamoto2011,Akamatsu2014} 
and is implemented in the state of the art relativistic computational fluid dynamics codes 
\cite{DelZanna2013,Karpenko2014,Shen2016}.

Despite a relative success in overcoming the non-causality and stability issues 
of the relativistic Navier-Stokes-Fourier equations, the mentioned formulations for 
dissipative fluids have several conceptual issues. For example, the main obstacle preventing 
their coupling with GR, in our opinion, is the lack of a variational formulation which does not 
allow to 
embed them into  
the Euler-Lagrange structure of Einstein's field equations. In other words, the 
viscous stress tensor has to be plugged into the canonical matter 
energy-momentum 
tensor in an \textit{ad hoc} manner.  Another difficulty 
concerns the multiphysics applications. For instance, it is not clear
how such theories can be coupled with electromagnetic fields, e.g. see a discussion 
in~\cite{Torrilhon2016}. 
%However, to be fair, it worth to recall that 
%despite a long history, the question of coupling of moving continuous media 
%and electromagnetic 
%field still does not have a generally accepted solution~\cite{DPRZ2017}. 
Furthermore, the 
connection of RET 
with the Boltzmann gas 
kinetic theory frequently emphasized as a strong argument in favor of such theories rises another 
question of how to deal with relativistic liquids and solids (e.g. the star 
interior, 
liquid-gas transition, also the outer crust of the neutron star 
is believed to be a crystalline solid~\cite{Chamel2008})
which apparently 
are 
not described by the kinetic theory of gases. Also, the infinite hierarchy of RET equations 
cannot be used in practice directly and requires to be restricted to a finite 
subsystem, that is high-order terms require a constitutive relations which 
express them only in terms of low-order moments. This constitutes the 
\textit{closure} problem of RET which is, in fact, the central problem of RET and is the 
topic of active research~\cite{MullerRuggeri1998,Torrilhon2016,Schaerer2017}. 
It is well known that the closure problem does not have a unique solution. In 
particular, recent study~\cite{Molnar2012} shows that there can be infinitely 
many choices for the closure procedure provided by Israel and Stewart which 
result in different explicit form and coefficients in equations of motion for 
the dissipative currents. Lastly, the physical meaning of 
high-rank 
tensors (higher than 2) remains unclear in RET. 
%Despite several open questions in RET and EIT type 
%theories, we believe 
%that the 
%main their issue  preventing them from coupling with the Einstein field 
%equations is the lack of 
%variational formulation of the former.

\subsection{Alternative geometric approach}

It is a direct goal of this paper to propose and discuss a new geometrical 
approach to formulating general relativistic equations for dissipative and non-dissipative dynamics 
of fluids and 
solids.
The proposed approach is a rather straightforward generalization of our unified 
formulation for 
Newtonian fluid and solid mechanics \cite{HPR2016,DPRZ2016,DPRZ2017,HYP2016,PRD-Torsion2019} which 
in turn relies on the non-linear Eulerian inelasticity theory by Godunov and 
Romenski~\cite{GodRom1974,God1978,Romenski1979,GodRom1995,GodRom1996,GodRom2003}, Besseling 
\cite{Besseling1968,Besseling1994}, and Rubin \cite{Rubin1987,Rubin2019}. In such a theory, the 
flowing medium 
is treated as a Riemann-Cartan manifold with non-zero torsion and with the main field being the 
non-holonomic local basis tetrad field, which we also shall call the 4-distortion field. 
The 4-distortion describes deformation and rotation of continuum particles which are assumed to 
have a finite length scale $ \ell $. The finiteness of the continuum particle length-scale is 
crucial in our 
theory for describing the ability of a medium to flow (fluidity). Because it is more convenient to 
work not 
with the length-scale but with the corresponding time-scale $ \tau $, we shall use the latter for 
characterizing the medium fluidity. Time $ \tau $ is a 
continuum interpretation of the seminal idea of the so-called \textit{particle settled life time} 
of 
Frenkel~\cite{Frenkel1955}, who applied it to describe the fluidity of liquids, e.g.
see~\cite{brazhkin2012two,bolmatov2013thermodynamic,bolmatov2015revealing,
Bolmatov2015a,Bolmatov2016}. In our continuum approach, the time $ \tau $ is 
called 
\textit{strain dissipation time} and it is the time taken 
by a given continuum particle (material element) to ``escape'' from the cage 
composed of its neighbor particles, i.e. the time taken to rearrange with one of its 
neighbors~\cite{HYP2016}. The more viscous a fluid is, the larger the time $ \tau $, i.e. the 
longer the continuum  particles stay in contact with each other. Moreover, the use of Frenkels'  
concept for time $ \tau $ allows for a unified mathematical description of the two main branches of 
continuum mechanics, fluid and solid 
dynamics, within a single system of governing equations, e.g. see 
\cite{HPR2016,DPRZ2016,DPRZ2017,HYP2016,Hyper-Hypo2019}.

From the mathematical viewpoint, the proposed theory has some important features. First of all,  
the non-relativistic counterpart of the new model belongs to the class of so-called 
Symmetric Hyperbolic Thermodynamically Compatible (SHTC) formulation of Newtonian continuum 
mechanics  
~\cite{SHTC-GENERIC-CMAT}, which originates from the works 
\cite{God1961,God1972,GodRom1995,Godunov1996,Rom1998,Rom2001} by Godunov and Romenski on the 
admissible structure of macroscopic thermodynamically consistent equations in continuum physics. 
Non-relativistic SHTC equations can be applied to describe all basic transport processes such as 
viscous momentum, heat, mass, and electric charge transfer 
\cite{DPRZ2016,DPRZ2017,SHTC-GENERIC-CMAT}.   
Moreover, as it follows from the name of SHTC formulation, all equations are 
\textit{hyperbolic} and hence have the well-posed initial value problem for arbitrary 
smooth initial data and all perturbations propagate at finite speeds even in the diffusive regime.
We expect that the relativistic version of all the SHTC equations including the one discussed in 
this paper preserve the property of being symmetric hyperbolic.  
However, we don't prove this 
rigorously in this paper. Nevertheless, we prove thermodynamic consistency of the model which is 
the key for recovering hyperbolicity in the SHTC framework. 

Secondly, from the point of view of formulating a general relativistic flow theory, 
i.e. 
compatible with GR, another important and very attractive feature of our new 
geometric approach 
is that 
it admits a variational formulation. More precisely, the overall time evolution of our theory is 
split into two parts, reversible and irreversible,
\begin{equation}\label{time.evolution.totla}
\left( \frac{\pd}{\pd t} \right)_{total} = \left( \frac{\pd}{\pd t} \right)_{revers} + \left( 
\frac{\pd}{\pd t} \right)_{irrevers}.
\end{equation}
It is the reversible part (all the differential terms) which incorporates most of the mathematical 
structure of the governing equations. This part admits a variational formulation and thus can be 
straightforwardly coupled with the Einstein field equations via the matter part of the 
Hilbert-Einstein action integral. In other words, the matter energy-momentum tensor of our theory 
has the convenient structure of the canonical matter energy-momentum tensor of GR, 
see Sec.\,\ref{sec.EulerPDEs}. 
Despite that the importance of the variational principle is well 
understood in relativistic physics, to the best of our knowledge, there were no 
much 
attempts to employ variational principle for deriving equations for relativistic dissipative 
continuum mechanics apart from the works by Carter, Comer, and 
Anderson~\cite{Carter1991,Andersson2015,Andersson2007a}. However, at the end of the day, the 
viscosity law is postulated but not derived in these papers. In contrast, our theory does employ 
the 
concept of viscosity at all. Nevertheless, an effective viscosity can be derived for our model in 
the 
so-called stiff relaxation limit (diffusive regime), see Sec.\,\ref{sec.asymptotic}.

Last but not least, we also remark the Hamiltonian nature of the non-relativistic SHTC equations  
\cite{SHTC-GENERIC-CMAT}, that is the reversible part of the time evolution of the SHTC equations 
can be generated by the corresponding Poisson brackets. This fact might be important for 
establishing connections of the theory with microscopic theories such as gas kinetic theory for 
example in the context of the Hamiltonian formulation of non-equilibrium thermodynamics known as 
General Equations for Non-Equilibrium Reversible-Irreversible Coupling 
(GENERIC)~\cite{GrmelaOttingerI,GrmelaOttingerII,Ottinger1998,Ottinger1998b,Ottinger2005,
PKG-Book2018}.
Moreover, equations for relativistic viscous heat conducting fluids were proposed by \"Ottinger in 
the GENERIC framework \cite{Ottinger1998b,Ottinger2019}. Because, as it was shown in 
\cite{SHTC-GENERIC-CMAT}, the non-relativistic SHTC equations are fully compatible with 
GENERIC, we expect to 
see many common features between \"Ottinger's and our formulation despite very different 
mathematical 
bases, i.e. Hamiltonian nature of GENERIC and variational nature of SHTC equations. For example, 
one can see that \"Ottinger's equation (9) \cite{Ottinger2019} is structurally equivalent to our 
equation \eqref{eqn.G.PDE}. The orthogonality conditions are different though, cf. condition (11) 
in \cite{Ottinger2019} and \eqref{eqn.G.PDE}$ _2 $ in this work.

The outline of the paper is the following. We first briefly discuss the main principles of the SHTC 
equations using the general relativistic Euler equations as an example in Section \ref{sec.SHTC}. 
In section \ref{sec.Motion}, we demonstrate that the classical Lagrangian formalism of Newtonian 
continuum mechanics can be generalized to the 4-dimensional formalism of GR. We introduce 
Lagrangian and Eulerian frames of reference and then derive equations of motion for the 4-continuum 
in the Lagrangian frame. The Lagrangian equations of motion are then transformed into the Eulerian 
frame and demonstrated to be covariant. In Section \ref{sec.Irreversible}, we introduce the main 
field of our theory, the 4-distortion field and formulate the final governing equations. The family 
of ADER Finite Volume and ADER Discontinuous Galerkin methods is briefly discussed in 
Section~\ref{sec.Numerics}, while the results of numerical simulations are presented in 
Section~\ref{sec.numerical.examples}. Eventually, we conclude with the final comments and discuss 
further developments of the theory in Section~\ref{sec.Conclusion}.

%We shall use the following index convention. Lowercase Greek letters ($ \alpha 
%$, $ \beta $, \ldots $ \lambda $, $ \mu $, 
%$ \nu $, etc.) denote spacetime indices, lowercase Latin 
%letters ($ a $, $ b $, $ c $, etc.) are used to denote 4-dimensional
%matter-time manifold objects, capital Latin letters ($ \sA 
%$, $ \sB $, $ \sC $, etc.) denote indices for the pure matter component of the 
%matter-time manifold 
%objects,  
%and lowercase Latin letters ($ i $, $ j $, $ k $) are used to denote pure 
%spatial components of the space-time objects.

\section{Main principles of the SHTC equations}\label{sec.SHTC}

Our unified formulation of Newtonian continuum mechanics 
\cite{HPR2016,DPRZ2016,DPRZ2017} has been developed in the framework of the 
SHTC equations. In this section, we shall briefly describe the main principles of 
the SHTC theory using the relativistic Euler equations for perfect 
fluids as an example. Also, most of the important details of the SHTC equations were recently 
summarized and revisited in~\cite{SHTC-GENERIC-CMAT}.

When one deals with a nonlinear time-dependent phenomenon and thus with an underlying nonlinear 
time-dependent PDE system, one has to be sure that such a system has the well-posed initial value 
problem (IVP), i.e. that, for arbitrary regular initial data, the solution exists locally in time, 
the 
solution is unique and stable. The well-posedness is not only a mathematical requirement but is a 
fundamental property of a PDE system representing a macroscopic physical system due to the 
deterministic nature of the macroscopic time evolution. Moreover, the well-posedness of the initial 
value problem is a fundamental property which allows us to solve such nonlinear PDE systems 
numerically. Thus, ill-posed problems suffer from unbounded growth of short-wavelength 
perturbations, which necessarily leads to grid-dependent numerical results that do not converge as 
the spatial resolution is enhanced. It is important to understand that not all 
physically sound mathematical 
models have the well-posed IVP. For example, the relativistic Navier-Stokes-Fourier equations have 
the ill-posed IVP~\cite{Hiscock1983,Hiscock1985,Hiscock1988}, as well as the Burnett equations 
which were derived from the Boltzmann equation via the Chapman-Enskog 
expansion~\cite{Bobylev1982,StruchtrupTorrilhonR13,Torrilhon2016}, also some Navier-Stokes-based 
non-linear viscosity models may have ill-posed IVP, e.g.~\cite{Schaeffer1987,Barker2015}.

Godunov was within the first who asked what physical principles may guaranty the well-posedness of 
a 
PDE system representing a continuum mechanics model~\cite{God1961}. In particular, he 
observed that if a first order system of conservation laws is compatible with the first law of 
thermodynamics then such a system can be cast into a symmetric hyperbolic form and thus has 
well-posed initial value problem. 
%Let us demonstrate the main features of the Godunov observation 
%and the main principles of the SHTC equations using the relativistic Euler 
%equations for perfect 
%fluids.

The relativistic Euler equations read
\begin{equation}\label{intro.rel.Euler}
\nabla_\mu T^\mu_{\ \nu} = 0, \qquad \nabla_\mu (\rho u^\mu) = 0, \qquad \nabla_\mu (s  u^\mu) 
= 
0,
\end{equation}
where the energy-momentum tensor $ T^\mu_{\ \nu} $ and the fluid pressure $ p $ are defined as
\begin{equation}\label{intro.T}
T^\mu_{\ \nu} := E \, u^\mu u_\nu + p h^\mu_{\ \nu}, \qquad p:= \rho E_\rho +s  E_s  
- E.
\end{equation}
Here, $ E = E(\rho,s ) $ is the energy density, $ \rho $ is the rest fluid density, $ s  $ 
is the entropy density, $ E_\rho = \frac{\pd E}{\pd\rho} $, $ E_s = \frac{\pd 
E}{\pd s} $, $ 
u^\mu $ is the 4-velocity satisfying the normalization condition $ 
u^\mu u_\mu = -1 $, and $ h^\mu_{\ \nu} = \delta^\mu_{\ \nu} + u^\mu u_\nu $.

System \eqref{intro.rel.Euler} is, in fact, an overdetermined system because there are one more 
equations than the unknowns due to the fact that $ u^0 $ is not an unknown because of  $ 
u^\mu u_\mu = -1 $. Godunov then suggested that in order such an 
over-determined system be compatible one of the 
equations should be a consequence of the others with some coefficients. Indeed, it can be shown 
that the zeroth equation ($ \mu = 0 $) of the energy-momentum conservation for $ E^\mu := T^\mu_{\ 
0} $ can be expressed as
\begin{equation}\label{intro.summation}
\nabla_\mu E^\mu = - \frac{u^i}{u^0} \nabla_\mu T^\mu_{\ i} - \frac{E_\rho}{u^0} \nabla_\mu (\rho 
u^\mu) - \frac{E_s }{u^0} \nabla_\mu (s  u^\mu).
\end{equation}
Then, in contrast to the non-relativistic settings where the thermodynamic potential and the 
conserved unknowns are directly available, in the general relativistic covariant settings, we still 
need somehow to specify a thermodynamic potential $ \mathscr{E}(q_\ell) $ and 
unknowns $ q_\ell 
$, 
see~\cite{Ruggeri1981Euler}. We 
chose 
\begin{equation}\label{intr.E.Q}
\mathscr{E}(q_\ell):= E^\mu u_\mu = -u_0 E, \qquad q_\ell := (T^\mu_{\ i} u_\mu, \rho u^\mu u_\mu, 
s  u^\mu u_\mu) = (-u_i E, -\rho,-s ), \ \ell = 1,2,\ldots,5.
\end{equation}
With this choice, we have the following thermodynamic identity
\begin{multline}\label{intro.dE}
\rmd (u_0 E) = u_0 \rmd E + E \rmd u_0 = u_0 \rmd E - E \frac{u^i}{u^0}\rmd u_i 
= \\ 
u_0 \rmd E - E 
\frac{u^i}{u^0} \left(\frac{\rmd(u_i E) - u_i \rmd E}{E}\right) = 
-\frac{E_\rho}{u^0}\rmd \rho - \frac{E_s }{u^0}\rmd s  - \frac{u^i}{u^0} \rmd (u_i E),
\end{multline}
or, in other words,
\begin{equation}\label{intro.E_q}
\mathscr{E}_{q_i} = \frac{\pd \mathscr{E}}{\pd (-u_i E) } = -\frac{u^i}{u^0}, \qquad 
\mathscr{E}_{q_4} = \frac{\pd \mathscr{E}}{\pd (-\rho)} = -\frac{E_\rho}{u^0}, \qquad 
\mathscr{E}_{q_5} = \frac{\pd \mathscr{E}}{\pd (-s )} = -\frac{E_s }{u^0}.
\end{equation}
Therefore, we have the 4-potential $ E^\mu(q_\ell) $, the scalar potential $ \mathscr{E}(q_\ell) $ 
and 
the conserved unknowns $ q_\ell $. Godunov then observed that if one introduces 
new (dual) 4-potential $ 
L^\mu(p^\ell) $, scalar potential $ L(p^\ell) $ and the new unknowns $ 
p^\ell $ as Legendre 
conjugates to the old ones
\begin{equation}\label{intro.p}
p^i := \mathscr{E}_{q_i} = -\frac{u^i}{u^0}, \qquad p^4 := \mathscr{E}_{q_4} = -\frac{E_\rho}{u^0}, 
\qquad p^5 := \mathscr{E}_{q_5} = -\frac{E_s }{u^0},
\end{equation}
\begin{equation}\label{intro.Legendre.L}
L := q_\ell \scE_{q_\ell} - \scE = q_\ell\, p^\ell - \scE,
\end{equation}
\begin{equation}\label{intro.Lmu}
L^\mu := p^i T^\mu_{\ i} + p^4 (\rho u^\mu) + p^5 (s  u^\mu) - E^\mu,
\end{equation}
then the system of relativistic Euler equations can be written in Godunov's canonical 
form~\cite{God1961,SHTC-GENERIC-CMAT}. 
Indeed, it follows from \eqref{intro.Lmu} that the partial derivatives of $ L^\mu $ are
\begin{equation}\label{intro.deriv.Lmu}
L^\mu_{p^i} = T^\mu_{\ i}, \qquad L^\mu_{p^4} = \rho u^\mu, \qquad L^\mu_{p^5} = s  u^\mu.
\end{equation}
Moreover, putting \eqref{intro.p} into \eqref{intro.Legendre.L} and \eqref{intro.Lmu} gives
\begin{equation}\label{intro.explicit.L}
%L = -\frac{1}{u^0 u_0}(\rho \scE_\rho + s  \scE_s  - \scE), \qquad 
L^\mu = - u^\mu L.
\end{equation}
Eventually, based on \eqref{intro.deriv.Lmu} and \eqref{intro.explicit.L}$ _2 $, one may conclude 
that 
the relativistic Euler equations  \eqref{intro.rel.Euler} can be cast into the 
canonical 
Godunov form
\begin{equation}\label{intro.God.form}
-\nabla_\mu (u^\mu L)_{p^\ell} = 0,
\end{equation}
and then into a symmetric quasilinear form
\begin{equation}\label{intro.symmetric}
\mathsf{A}^\mu_{\ell m} \nabla_\mu p^m = 0,
\end{equation}
where $ m,\ell = 1,2,\ldots,5 $, and matrices $ \mathsf{A}^\mu_{\ell m} = -(u^\mu L)_{p^\ell p^m} $ 
are obviously symmetric as they are the second order derivatives of the potentials $ L^\mu =-u^\mu 
L$. 
Moreover, if to assume that the potential $ \scE(q_{\ell}) $ is convex, and hence $ L(p^\ell) $ is 
convex 
as well (due to the property of the Legendre transformation to preserve the convexity), then for 
any time-like 4-vector $ 
\zeta_\mu $ (independent of $ p^\ell $) the matrix $ \mathsf{A}_{\ell m} := \mathsf{A}^\mu_{\ell m} 
\zeta_\mu = L_{p^\ell p^m}$ is positive 
definite. Therefore, quasilinear system \eqref{intro.symmetric}, as well as the original Euler 
system~\eqref{intro.rel.Euler}, is symmetric hyperbolic and hence, has 
well-posed initial value 
problem (locally in any time-like direction) as well as the finite speeds for 
perturbations 
propagation. In other words, Godunov's 
observation~\cite{God1961} establishes an intimate 
connection 
between the well-posedness and causality of a nonlinear mathematical model 
expressed as an over-determined 
system of
conservation laws and the thermodynamics.

\section{Motion of the continuum in the 4D Lagrangian 
formalism}\label{sec.Motion}

We recall that the reversible and 
irreversible (dissipative) parts of the time evolution are treated separately 
in our theory, see \eqref{time.evolution.totla}. Therefore, our way to 
derive governing equations is as follows. We first derive the reversible part 
from 
a variational principle and then by preserving its structure we add low order 
(algebraic) terms based on the second law of thermodynamics.
The reversible part of the time evolution describes reversible deformations of 
the continuum, i.e. elasticity (after adding dissipative terms, it becomes 
local elasticity). This part is rather 
kinematical and, in the absence of irreversible processes, is described by the 
one-to-one mapping $ x^\mu(\myxi^a) $ 
between the Lagrangian, $ \myxi^a $, and Eulerian, $ x^\mu $, coordinates of 
the continuum particles. The mapping $ x^\mu(\myxi^a) $ completely defines the 
motion of the non-dissipative continuum and as we shall see satisfy 
second-order (w.r.t. $ x^\mu(\myxi^a) $) Euler-Lagrange 
equations which however we shall treat as an enlarged system of first-order 
equations for the gradient $ \frac{\pd x^\mu}{\pd \myxi^a} = \F{\mu}{a}$. This 
gradient can be viewed as \textit{holonomic} basis tetrad, i.e. its torsion $ 
\pd_a 
\F{\mu}{b} - \pd_b\F{\mu}{a}$ is zero. On the other hand, the irreversibility 
of deformation implies that the mapping $ x^\mu(\myxi^a) $ becomes multivalued 
and 
meaning of $ \F{\mu}{a} $ as being gradient of $ x^\mu(\myxi^a) $ becomes 
questionable which is expressed in the non-vanishing torsion $ \pd_a 
\F{\mu}{b} - \pd_b\F{\mu}{a} \neq 0$. This means that the tetrad $ \F{\mu}{a} $ 
becomes \textit{non-holonomic}. The irreversible part of the time evolution is 
then 
added to the reversible equations as low order terms (relaxation terms) which 
acts as the source of non-holonomy for the tetrad $ \F{\mu}{a} $.

Nevertheless, a completely different route to the governing equations of our 
theory is possible. This route does not assume existence of the mapping $ 
x^\mu(\myxi^a) $ but treats tetrad $ \F{\mu}{a} $ as non-holonomic from the 
very beginning. This way of deriving governing equations in the framework of 
the Riemann-Cartan geometry was discussed recently in \cite{PRD-Torsion2019} in 
the non-relativistic settings. However, for the first attempt to obtain 
relativistic version of the SHTC equations we follow the first route explained 
above as being more simple.

\subsection{Eulerian and Lagrangian { viewpoints}} 
\label{sec.EulLagFrames}
Let us consider a spacetime manifold $\spacetime$ equipped with an arbitrary 
curvilinear coordinate
system $ x^\mu $
and a Riemannian  metric $ 
g_{\mu\nu}(x^\lambda)$ with a signature $ (-,+,+,+) $, i.e. it is implied that the zeroth 
coordinate $ x^0 $ is the coordinate time %(i.e. its coordinate basis vector is timelike)
and also will be denoted as $ x^0 = t $ (we adopt a timescale for which the light speed is  $ c = 1 
$). The objects 
related to $ \spacetime $ have indices which are Greek alphabet letters $\alpha,\beta,\dots, 
\lambda, \mu,\nu, \dots $.
%
%
%Let us 
%also consider a continuum which occupies,
%instant $ \ox{t}{} $, a three-dimensional \textit{matter} manifold with 
%coordinates  $ \ox{x}{A} 
%$, $ A=1,2,3 $. 

Let us then consider the 4-continuum which is the collection of all the material particle 
worldlines. Thus, the 4-continuum is a 4-dimensional \textit{matter-time}
manifold $ \mattertime $ \textit{embedded} into $ \spacetime $. The objects related to $ 
\mattertime $ have indices 
which are Latin letters $ a,b,c, \ldots =0,1,2,3$, while capital Latin letters, 
e.g. $ \sA,\sB,\sC,\ldots = 1,2,3 $, denote pure material components of the 
matter-time tensors. The 4-continuum $ 
\mattertime $ is  
parametrized by its own (different and independent from $ x^\mu $) 
coordinate system $ 
\myxi^a 
=(\myxi^0,\myxi^\sA) := 
(\proptime ,\myxi^\sA)$, where the three scalars $\myxi^{\sA} $ \textit{label} the 
matter particles and hence label the particle worldlines, while $ \myxi^0 :=\proptime  $ is defined 
to 
be the { matter} \textit{proper time}, that is the time of 
the Lagrangian observer 
{  which is 
comoving with the matter} as measured from his comoving clock
(should not be confused with the strain dissipation time $\tau$ used in Section~\ref{sec.Intro} and other 
sections), i.e.
\begin{equation}\label{eqn.proper.time}
-\rmd \proptime ^2 = g_{\mu\nu} \rmd x^\mu \rmd x^\nu. 
\end{equation}

In continuum mechanics, the coordinates $ \myxi^a $ on the matter-time manifold 
$ \mattertime $ are called 
the \textit{Lagrangian coordinates} of the 4-continuum and are associated to a 
\textit{Lagrangian observer}, which is 
\textit{comoving and co-deforming} with the medium. On the other 
hand, the coordinate system $ x^\mu $ of the spacetime manifold $ \spacetime $ 
is 
associated to an observer which is 
\textit{not comoving with the matter}. Such a
coordinate system is called \textit{Eulerian coordinate system}.
%\sout{{ in the Newtonian 
%continuum mechanics}}
%\footnote{Not only in Newtonian mechanics.    \sout{In the relativistic 
%literature, the Eulerian observer is associated with an observer which moves 
%relative to both, the Lagrangian coordinate system $ \myxi^a $ and the 
%Eulerian coordinate system $ 
%x^\mu $. Such an observer is not required for the theory formulation but will 
%emerge in the 3+1 numerical formalism in Section~\ref{sec.Numerics}. Until 
%then, we shall use only two types of observers, the Lagrangian which is 
%identified with the coordinates system $ \myxi^a $ and the Eulerian which is 
%identified with the Eulerian system $ x^\mu $.}}. 
The 
Lagrangian and 
Eulerian coordinates of continuum particles are related to each other in a 
\textit{one-to-one manner} by 
the mappings, e.g.~\cite{Sedov1971eng,Maugin1974},
\begin{equation}\label{eqn.motion.laws1}
x^\mu = x^\mu(\myxi^a), \qquad \myxi^a = \myxi^a(x^\mu).
\end{equation}

This conventional formulation of continuum mechanics is essentially a 
deformation theory, that is the stresses in matter depend on the strain state 
of the medium. In order to measure strains 
in a material body, one has to be able to measure 
distances between {  labeled }
material points, and hence one needs a {  material} 
metric. Thus, it is necessary to remark that the role of 
the fields $ \myxi^a $ is to merely label the trajectories and they do not 
relate to the geometry of the spacetime $ \spacetime $. This means that we are 
free to choose the metric $ \Kab{ab} $ of the matter-time manifold $ 
\mattertime 
$ which  is a non-dynamical parameter of the theory. For example, $ \Kab{ab} $ 
can be set to be flat even though the spacetime metric $ g_{\mu\nu} $ has a 
non-vanishing curvature.
Moreover, despite using the 4-dimensional formalism, the 4-continuum $ 
\mattertime $ should not 
be treated as a general spacetime but it has a very certain structure. 
In particular, the most important feature of its structure is that the time and 
matter 
dimensions of $ \mattertime $ cannot be mixed. Each 3-dimensional slice 
of the matter-time manifold corresponding to $ 
\proptime  = const $ represents a 3-dimensional matter manifold $ \matter $ 
consisting of exactly the same particles (molecules) which have constituted the 
matter at $ \tau = 0 $. In addition, because the time in $ \mattertime $ is the 
proper time 
$ \tau $, the time dimension $ \myxi^0 $ is not curved (the time is absolute in 
$ \mattertime $). For example, it is 
usually convenient and natural to define $ \Kab{ab} $ to be globally flat 
(i.e. $ \myxi^a $ is a Cartesian coordinate system), as we assume 
in this paper.
However, in general, the matter 3-metric $ \KAB_{\sA\sB} $ (the matter 
components of $ 
\Kab{ab} $) can be non-Euclidean, e.g. in the case if the matter is an 
elastoplastic solid that suffered from plastic deformations in the 
past~\cite{GodRom1972,God1978,GodRom2003,Katanaev1992}, or can be Euclidean 
(spatially flat) but 
non-constant which 
can be conditioned by the geometry of the problem, see numerical examples in 
Section~\ref{sec.numerical.examples}. 
%Thus, we shall 
%formally demonstrate later, in Section~\ref{sec.EulerPDEs}, 
%that the developed theory is 
%invariant with respect to the change of the Lagrangian coordinates $ \myxi^a 
%\rightarrow 
%\myxi^{a'}$ preserving 4-volume ($ \det(\pd \myxi^a/\pd \myxi^{a'}) = 1 $) and 
%hence with respect to the matter-time metric $ \Kab{ab} $. 
Thus, in 
general, the metric $ \Kab{ab} $ may vary from point to point, i.e. $ 
\Kab{ab} 
= \Kab{ab}(\myxi^c) $,
%\footnote{\Francesco{
%should we make explicit the coordinate dependence of $\Kab{ab} = 
%\Kab{ab}(\xi^\lambda)$?
%\IP{Yes, it can vary from point to point in general. So, I added the 
%dependence.}
%}}
and the most general admissible structure of $ \Kab{ab} $ is
\begin{equation}\label{eqn.gab}
\Kab{ab} = \left(\begin{array}{rccc}
-1 & 0 & 0 		& 0\\ 
 0 &   &        &  \\ 
 0 &   & \KAB_{\sA\sB} & \\
 0 &   &   		& 
\end{array}\right).
\end{equation}

{ The } local { Lagrangian  observer is not able to recognize if the 
matter is deforming or not because the Lagrangian lengths given by the metric $ 
\Kab{ab} $ are constant along the trajectories. Therefore, any 
deformation 
theory needs at least two observers, the} local comoving {  (Lagrangian) and} a 
non-comoving {, 
\eg the Eulerian observer, in order to measure the relative length 
changes. In the following section, we thus proceed with the introduction of 
fields which allow us to make such measurements.
}

\subsection{4-Jacobians, 4-velocities, frame of reference, proper strain 
measure}\label{sec.4Jac}

\subsubsection{4-Jacobians}

Let us now introduce the very important fields of our theory, the 4-Jacobians,
\begin{equation}\label{eqn.Jacobians}
\F{\mu}{a} := \frac{\pd x^\mu}{\pd \myxi^a}, \qquad \cgrad{a}{\mu} := \frac{\pd 
\myxi^a}{\pd x^\mu} 
\end{equation}
with the obvious orthogonality properties:
\begin{equation}\label{eqn.Jac.mult}
\cgrad{a}{\mu} 
\F{\mu}{b} = \delta^a_{\ b}, \qquad \F{\mu}{a}\cgrad{a}{\nu} = \delta^\mu_{\ 
\nu},
\end{equation}
where $ \delta^a_{\ b} $ and $ \delta^\mu_{\ \nu} $ are the Kronecker deltas in 
the Lagrangian and Eulerian frames, accordingly. In the relativistic elasticity 
literature~\cite{Kijowski1992,Wernig-Pichler2006,Broda2008,Gundlach2012}, $ 
\myxi^a(x^\mu) $ is called the \textit{configuration } and the 
Jacobian $ \cgrad{a}{\mu} $ is the \textit{configuration gradient}. We shall 
also use this name 
here. { Usually, it is convenient to define the coordinate system $ 
\myxi^a $ identical to $ x^\mu $ so that initially, at $ \tau = 0 $, one has $ 
\F{\mu}{a} = \text{diag}(1,1,1,1) $  and $ \cgrad{a}{\mu} = 
\text{diag}(1,1,1,1) $. We emphasize that even in this case, we are free to set 
$ \Kab{ab} $ to be flat despite $ g_{\mu\nu} $ may have a non-vanishing 
curvature. The \textit{relaxed} or \textit{unstressed} state of a material element is then can be 
identified with $ \cgrad{a}{\mu}(x^\mu) = R^a_{\ \mu} $, where $ R^a_{\ \mu} = 
const $ is an 
orthogonal spatial transformation with respect to $ x^\mu $.}

\subsubsection{4-velocities}

Furthermore, it is implied that the Lagrangian coordinates $ \myxi^{\sA }$ and $ 
\proptime  $ are 
independent variables which expressed in that the Lagrangian 4-velocity (the velocity as measured 
with respect to the Lagrangian coordinate system $ \myxi^a $) is given by
\begin{equation}\label{eqn.Lagr.4vel}
\uLag^a := \frac{\pd \myxi^a}{\pd \tau} = (1,0,0,0).
\end{equation}
Also, we define the 4-velocity of the material elements with 
respect to the Eulerian coordinate system $ x^\mu $ as
\begin{equation}\label{eqn.4vel.def}
u^\mu := \F{\mu}{0} = \frac{\pd x^\mu}{\pd \myxi^0} = \frac{\pd x^\mu}{\pd \proptime },
\end{equation}
i.e. it is the tangent to the worldline of the Eulerian 
observer span by the 
parameter $ \proptime  $.  Note that the Lagrangian 4-velocity $ \uLag^\mu = 
\F{\mu}{a}\uLag^a $, if written in the Eulerian frame, and $ u^\mu $ 
are related by the identity
\begin{equation}\label{eqn.4vel.relation}
u^\mu \equiv \uLag^\mu.
\end{equation}
Therefore, in the rest of the paper, we shall not distinguish between them and 
will always write $ u^\mu $.

The situation is different with respect to the covariant components of the 
4-velocity. The covariant components can be introduced in two ways. The 
Lagrangian definition is that the 4-velocity $ \bm{u} $ is decomposed in the 
Lagrangian cobasis $ \bm{\rmd \myxi}^a $
\begin{equation}\label{U.cov.Lag}
	\bm{u} = U_a \bm{\rmd \myxi}^a, 
	\qquad 
	\text{or} 
	\qquad 
	\uLag_a := \Kab{ab}\uLag^b = (-1,0,0,0),
\end{equation}
or, if written in the Eulerian frame,
\begin{equation}\label{U.cov.Eul}
\uLag_\mu = \cgrad{a}{\mu}\uLag_a = -\cgrad{0}{\mu}.
\end{equation}

On the other hand, the standard spacetime definition (Eulerian definition) is 
that the 4-velocity $ \bm{u} $ is decomposed in the spacetime cobasis $ 
\bm{\rmd x}^\mu $
\begin{equation}\label{u.cov.Eul}
	\bm{u} = u_\mu \bm{\rmd x}^\mu, 
	\qquad 
	\text{or} 
	\qquad u_\mu := g_{\mu\nu} u^\nu.
\end{equation}
This gives covariant components $ u_\mu $ that, in general, are not equal to $ 
\uLag_\mu $ because, one may write
\begin{equation}\label{U.cov.Eul2}
\uLag_\mu = \cgrad{a}{\mu}\uLag_a = \cgrad{a}{\mu}\Kab{ab}\uLag^b = 
\cgrad{a}{\mu}\Kab{ab}\cgrad{b}{\nu}\uLag^\nu = \Kab{\mu\nu}\uLag^\nu =
\Kab{\mu\nu} u^\nu.
\end{equation}

We note that despite the ambiguity in the definition of the covariant 
components of the 4-velocity, both definitions satisfy the normalization 
condition
\begin{equation}\label{eqn.4vel.normal}
u^\mu u_\mu = -1,
\qquad
\uLag^\mu \uLag_\mu = -1
\end{equation}
i.e. $ u^\mu $ ($= \uLag^\mu $) is normalized and timelike. 
Nevertheless, because we are building an Eulerian description of the continuum, 
we shall use the Eulerian definition \eqref{u.cov.Eul} for the covariant 
components $ u_\mu $ of the 4-velocity in the rest of the 
paper.

%the definition of the proper time~\eqref{eqn.proper.time} and 4-velocity $ 
%u^\mu $, it also follows 
%that 

It is useful to write the Jacobians $ \F{\mu}{a} $ and $ \cgrad{a}{\mu} $  
explicitly in order to emphasize that the first 
column of $ \F{\mu}{a } $ and the first row of $ \cgrad{a}{\mu} $ are 
$ u^\mu $ and $ -\uLag_\mu $, accordingly:
\begin{equation}\label{eqn.Jac}
\F{\mu}{a} =  \left( \begin{array}{cccc}
 u^0 & \F{0}{1} & \F{0}{2} & \F{0}{3} \\ 
 u^1 & \F{1}{1} & \F{1}{2} & \F{1}{3} \\ 
 u^2 & \F{2}{1} & \F{2}{2} & \F{2}{3} \\ 
 u^3 & \F{3}{1} & \F{3}{2} & \F{3}{3}
\end{array} 
\right),
\quad
\cgrad{a}{\mu} =  \left( \begin{array}{rrrr}
-\uLag_0 & -\uLag_1 & -\uLag_2 & -\uLag_3 \\ 
\cgrad{1}{0} & \cgrad{1}{1} & \cgrad{1}{2} & \cgrad{1}{3} \\ 
\cgrad{2}{0} & \cgrad{2}{1} & \cgrad{2}{2} & \cgrad{2}{3} \\ 
\cgrad{3}{0} & \cgrad{3}{1} & \cgrad{3}{2} & \cgrad{3}{3}
\end{array} 
\right),
\end{equation}
where, in general, $ \F{0}{\sA} = \frac{\pd x^0}{\pd \myxi^{\sA}} = \frac{\pd 
t}{\pd 
\myxi^{\sA}} \neq 0 
$ which expresses the non-absoluteness of the  coordinate
time
%
%\footnote{\Francesco{ We probably should define somewhere 'Eulerian time' as the '\textbf{time of 
%the coordinates}'. We may solve this ambiguity defining at the beginning of the text or in the 
%footnotes:\\
% %
% i ) 'Eulerian time' is intended as the time coordinate in the standard $3+1$ foliation with 
%Eulerian observers. It has to be distinguished from the 'proper time of the Eulerian observer' but 
%also from the 'Lagrangian time coordinate' (for the Lagrangian coordinates we have 
%$\bm{t}_\Lag\equiv \bm{u}$ (and $t\equiv \tau \equiv \tau_0$));
% %
%} \IP{May be we should call $ x^0 =t $ the \textit{coordinate time}, i.e. it 
%is just the 
%timelike component
%within the $ x^\mu $? I also added a comment about this in the first paragraph
%of Section~\ref{sec.EulLagFrames}}
%}
%
$ x^0 = t $ for different Lagrangian observers $ \myxi^{\sA} $.

An important \textit{orthogonality condition} for the matter components $ 
\cgrad{\sA}{\mu} 
$ of the 4-Jacobian $ \cgrad{a}{\mu} $ immediately follows from 
\eqref{eqn.Jac.mult}$ _1 $ 
and the definition of the 4-velocity \eqref{eqn.4vel.def}
\begin{equation}\label{eqn.orthogon.cond}
\cgrad{\sA}{\mu} u^\mu = 0.
\end{equation}
%Note that there are no reasons 
%for the orthogonality of $ \F{\mu}{\sA} $ and $ u_\mu $ because $ \F{\mu}{\sA} 
%u_{\mu} = 
%\F{\mu}{\sA} 
%g_{\mu\lambda} u^\lambda = \F{\mu}{\sA} g_{\mu\lambda} x^\lambda_{\ 0}$, where 
%the right-hand side 
%is 
%not necessarily zero in general.

\subsubsection{Local relaxed reference frame}

{ We are now in the position to introduce a material 
strain measure in order to measure relative changes in the } material distances.
%\Francesco{Actually, 
%particles do not belong to the same time-slice $\Sigma_\tau$, but the distance is projected along 
%$\Sigma_\tau$.} 
In fact, the 4-Jacobians $ \F{\mu}{a} $ and $ \cgrad{a}{\mu} $ already contain all
the necessary information about the relative change of the Eulerian displacements $ \rmd x^\mu = 
\F{\mu}{a} \rmd \myxi^a$ 
with respect 
to the Lagrangian coordinate increments $ \rmd \myxi^a $ which is pretty enough to build a 
deformation theory. 
However, we still need to define a proper strain measure because, as will be discussed later, the 
energy potential of the matter should be a Lorentz scalar and thus, it may depend only on 
invariants of a rank 2 spacetime tensor properly constructed from $ 
\cgrad{a}{\mu} $ or $ \F{\mu}{a} $ 
(recall that the 
Jacobians $ \cgrad{a}{\mu} $ and 
$ \F{\mu}{a} $ transforms as covariant and contravariant spacetime vectors, 
respectively). 
Moreover, while measuring the 
material lengths with respect to an observer which is not co-moving with the matter, one needs also 
to avoid the 
effect of the Lorentz length 
contraction and hence, such measurements should be performed in the 
\textit{material element rest frame}.
In the relativistic dynamics of perfect fluids, the material element rest frame is defined as a 
frame, e.g. a \textit{basis tetrad} of four vectors $ \{\boldsymbol{e}_\mu\} $, whose 
space-like 
vectors $ \boldsymbol{e}_i $ ($ i=1,2,3 $)
are arbitrary but orthogonal to the material 4-velocity $ u^\mu $ and the 
time-like vector $ \boldsymbol{e}_0 $ is tangent to and co-directional with $ u^\mu $. 
However, in a strain-based theory, such an arbitrariness in the choice of the space-like vectors of 
the rest frame can be naturally overcome. In fact, the triad $ \boldsymbol{e}_i 
$ has definite directions 
which are conditioned by the choice of the so-called \textit{local relaxed 
reference frame} (LRRF) $
\{\boldsymbol{e}_a\} $, which is an orthonormal basis tetrad attached to each material 
element\footnote{It should be well 
understood that the prescription of the directions via the choice of $
\{\boldsymbol{e}_a\} $ is possible without the loss of generality. In general, the orientations of 
the directions of $\{\boldsymbol{e}_\mu\} $ are different from $\{\boldsymbol{e}_a\} $ but the 
rotation embedded in $ \F{\mu}{a} $ take this into account.}. Such a frame is 
associated with the relaxed (stress-free) state of the matter. In the simplest 
case, the LRRF 
$ 
\boldsymbol{e}_a $ can be identified with the coordinate basis, i.e. $ \boldsymbol{e}_a = 
\frac{\boldsymbol{\pd}}{\boldsymbol{\pd}\myxi^a} = \boldsymbol{\pd}_a$. The rest frame $ 
\boldsymbol{e}_\mu $ is then given by $ \boldsymbol{e}_\mu = \cgrad{a}{\mu} \boldsymbol{e}_a $. 
However, the association of the LRRF $ \boldsymbol{e}_a $ with the coordinate 
basis $ 
\boldsymbol{\pd}_a $ relies on the definition of the Lagrangian coordinates and hence, it 
is very restrictive if one wants to deal with irreversible deformations. 
Therefore, the concept of the coordinate associated \textit{holonomic} frame, 
i.e. $ \boldsymbol{e}_a = 
\boldsymbol{\pd}_a$, 
will be replaced by the concept of \textit{non-holonomic} frame in 
Section~\ref{sec.Irreversible} 
which cannot be associated to a global coordinate system $ \myxi^a $. 
%For further 
%understanding of modeling of irreversible deformations in 
%Section~\ref{sec.Irreversible}, it should 
%be now well understood that there is no need to explicitly introduce the frame 
%$ 
%\{\boldsymbol{e}_a\} $ because it is implicitly introduced via 
%the definition of the fields $ \F{\mu}{a} $ or $ \cgrad{a}{\mu} $ which thus 
%will be also 
%referenced to as the tetrad fields.

\subsubsection{Proper strain measure}

The material strain measure can be introduced independently of whether holonomic or 
non-holonomic LRRF is used. Thus, we proceed by defining the Lagrangian 
matter 
metric $\LagrProj_{ab} $ on $ \mattertime $ as the projection of the Lagrangian 
matter-time metric $ \Kab{ab} $ onto the three-dimensional matter space:
\begin{equation}\label{eqn.projectorMT}
\LagrProj_{ab} := \Kab{ab} + \uLag_a \uLag_b =
%\left(\begin{array}{rccc}
%-1 & 0 & 0 & 0\\ 
% 0 &   &   &  \\ 
% 0 &   & \KAB_{\sA\sB} & \\
% 0 &   &   &  
%\end{array}\right) 
%+ 
%\left(\begin{array}{rccc}
% 1 & 0 & 0 & 0 \\ 
% 0 & 0 & 0 & 0 \\ 
% 0 & 0 & 0 & 0 \\
% 0 & 0 & 0 & 0 
%\end{array}\right) 
%= 
\left(\begin{array}{rccc}
 0 & 0 & 0 & 0\\ 
 0 &   &   &  \\ 
 0 &   & \KAB_{\sA\sB} & \\
 0 &   &   &  
\end{array}\right),
\end{equation}
%where $ u_a := \Kab{ab} u^b = (-1,0,0,0) $, $ u^a := 
%\pd_\tau \myxi^a = (1,0,0,0)$ is the\textit{ Lagrangian material
%4-velocity} introduced in \eqref{eqn.Lagr.4vel}.
%\footnote{ The 
%Lagrangian 
%material 4-velocity relates to the 4-velocity $ u^\mu $ of the same Lagrangian 
%observer with respect to the Eulerian coordinate system as $ u^\mu = \F{\mu}{a} 
%u^a $}.  
or, in the Eulerian frame,
\begin{equation}
\G_{\mu\nu}(x^\lambda) := \LagrProj_{ab} \cgrad{a}{\mu}\cgrad{b}{\nu} = 
\KAB_{\sA\sB} 
\cgrad{\sA}{\mu}\cgrad{\sB}{\nu}, \qquad 
\det(\G_{\mu\nu}) = 0, \qquad \G_{\mu\nu} u^\mu = 0. \label{eqn.G}
\end{equation}
The property \eqref{eqn.G}$ _3 $ follows from \eqref{eqn.orthogon.cond} { and it says 
that $ 
\G_{\mu\nu} 
$ is the proper strain measure because it gives the material length as measured in the local 
material 
element rest frame.} In the same way, the material metric will be introduced 
in Sec.\,\ref{sec.Irreversible} with the only difference that the holonomic 
tetrad 
$ \cgrad{a}{\mu} $ will be replaced by non-holonomic tetrad (4-distortion).

Also, note that despite $ \G_{\mu\nu} $ is orthogonal 
to the 4-velocity $ u^\mu $, it cannot be used to project spacetime tensors 
onto the material element rest frame because it gives not the spacetime spatial 
distances but the material distances. Recall that in order to measure the 
length of a spacetime 4-vector in the rest frame, one has to use the \textit{comoving-spatial 
metric}
\begin{equation}\label{eqn.h}
h_{\mu \nu } := g_{\mu \nu }+u_{\mu }u_{\nu }, \qquad g^{\mu\nu}h_{\mu\nu} = 3, \qquad h_{\mu\nu} 
u^\mu = 0,
\end{equation}
which is induced by the spacetime metric $ g_{\mu\nu} $.
{ The rest frame spatial distances measured with $ h_{\mu\nu} $ and $ 
\G_{\mu\nu} $ are different in general. Thus, by comparing $ h_{\mu\nu} $ and $ 
\G_{\mu\nu} $, \ie $ (h_{\mu\nu} - \G_{\mu\nu})/2 $,}
%\begin{equation}\label{eqn.strain.measure}
%\frac{1}{2}(h_{\mu\nu} - \G_{\mu\nu}),
%\end{equation}
one { may conclude about the relative length changes in the matter.}
The bigger this difference, the bigger the strain in the matter. 

In what follows, the projectors
\begin{subequations}\label{eqn.hmunu}
	\begin{align}
		h^\mu_{\ \lambda} &:= g^{\mu\lambda}h_{\lambda\nu} = \delta^\mu_{\ \nu} + u^\mu 
		u_\nu,\\[2mm]
		h^{\mu\nu} &:= 
		g^{\mu\lambda} h_{\lambda\gamma} g^{\gamma\nu} = g^{\mu\nu} + u^\mu u^\nu
	\end{align}
\end{subequations}
with the properties
$%\begin{equation}\label{eqn.h.operators}
h^\mu_{\ \lambda }h^\lambda_{\ \nu }= h^\mu_{\ \nu}$, $ h^{\mu\lambda} 
h_{\lambda\nu} = 
h^\mu_{\ \nu}
$%\end{equation}
will be also used along side with the tensor $ h_{\mu\nu} $.

%Also, note that the projectors $ h_{\mu\nu} $, $ h^{\mu\nu} $ and $ h^\mu_\nu $ 
%can be used in a line with $ g_{\mu\nu} $ for 
%rising  and lowering indices of $ \Gmunu $ and to construct invariants 
%because of 
%\begin{equation}\label{eqn.orth.G}
%\Gmunu u^\nu = 0,
%\end{equation}
%which automatically follows from~\eqref{eqn.G} and \eqref{eqn.orthogon.cond}. Therefore, 
%one may write
%\begin{equation}\label{eqn.hg.rising}
%\G^\lambda_\nu =  g^{\lambda\mu}\Gmunu = h^{\lambda\mu}\Gmunu.
%\end{equation}

\subsection{Equations of motion in the Lagrangian frame}\label{sec.LagrPDE}

By the \textit{equations of motion} we understand not only the Euler-Lagrange 
equations but an extended system of PDEs whose 
solution is one of the 4-Jacobians \eqref{eqn.Jacobians} of the 
mapping~\eqref{eqn.motion.laws1}. Such a system, thus, completely specifies the 
motion of the continuum.

In this section, we derive the equations of motion 
from Hamilton's principle of stationary action. However, having two reference 
frames at 
hand, the Lagrangian $ \myxi^a $ and the Eulerian $ x^\mu $ one, it is 
naturally to question in 
which frame we should perform the variation? Thus, keeping in mind that our 
ultimate goal is to obtain the equations of motion in the Eulerian frame $ 
x^\mu $, one may think about two routes to get these equations. The first 
route, 
is to perform variation directly in the Eulerian settings, while the second 
route is to perform variations in the Lagrangian frame $ \myxi^a $ and then 
change the variables $ \myxi^a \rightarrow x^\mu $. Which route should we  
chose? Or are they equivalent?

It is known that the PDE for the 
energy-momentum tensor which emerges as the Euler-Lagrange equation are indeed 
equivalent in both cases. However, energy-momentum conservation give us only 
four equations which is not enough to define 16 components of $ \F{\mu}{a} $. 
Thus, we also need 12 more 
PDEs which are not the result of the variation
but usually are \textit{integrability conditions}. The fact is that, in the 
Eulerian frame, such a 
PDEs \textit{cannot} be rigorously derived from the definitions of the 
potentials $ 
\myxi^a(x^\mu) $ and usually is \textit{postulated} based on some extra 
reasoning or introduced ``by hand'', e.g. see 
\cite{Gundlach2012,Gavrilyuk2011}. 
On the other side, the 
second 
route, i.e performing the variations in the Lagrangian frame and then applying 
the change of variables $ \myxi^a \rightarrow x^\mu $, allows for an unambiguous
derivation of the PDE for the configuration gradient $ \cgrad{a}{\mu} $ only 
from the definition of the potentials $ x^\mu(\myxi^a) $ and without the 
attraction 
of extra reasoning. Therefore, in what follows, as well as in our Newtonian 
paper~\cite{DPRZ2017,SHTC-GENERIC-CMAT}, we follow the latter route. 

%Thus, in 
%order to derive the equations of motion in the Lagrangian frame $ 
%\myxi^a $, we shall use Hamilton's principle to obtain the Lagrangian 
%evolution 
%equation for the so-called relativistic Piola-Kirchhoff stress (Lagrangian 
%4-stress) and for the Jacobian $ \F{\mu}{a} $. Then, these Lagrangian 
%equations 
%will be transformed into the Eulerian frame by means of the coordinate change 
%$ 
%\myxi^a \rightarrow x^\mu $ in order to obtain the conventional conservation 
%law for the energy-momentum and a PDE for the configuration gradient $ 
%\cgrad{a}{\mu} $.

\subsubsection{Hamilton's principle in the Lagrangian frame}\label{sec.Hamilton.Lagr}

Let us consider the action integral in the Lagrangian frame $ \myxi^a $
\begin{equation}\label{Action.Lagr}
S^{\sLrm} = \int \sqrt{-\kappa}\,
\tilde{\Lambda} \rmd \boldsymbol{\myxi}, \qquad \kappa = \det(\Kab{ab}),
\end{equation}
where the Lagrangian density $ \hat{\Lambda} = \sqrt{-\kappa}\tilde{\Lambda} $
does not depend explicitly on the 4-potential $ x^\mu(\myxi^a) $ 
and coordinates $ \myxi^a $ (due to the requirement of the translational invariance) 
but only on the partial derivatives $ \F{\mu}{b}(\myxi^a) = 
\pd_b\, x^\mu $:
\begin{equation}\label{eqn.Lagrangian}
\hat{\Lambda}(\myxi^a, x^\mu(\myxi^a),\F{\mu}{b}(\myxi^a)) = 
\Lambda(\F{\mu}{b}(\myxi^a)) .
\end{equation}
Hence, the first variation of 
the action $ S^\sL $ gives the Euler-Lagrange 
equations
\begin{equation}\label{eqn.Euler-Lagrange}
\frac{\pd \Lambda_{\F{\mu}{a}}}{\pd \myxi^a} = 0 \qquad \text{or} \qquad
\pd_a\,\Lambda_{\F{\mu}{a}} = 0,
\end{equation}
whose meaning will be clarified below. 
Here, as well as throughout the paper, we use the notation $ 
\Lambda_{\F{\mu}{a}} = \frac{\pd \Lambda}{\pd \F{\mu}{a}} $.

%\IP{Not clear how to insert metric here, which depends on $ x^\mu $... while $ 
%\myxi^a $ represent only 
%{\bf part} of the spacetime occupied by the matter manifold.}

%\IP{Therefore, one may remark that despite we build a fully covariant theory which supposed to be invariant with 
%respect to any coordinate systems and then a question arises: it looks like that the equations written down in the  
%Lagrangian coordinates $ \xi^a $ should have the same form in the Eulerian coordinates. But this is  }

%\IP{It remains to transform \eqref{eqn.Euler-Lagrange} to the Eulerian 
%coordinates $ x^\mu $. How do the covarriant 
%derivatives $ \nabla_\mu $ should appear here not clear though.}

There are only 4 conservation laws in \eqref{eqn.Euler-Lagrange} for 16 
unknowns $ \F{\mu}{a} $. Hence, we need 12 more equations do define all sixteen 
fields $ \F{\mu}{a} $. The remaining 12 evolution equations can be obtained 
from the \textit{integrability conditions} 
\begin{equation}\label{eqn.integrability}
\pd_b\, \F{\mu}{a} - \pd_a\, \F{\mu}{b} = 0
\end{equation}
which are trivial consequences of the definition of $ \F{\mu}{b}(\myxi^a)$. 
There are 24 equations within \eqref{eqn.integrability}, while only 12 of them 
are \textit{evolution} equations, i.e. those for $ a,b = 0 $. The rest 12 are 
pure 
spatial 
constraints and are conserved along the trajectories, they are the so-called 
\textit{involution constraints}, e.g. see~\cite{SHTC-GENERIC-CMAT}. 
Conservation laws 
\eqref{eqn.Euler-Lagrange} together with the integrability conditions 
\eqref{eqn.integrability} form a closed system of 16 PDEs (if $ \Lambda $ is 
specified) for 16 fields $ \F{\mu}{a} $ which we shall 
call the 
\textit{equations of motion} written in the Lagrangian coordinates $ \myxi^a$. 
%Following the 
%Newtonian continuum mechanics, $ \Lambda_{\F{\mu}{a}} $ can be called the 
%\textit{relativistic 
%first Piola-Kirchhoff stress}~\cite{GodPesh2010,Kijowski1998}.

%It should be kept in mind that the object $ \Lambda_{\F{\mu}{a}} $ is not a 
%tensor but transforms like a 4-vector with respect to transformations of the 
%Eulerian coordinates $ x^\mu $. Following the Newtonian continuum 
%mechanics, we shall call it the \textit{relativistic first Piola-Kirchhoff 
%stress} and 
%denote as 
%\begin{equation}\label{eqn.Piola}
%\Piolamix{a}{\mu} :=-\Lambda_{\F{\mu}{a}}.
%\end{equation}

\subsubsection{Symmetric Hyperbolicity of the Lagrangian equations of 
motion}\label{sec.SymHypLagr}

Depending on the Lagrangian $ \Lambda $, system \eqref{eqn.Euler-Lagrange},
\eqref{eqn.integrability} can be a highly nonlinear PDE system. Hence, to be 
physically meaningful, one has to assure that this system is also 
mathematically \textit{well-posed} (Hadamard stability), i.e. solution exists, 
is unique and continuously depends on input data (initial conditions, material 
parameters, etc.). Since we are interested in time evolution of the matter, we 
shall consider the well-posedness property only in the time direction 
(evolutionarity). In the PDE theory, the time dependent PDE systems which are 
well-posed for arbitrary but regular initial data (at $ \proptime  = 0 $) are called 
\textit{t-hyperbolic} systems, i.e. hyperbolic in the time direction.

In what follows, we demonstrate even a stronger well-posedness property of 
system 
\eqref{eqn.Euler-Lagrange},
\eqref{eqn.integrability}, that is this system is, in fact, \textit{symmetric 
t-hyperbolic} in the sense of Friedrichs~\cite{Friedrichs1958} if the 
Lagrangian $ \Lambda(\F{\mu}{a}) $ is a convex potential.
Thus, in order to see this we need to
separate the 
proper time derivative $ \pd_\proptime  = \pd_0 $ from the spatial (matter) 
derivatives $ 
\pd_{\sA} $, $ \sA=1,2,3 $. After that, 4 equations 
\eqref{eqn.Euler-Lagrange}, and those 12 equations \eqref{eqn.integrability} which are evolution 
equations can be written as
\begin{equation}\label{eqn.Lagr.motion}
\pd_\proptime \, \Lambda_{u^\mu} + \pd_{\sA}\, \Lambda_{\F{\mu}{\sA}} = 0, \qquad
\pd_\proptime \, \F{\mu}{\sA} - \pd_{\sA}\, u^\mu = 0.
\end{equation}
Let us now 
introduce new variables $ m_\mu := \Lambda_{u^\mu} $ and a 
new 
potential $ U(m_\mu,\F{\mu}{\sA}) $ as the 
partial Legendre transformation of 
$ \Lambda(u^\mu,\F{\mu}{\sA}) = \Lambda(\F{\mu}{0},\F{\mu}{\sA}) 
=\Lambda(\F{\mu}{a})$ with respect to the 4-velocity, i.e. $ U := u^\mu 
\Lambda_{u^\mu} - \Lambda = u^\mu m_\mu - \Lambda $. Hence, we have 
\begin{equation}
U_{m_\mu} = u^\mu, \qquad U_{\F{\mu}{\sA}} = -\Lambda_{\F{\mu}{\sA}},
\end{equation}
while system \eqref{eqn.Lagr.motion} now reads
\begin{equation}\label{eqn.Lagr.motion.U}
\pd_\proptime \, m_\mu - \pd_{\sA}\, U_{\F{\mu}{\sA}} = 0, \qquad
\pd_\proptime \, \F{\mu}{\sA} - \pd_{\sA}\, U_{m_\mu} = 0.
\end{equation}

As in the non-relativistic case~\cite{DPRZ2017,SHTC-GENERIC-CMAT}, if we multiply 
\eqref{eqn.Lagr.motion.U}$ _1 $ by $ U_{m_\mu} $ and \eqref{eqn.Lagr.motion.U}$ 
_2 $ by $ U_{\F{\mu}{A}} $ and then sum up the results we obtain an 
\textit{extra 
conservation law} 
\begin{equation}\label{eqn.Energy.cons.law}
\pd_\proptime  U - \pd_{\sA} ( U_{m_\mu} U_{\F{\mu}{\sA}}) = 0
\end{equation}
which is fulfilled on solutions to \eqref{eqn.Lagr.motion.U}. Therefore, system 
\eqref{eqn.Euler-Lagrange}, \eqref{eqn.integrability} is the system of 
\textit{conservation laws} accompanied with another conservation 
law~\eqref{eqn.Energy.cons.law}. Hence, according to the Godunov-Boilat 
theorem~\cite{God1961,Boillat1974,SHTC-GENERIC-CMAT}, it is a symmetric 
hyperbolic system if $ U $ is a convex potential. Indeed, in terms of the 
variables $ u^\mu = U_{m_\mu} 
$ and $ \pi^{\sA}_{\ \mu} = U_{\F{\mu}{\sA}} $ and the potential $ L(u^\mu,\pi^{\sA}_{\ 
\mu}) := m_\mu U_{m_\mu} + \F{\mu}{\sA} U_{\F{\mu}{\sA}} - U = m_\mu u^\mu + \F{\mu}{\sA} 
\pi^{\sA}_{\ \mu} - U$, it can 
be 
rewritten~\cite{SHTC-GENERIC-CMAT,GodPesh2010} in the Godunov form similar to 
\eqref{intro.God.form} and then in a symmetric quasilinear form.
Hence, we can be sure that despite the nonlinearity of system 
\eqref{eqn.Euler-Lagrange}, \eqref{eqn.integrability}, the initial value 
problem (the Cauchy problem) for it is well-posed locally in time if $ U(m_\mu,\F{\sA}{\mu}) $ is 
convex.

Finally, we note that the physical meanings of the Lagrangian~$ \Lambda $ and the potential $ U $ 
are 
yet hidden. So far, we may only 
try to 
establish a connection between $ \Lambda $ and $ U $ using an analogy with the non-relativistic 
continuum mechanics~\cite{DPRZ2017,SHTC-GENERIC-CMAT}. Thus, the 4-vector $ m_\mu $ is Legendre 
conjugate to the 4-velocity $ u^\mu $ and hence it can be given the meaning of the 4-momentum, i.e. 
$ m_\mu = \rho_0 u_\mu $, where $ \rho_0 $ is the axiomatically given \textit{reference matter} 
density (should not be confused with the rest matter density $ \rho $ introduced in 
Section~\ref{sec.matter.current}),
then, from the definitions of $ \Lambda $ and $ U $, 
we obtain 
that
$ U = u^\mu \Lambda_{u^\mu} - \Lambda =  u^\mu \rho_0 u_\mu - \Lambda = -\rho_0 
- \Lambda $. Therefore, they relate as
%\begin{equation}\label{eqn.Lambda.meaning}
$-\Lambda = \rho_0 + U$.
%\end{equation}

%and hence, the Lagrangian $ -\Lambda = \rho_0 + U $ can be identified with the \textit{potential 
%energy}, while the first term on the right-hand side of \eqref{eqn.Lambda.meaning} can be 
%associated with the \textit{rest energy}. Thus, the conservation law \eqref{eqn.Energy.cons.law} 
%should be treated as the total energy conservation while the zeroth ($ \mu = 0 $) conservation law 
%in \eqref{eqn.Lagr.motion.U}$ _1 $ should be treated as the conservation of the rest energy.

%\IP{Is it correct to call $ \Lambda $ the potential energy? I don't see what 
%else it can 
%be...}

\subsection{Governing equations in the Eulerian frame, Lagrange-to-Euler transformation} 
\label{sec.Lagr2Euler}

The most striking feature of the Lagrangian formalism is that it is a pure 
\textit{material description}
%\footnote{We recall that the partial derivatives $ 
%\pd_a $ are pure material derivatives but not the spacetime derivatives.}
and the Lagrangian observer, living in $ \mattertime $, experiences the gravity effect in 
a very peculiar way. Indeed, in the Einstein general relativity theory, the gravity is associated 
with the curvature of the spacetime which results in the use of the covariant partial derivatives $ 
\pd_\mu \rightarrow \nabla_\mu $ in the Eulerian field equations. On the other hand, the material 
(Lagrangian) coordinates $ \myxi^a $ and partial derivatives $ \pd_a $ are not 
related to the 
spacetime at all. This, 
in particular, means that the Lagrangian metric $ \Kab{ab} $ can be taken flat 
even though the 
spacetime has a non-vanishing curvature. So, how the Lagrangian observer actually feel the gravity 
being in a flat material manifold $ \mattertime $? In fact, the gravity field $ g_{\mu\nu} $ 
is taken into account in the Lagrangian description implicitly in $ \Lambda_{\F{\mu}{a}} $ as 
explained in 
Appendix~\ref{app.Lagr.gravity}. In order to appreciate the gravity effect in an explicit manner, 
we need to leave the material manifold and observe the motion from an Eulerian 
observer standpoint. 

Thus, in this section, we shall transform equations of 
motion~\eqref{eqn.Euler-Lagrange} 
and \eqref{eqn.integrability} written for the Lagrangian observer
into the frame of an Eulerian observer. We shall also see that 
the 
Lagrangian conservation laws 
\eqref{eqn.integrability} transform into the covariant conservation laws for 
the canonical energy-momentum 
tensor appearing as the result of variation of the matter part of the  
Hilbert-Einstein action.

\subsubsection{Matter current and rest matter density}\label{sec.matter.current}

In relativistic elasticity literature, 
e.g.~\cite{Kijowski1990,Kijowski1992,Kijowski1998,Wernig-Pichler2006,Broda2008,Gundlach2012}, the 
matter current $ J^\mu $ is introduced first, 
while the 4-velocity is defined as a vector proportional to $ J^\mu $. 
This is because 
the authors of the mentioned papers prefer to use not the full configuration 
gradient $ \cgrad{a}{\mu} $ but only its matter components $ \cgrad{A}{\mu} $. 
However, the use of the full 4-Jacobians $ \cgrad{a}{\mu} $ and 
$ \F{\mu}{a} $ provides a more natural and unified (with Newtonian mechanics) 
way to introduce the 
4-velocity as a time derivative of the motion, $ u^\mu = \pd_\proptime  x^\mu $, 
see~\eqref{eqn.4vel.def}. The matter 
current, and mass density then can be introduced via the 4-velocity. We now 
show that, in fact, both routes are equivalent.

Using the formulas $ w:=\det(\cgrad{a}{\mu}) $, $ \pd w/\pd \cgrad{a}{\mu} = w \, \F{\mu}{a} $, $ 
\pd 
\F{\mu}{a}/\pd \cgrad{b}{\lambda} = -\F{\mu}{b} \F{\lambda}{a} $, and $ \pd_\nu \cgrad{a}{\mu} - 
\pd_\mu \cgrad{a}{\nu} = 0$, one can prove that 
\begin{equation}\label{eqn.divF}
\pd_\mu(w \, \F{\mu}{a}) = w \, \F{\mu}{a} \F{\lambda}{b} (\pd_\mu 
\cgrad{b}{\lambda} - 
\pd_\lambda 
\cgrad{b}{\mu}) = 0, 
\end{equation}
and in particular that (for $ a = 0 $)
\begin{equation}\label{eqn.mass.cons1}
\pd_\mu (w u^\mu) = 0.
\end{equation}
The latter equation together with $ u^\mu \pd_\mu \rho_0 = \pd_\tau \rho_0 = 0$ implies the 
conservation law
\begin{equation}\label{eqn.mass.cons2}
\pd_\mu(\sqrt{-g}\rho u^\mu) = 0,
%\quad \text{\Francesco{here we should want  $\pd_\mu (\sqrt{-g} \rho u^\mu)$}}
\end{equation}
where we have define $ \rho_0 $ as the material constant with the meaning of the reference mass 
density (moles                    
per unit volume of the relaxed material), while we also have defined $ \rho := 
\rho_0 w/\sqrt{-g}$ as 
the 
rest frame matter density, and $ g 
= \det(g_{\mu\nu}) $ is the determinant of the spacetime metric. Note that $ \rho $ transforms as a 
scalar but not as a scalar density under the coordinate change $ x^\mu\rightarrow x^{\mu'}$ 
because $ w $ and $ \sqrt{-g} $ are scalar densities of weight $ W=1 $. Hence,
the matter current
\begin{equation}\label{matter.current}
J^{\mu}:=\sqrt{-g}\rho u^\mu
\end{equation}
transforms as a tensor density of weight $ W=1 $.
%
%Note that $ w = \det(\cgrad{a}{\mu}) $ is scalar density of weight $ W = 1 $. Hence, $ w 
%\F{\mu}{a} 
%$ transforms 
%as a vector density of weight $ W=1 $. 
Recall that the covariant derivative of an arbitrary vector 
density $ A^\mu $ of weight $W$ is 
\begin{equation}
\nabla_\nu A^\mu = \pd_\nu A^\mu + \Gamma^\mu_{\ \nu\lambda}A^\lambda - W \Gamma^\lambda_{\ 
\nu\lambda} A^\mu,
\end{equation}
where $ \nabla_\mu $ is the conventional covariant derivative associated with the Levi-Civita 
connection.
In particular, for $ W=1 $, due to the fact that the Levi-Civita is 
torsion-free connection, i.e. $ 
\Gamma^\mu_{\ \lambda\nu} = \Gamma^\mu_{\ 
\nu\lambda} $, one obtains that the covariant divergence $ \nabla_\mu A^\mu $ equals to the 
ordinary 
divergence 	
\begin{equation}\label{cov.div.gen}
\nabla_\mu A^\mu = \pd_\mu A^\mu.
\end{equation}
Therefore, \eqref{cov.div.gen} and \eqref{eqn.mass.cons2} imply the covariant rest mass density
conservation law $ \nabla_\mu(\sqrt{-g}\rho u^\mu) = 0 $, and eventually thanks to the fact that th 
Levi-Civita connection is metric compatible $ 
\nabla_\lambda g_{\mu\nu} = 0 $, one obtains the conventional form of the rest mass density 
conservation 
law
\begin{equation}\label{eqn.mass.cons3}
    \nabla_\mu(\rho u^\mu) = 0.
\end{equation}

Finally, we note that the rest mass density conservation law \eqref{eqn.mass.cons3} was derived 
based on the assumption that $ \cgrad{a}{\mu} $ is torsion-less, i.e. $ \pd_\nu\cgrad{a}{\mu} - 
\pd_\mu\cgrad{a}{\nu} = 0$. However, our goal is to deal with irreversible deformations and this 
condition has to be relaxed. Therefore, later we will also show that equations 
\eqref{eqn.mass.cons3} is in fact the consequence of the 4-distortion evolution equation, see 
Section\,\ref{sec.gov.pdes}.

%\IP{What remains unclear for me in the definitions of $ \rho $ in 
%\eqref{eqn.rho.definition} is 
%that, on one hand, $ \rho $ 
%depends on all $ \cgrad{a}{\mu} $ via $ w $, on the other hand, the right hand 
%side of 
%\eqref{eqn.rho.definition} depends only on spatial components $ \cgrad{A}{\mu} 
%$ via $ J^\mu $. Perhaps it is the consequence of $ u^\mu u_\mu = -1 $....
%}

%From this definition and from the fact that $ 
%\epsilon^{\mu\alpha\beta\gamma}\cgrad{1}{\alpha}
%\cgrad{2}{\beta}\cgrad{3}{\gamma} \cgrad{A}{\mu}$ vanishes as the determinant 
%of a matrix with two identical columns, one may conclude that the matter 
%current is orthogonal to $ \cgrad{A}{\mu} $,
%\begin{equation}\label{eqn.orhtog.current}
%J^\mu \cgrad{A}{\mu} = 0.
%\end{equation}
%Therefore, because of this and \eqref{eqn.orthogon.cond}, the matter current 
%is 
%proportional to the 4-velocity:
%\begin{equation}\label{eqn.Ju}
%J^\mu = \rho\sqrt{-g}\, u^\mu, \qquad g = \det(g_{\mu\nu}),
%\end{equation} 
%where

\subsubsection{Energy-Momentum transformation}\label{sec.EulerPDEs}

We now transform the Lagrangian conservation laws \eqref{eqn.Euler-Lagrange} into the Eulerian 
frame. After the change of the Lagrangian partial 
derivatives with $ \pd_a = 
\F{\nu}{a}\pd_\nu $, \eqref{eqn.Euler-Lagrange} 
becomes
\begin{equation}\label{eqn.EM.transform}
\F{\nu}{a}\pd_\nu \Lambda_{\F{\mu}{a}} = 0.
\end{equation}
Then, identity \eqref{eqn.divF} can be used to rewrite \eqref{eqn.EM.transform} in a conservative 
form
%Indeed, by 
%multiplying  \eqref{eqn.EM.transform} by $ w $ and adding $ 0 = \Lambda_{\F{\mu}{a}}\pd_\nu (w 
%\F{\nu}{a})  $, we obtain that \eqref{eqn.EM.transform} is equivalent to 
\begin{equation}\label{eqn.ener.momen.0}
\pd_\nu (w \, \F{\nu}{a} \Lambda_{\F{\mu}{a}}) = 0.
\end{equation}

Finally, using that $ \pd \cgrad{b}{\lambda}/\pd \F{\mu}{a} = -\cgrad{b}{\mu} \cgrad{a}{\lambda} $,
we can write $  \Lambda_{\F{\mu}{a}} = -\Lambda_{\cgrad{b}{\lambda}} 
\cgrad{b}{\mu} 
\cgrad{a}{\lambda} $, and subsequently one obtains that the Lagrangian equations of 
motion~\eqref{eqn.Euler-Lagrange}, in the Eulerian coordinates $ x^\mu $ 
transforms into %(we have omitted the minus sign here)
\begin{equation}
-\pd_\nu(w \, \cgrad{b}{\mu} \Lambda_{\cgrad{b}{\nu}}) = 0.\label{eqn.divEnMom}
\end{equation}

%\begin{equation}\label{eqn.ener.momen.1}
%\F{\nu}{a}\pd_\nu (\Lambda_{\cgrad{b}{\lambda}} \cgrad{b}{\mu} \cgrad{a}{\lambda}) = 0.
%\end{equation}

We shall see now that the tensor in the brackets in \eqref{eqn.divEnMom} is, 
in fact, equal to the canonical matter energy-momentum tensor of GR. We shall also show that the 
partial derivatives $ \pd_\nu $ in \eqref{eqn.divEnMom} can be replaced with the covariant 
derivatives $ \nabla_\nu $, which are 
the standard covariant derivatives with respect to the torsion-free metric compatible ($ 
\nabla_\lambda g_{\mu\nu} = 0 $) Levi-Civita connection, which can be 
uniquely expressed in terms of the Christoffel symbols $ \Gamma^{\lambda}_{\ 
\mu\nu} 
= \frac{1}{2} g^{\lambda\alpha}\left( \pd_\mu 
g_{\alpha\nu} + 
\pd_\nu g_{\alpha\mu} - \pd_\alpha g_{\mu\nu}\right)
$.

First of all, we recall that the action \eqref{Action.Lagr} under the Lagrange-to-Euler coordinate 
transformation $ \myxi^a \rightarrow x^\mu $ transforms as
\begin{align}\label{Action.Euler}
S^{\sLrm} = \int \Lambda(\myxi^a,x^\mu(\myxi^a),\pd_a x^\mu) \rmd \bm{\xi} & = \nonumber\\
\int w \Lambda (\xi^a(x^\mu),x^\mu,\pd_\mu \myxi^a) \rmd \bm{x} & := S^{\sErm}.
\end{align}
% (for simplicity we assume that the material metric $ g_{\sA\sB} $ has $ 
%\kappa^{\rm m} = \det(g_{\sA\sB})= 1$) 
Motivated by this action transformation, we introduce the potential (scalar-density because $ w $ 
transforms as a scalar-density)
\begin{equation}\label{eqn.E.Lambda}
\LED := w \, \Lambda
\end{equation} 
and hence, \eqref{eqn.divEnMom} now reads
\begin{equation}\label{eqn.divEnMom1}
\pd_\nu \EMTD{\nu}{\mu} = 0, \qquad 
\EMTD{\nu}{\mu} := -\left (\cgrad{a}{\mu} \LED_{\cgrad{a}{\nu}} - \delta^{\nu}_{\ \mu} \LED \right),
\end{equation}
where $ \EMTD{\nu}{\mu} $ is the tensor-density of weight $ W = 1 $ which we shall call 
\textit{energy-momentum tensor-density} of our theory.

On the other hand, if 
\begin{equation}\label{Action.GR}
S^{\sGrm\sRrm} = \int \mathcal{L}^{\sGrm} + \mathcal{L}^{\sMrm\sFrm} \rmd \bm{x}
\end{equation}
is the Hilbert-Einstein action of GR then the \textit{canonical} energy-momentum tensor-density 
reads 
\cite{Maugin1972,Kijowski1998}
\begin{equation}\label{energy.momentum.GR}
T_{\mu\nu} := -2 \frac{\pd \mathcal{L}^{\sMrm\sFrm}}{\pd g^{\mu\nu}} = -\sqrt{-g}\left [2 
\frac{\pd  \E }{\pd g^{\mu\nu}} - g_{\mu\nu} \E \right ]
\end{equation}
where $ \mathcal{L}^\sGrm $ being the gravity part of the total
Lagrangian density, $ \mathcal{L}^{\sMrm\sFrm} := \sqrt{-g}\,  \E $ being the matter-field part, 
and $ 
 \E  $ is the total energy of the matter and fields (per unit volume).

Our first task, therefore, is to show that $ \EMTD{\mu}{\nu}$  and the canonical tensor-density $ 
\EMTGR{\mu}{\nu} = g^{\mu\lambda} T_{\lambda\nu} = -\sqrt{-g}(2 g^{\mu\lambda}e_{g^{\lambda\nu}} - 
\delta^{\mu}_{\ \nu}  \E ) $ are coincide. Our second task then is to show that the 
energy-momentum $ \EMTD{\mu}{\nu} $ is covariantly conserved, $ \nabla_\mu\EMTD{\mu}{\nu} = 0 $.
Thus, by comparing the actions \eqref{Action.Euler} and \eqref{Action.GR}, we conclude that at 
least we have to assume that
\begin{equation}\label{E.e}
\LED = \sqrt{-g}\,\calE.
\end{equation}
Furthermore, since $  \E  $ is assumed to be a relativistic scalar, it can depend
on $ \cgrad{a}{\mu} $ only via its invariants, which are formed with the help of the 
spacetime metric $ g_{\mu\nu} $, see Sec.\,\ref{sec.EOS}. Hence, in fact, $  \E  = 
 \E (\cgrad{a}{\mu},g_{\mu\nu}) $. This can be used to show that (e.g. see 
\cite{Beig2003,Wernig-Pichler2006,Broda2008})
\begin{equation}\label{xi.to.g}
	\cgrad{a}{\mu} \frac{\pd  \E }{\pd\cgrad{a}{\nu}} = 2 g^{\nu\lambda} \frac{\pd  \E }{\pd 
	g^{\lambda\mu}}.
\end{equation}
Therefore, combining \eqref{E.e} and \eqref{xi.to.g} and that $ \pd\sqrt{-g}/\pd\cgrad{a}{\nu} = 0 
$, 
we conclude that
\begin{equation}
\EMTD{\mu}{\nu} = -\left (\cgrad{a}{\mu}  \LED _{\cgrad{a}{\nu}} - \delta^{\nu}_{\ \mu} 
\LED \right ) 
 = 
-\sqrt{-g}\left (2 g^{\mu\lambda}\E_{g^{\lambda\nu}} - \delta^{\mu}_{\ \nu}  \E  \right )  = 
\EMTGR{\mu}{\nu}.
\end{equation}

Finally, it remains to prove that the tensor-density $ \EMTD{\mu}{\nu} $ is covariantly conserved:
\begin{equation}\label{EMD.conserv.cov }
\nabla_\mu\EMTD{\mu}{\nu} = 0.
\end{equation}

Firstly, let us prove the identity
\begin{equation}\label{identity1}
2\sqrt{-g}g^{\mu\lambda} \E_{g^{\lambda\nu}} = 2 g^{\mu\lambda} \LED _{g^{\lambda\nu}} + 
\delta^{\mu}_{\ 
\nu} \LED
\end{equation}
which we shall need later in an equivalent form
\begin{equation}\label{identity1.modif}
%\sqrt{-g}(2g^{\mu\lambda}e_{g^{\lambda\nu}} - \delta^\mu_{\ \nu}  \E ) = 2 
%g^{\mu\lambda} \LED _{g^{\lambda\nu}} \quad \text{or} \quad 
\cgrad{a}{\nu} \LED _{\cgrad{a}{\mu}} - \delta^\mu_{\ \nu}  \LED  = 2 g^{\mu\lambda} \LED 
_{g^{\lambda\nu}},
\end{equation}
where we have used \eqref{xi.to.g} and that $ 
\pd\sqrt{-g}/\pd\cgrad{a}{\nu} = 0 $.
Identity \eqref{identity1} is obtained from  
\begin{equation}
2\sqrt{-g}g^{\mu\lambda}\E_{g^{\lambda\nu}} = 2\sqrt{-g}g^{\mu\lambda}\left 
(\frac{1}{\sqrt{-g}} \LED\right )_{g^{\lambda\nu}} = %  \\
2\sqrt{-g}g^{\mu\lambda}\frac{1}{\sqrt{-g}}\left(
\frac{1}{2}g_{\lambda\nu} \LED  +  \LED _{g^{\lambda\nu}} \right) = 
\delta^\mu_{\ \nu}  \LED  + 2 g^{\mu\lambda} \LED _{\lambda\nu}.
\end{equation}

Secondly, we note that the covariant divergence of a tensor-density $ A^{\mu}_{\ \nu} $ of weight $ 
W = 1 $ reads (for the Levi-Civita connection)
\begin{equation}\label{cov.div.tens.dens}
\nabla_\mu A^{\mu}_{\ \nu} = \pd_\mu A^{\mu}_{\ \nu} + \Gamma^{\mu}_{\ \mu\lambda} A^{\lambda}_{\ 
\nu}
- \Gamma^\lambda_{\ \mu\nu} A^{\mu}_{\ \lambda}
- W \Gamma^\lambda_{\ \mu\lambda} A^{\mu}_{\ \nu}
=  % \\
\pd_\mu A^{\mu}_{\ \nu} - \Gamma^\lambda_{\ \mu\nu} A^{\mu}_{\ \lambda}.
\end{equation}
Therefore, in order to prove \eqref{EMD.conserv.cov } we need to show that
\begin{equation}
\pd_\mu\EMTD{\mu}{\nu} - \Gamma^\lambda_{\ \mu\nu}\EMTD{\mu}{\lambda} = 0.
\end{equation}
We proceed with calculating the divergence $ \pd_\mu\EMTD{\mu}{\nu} $:
\begin{multline}\label{prove1}
\pd_\mu\EMTD{\mu}{\nu} = - \LED _{\cgrad{a}{\mu}}\pd_\mu\cgrad{a}{\nu} - \cgrad{a}{\nu}\pd_\mu 
 \LED _{\cgrad{a}{\mu}} + \pd_\nu  \LED  = %
- \LED _{\cgrad{a}{\mu}}\pd_\mu\cgrad{a}{\nu} - 
\cgrad{a}{\nu}\pd_\mu 
 \LED _{\cgrad{a}{\mu}} +  \LED _{\cgrad{a}{\lambda}}\pd_\nu\cgrad{a}{\lambda} + 
 \LED _{g^{\lambda\kappa}}\pd_\nu g^{\lambda\kappa}.
\end{multline}
We then use the fact that $ \pd_\mu\cgrad{a}{\nu} = \pd_\nu\cgrad{a}{\mu} $ and that $ 
 \LED _{g^{\lambda\kappa}} = \frac12 g_{\lambda\eta}(\cgrad{a}{\kappa} \LED _{\cgrad{a}{\eta}} - 
\delta^\eta_{\ \kappa} \LED) = -\frac12 g_{\lambda\eta} \EMTD{\eta}{\kappa}$, see 
\eqref{identity1.modif}. Therefore, \eqref{prove1} now reads
\begin{equation}\label{prove2}
\pd_\mu\EMTD{\mu}{\nu} = - \cgrad{a}{\nu}\pd_\mu  \LED _{\cgrad{a}{\mu}} -\frac12 g_{\lambda\eta} 
\EMTD{\eta}{\kappa}\pd_\nu g^{\lambda\kappa},
\end{equation}
which, thanks to that $ \pd_\mu  \LED _{\cgrad{a}{\mu}} = 0 $ which is the Euler-Lagrange equation 
for 
$ 
S^\sErm $, see \eqref{Action.Euler}, reduces to 
\begin{equation}\label{prove3}
\pd_\mu\EMTD{\mu}{\nu} - Z^\kappa_{\ \nu\eta} \EMTD{\eta}{\kappa} = 0, \qquad Z^\kappa_{\ \nu\eta} 
:= 
\frac12 g^{\lambda\kappa} \pd_\nu g_{\lambda\eta} ,
\end{equation}
where we have used $ g_{\lambda\eta} \pd_\nu g^{\lambda\kappa} = - g^{\lambda\kappa} \pd_\nu 
g_{\lambda\eta} $.

Finally, using the metric compatibility property of the Levi-Civita connection, one may express $ 
Z^\kappa_{\ \nu\eta} $ as
\begin{equation}
Z^\kappa_{\ \nu\eta} = \frac12 g^{\kappa\sigma}\Gamma^\lambda_{\ \nu\sigma}g_{\lambda\eta} + 
\frac12 \Gamma^\kappa_{\ \nu\eta},
\end{equation}
which after plugging into \eqref{prove3}$ _1 $ gives
\begin{equation}
0 = \pd_\mu\EMTD{\mu}{\nu} - Z^\kappa_{\ \nu\eta} \EMTD{\eta}{\kappa} = \pd_\mu\EMTD{\mu}{\nu} - 
\Gamma^\kappa_{\ \nu\eta} \EMTD{\eta}{\kappa}
\end{equation}
which together with the divergence formula for the tensor-density \eqref{cov.div.tens.dens} and 
symmetry $ \Gamma^\kappa_{\ \nu\eta} = \Gamma^\kappa_{\ \eta\nu} $ give 
covariant conservation law \eqref{EMD.conserv.cov } for the energy-momentum tensor-density $ 
\EMTD{\mu}{\nu} $.
For the rest of the paper, it is convenient to switch from the energy-momentum tensor-density $ 
\EMTD{\mu}{\nu} $ to 
the pure 
energy-momentum tensor $ \EMT{\mu}{\nu} $
\begin{subequations}\label{EM.conserv.cov}
\begin{equation}
\EMT{\mu}{\nu} := \frac{1}{\sqrt{-g}}\EMTD{\mu}{\nu} = -(\cgrad{a}{\nu}\E_{\cgrad{a}{\mu}} - 
\delta^\mu_{\ \nu} \E),
\end{equation}
which thanks to the metric compatibility property of the Levi-Civita 
connection also covariantly conserved
	\begin{equation}
			\nabla_{\mu}\EMT{\mu}{\nu} = 0.
	\end{equation}
\end{subequations}

We have shown that the energy-momentum tensor-density of our theory is covariantly conserved. 
However, since our formulation
is intrinsically based on two coordinate systems, the Eulerian $ x^\mu $ and 
the Lagrangian $ 
\myxi^a $, one also has to be sure that the change of the Lagrangian 
coordinates $ \myxi^a 
\rightarrow \myxi^{a'} $ will not affect the form of the energy-momentum conservation.

Therefore, let us consider a change of variables $ \myxi^a 
\rightarrow \myxi^{a'} $ which results in 
the multiplicative decomposition of 
the configuration gradient $ \cgrad{a'}{\mu} = \cgrad{a'}{a} \cgrad{a}{\mu} $ 
and hence, $ w' = \xi' w $, $ \xi' := \det( \cgrad{a'}{a}) $. 
Also, matter-time metric $ \Kab{ab} $ transforms as usual
$ \Kab{a'b'}= \Kab{ab} \cgrad{a}{a'}\cgrad{b}{b'}$ and hence, $ \sqrt{-\kappa'} = \sqrt{-\kappa} 
/\xi' $
Then, using the 
chain rule, \eqref{eqn.divEnMom} can be written as
\begin{equation}\label{eqn.chainrule}
-\pd_\nu \left (\frac{w'}{\xi'}  
\cgrad{b}{b'}\cgrad{b'}{\mu}\cgrad{a'}{b}\xi'\sqrt{-\kappa}\tilde{\Lambda}_{\cgrad{a'}{\nu}}\right 
) = 0.
\end{equation}
where we have explicitly written the Lagrangian density as $ \Lambda = \sqrt{-k}\tilde{\Lambda} $, 
see \eqref{Action.Lagr}, \eqref{eqn.Lagrangian}. Therefore, we arrive at
\begin{equation}\label{eqn.chainrule2}
-\pd_\nu \left (w' \cgrad{a'}{\mu}\Lambda_{\cgrad{a'}{\nu}}\right ) = 0,
\end{equation}
which has the same form as \eqref{eqn.divEnMom}.

%It is implied that the covariant derivative $ \nabla_\nu 
%$ for an arbitrary contravariant vector field $ V^\mu $ is defined in a 
%standard way, i.e.
%\begin{equation}\label{eqn.cov.der}
%\nabla_\nu V^{\mu} := \pd_\nu V^{\mu} + \Gamma^\mu_{\nu\lambda} V^{\mu},
%\end{equation}

%\IP{The only problem here is that everybody writes $ T^\mu_{\ \nu} = 2 g 
%^{\mu\lambda} 
%\ED_{g^{\lambda\nu}} - \ED \delta^\mu_{\ \nu} $ while if I do variation with the 
%Lagrangian $ 
%\sqrt{-g}\ED 
%$ then I get $ T^\mu_{\ \nu} = 2 g ^{\mu\lambda} 
%\ED_{g^{\lambda\nu}} + \ED \delta^\mu_{\ \nu} $, i.e. there is ``+''.}

\subsubsection{Configuration gradient evolution}
We now derive the evolution 
equation for the 
configuration gradient $ \cgrad{a}{\mu} $ in the Eulerian coordinates $ 
x^\mu $. We emphasize that this Eulerian equation can not be derived but only postulated if the 
variational principle is formulated in the Eulerian frame. 

In order to obtain the evolution equation for $ \cgrad{a}{\mu} $, we consider the integrability 
condition~\eqref{eqn.integrability} 
corresponding to $ b = 
0 $ (or equivalently to $ a = 0 $, the rest equations $ a\neq 0 $ and $ b\neq 0 $ will be pure 
spatial constraints) which can be written 
\begin{equation}\label{GRGPR.A0}
\pd_\proptime  \F{\mu}{a} - \pd_a u^\mu = 0, 
\end{equation}
where we use the definition of the 4-velocity~\eqref{eqn.4vel.def} and that $ \myxi^0 = \proptime  
$.
Using definitions~\eqref{eqn.Jacobians}, we can express Lagrangian partial derivatives $ 
\pd_\proptime  $ 
and $ \pd_a $ as 
$ \pd_\proptime  = u^\nu \pd_\nu$ and $ \pd_a = \F{\lambda}{a}\pd_\lambda$, after which, 
\eqref{GRGPR.A0} becomes
\begin{equation}\label{GRGPR.A1}
u^\nu \pd_\nu \F{\mu}{a} - \F{\lambda}{a}\pd_\lambda u^\mu = 0.
\end{equation}
Then, based on the identity $ 0 \equiv u^\nu\pd_\nu \delta^\mu_{\ \eta} = u^\nu\pd_\nu 
(\F{\mu}{a}\cgrad{a}{\eta})  
$, the derivative $u^\nu \pd_\nu \F{\mu}{a}$ can be substituted by $ u^\nu \pd_\nu \F{\mu}{a} = - 
\F{\eta}{a} 
\F{\mu}{b} u^\nu \pd_\nu\cgrad{b}{\eta} $. Thus, we have
\begin{equation}\label{GRGPR.A2}
\F{\eta}{a} 
\F{\mu}{b} u^\nu \pd_\nu\cgrad{b}{\eta} +  \F{\lambda}{a}\pd_\lambda u^\mu = 0.
\end{equation}
Eventually, after multiplying this equation by $ \cgrad{a}{\gamma} $ and then 
by $ \cgrad{a}{\mu} $ 
we obtain the sought equation for $ \cgrad{a}{\mu} $
\begin{equation}\label{eqn.A.Lie}
u^\nu \pd_\nu \cgrad{a}{\mu} + \cgrad{a}{\nu} \pd_\mu u^\nu = 0,
\quad\text{or}\quad
\Lie{\boldsymbol{u}}\cgrad{a}{\mu} = 0
\end{equation}
where $ \Lie{\boldsymbol{u}} $ is the Lie derivative\footnote{It is necessary to keep in mind that 
$ 
\cgrad{a}{\mu} $ is 
not a tensor but it transforms as a 4-vector under the coordinate change $ 
x^\mu \rightarrow x^{\mu'} $.} in the direction of the 4-velocity $ u^\nu $. 

Alternatively, using the identity $ \cgrad{a}{\nu} \pd_\mu \F{\nu}{b} = -\F{\nu}{b} \pd_\mu 
\cgrad{a}{\nu}$, one can rewrite \eqref{eqn.A.Lie} as
\begin{equation}\label{eqn.PDE.cgrad}
u^\nu (\pd_\nu\cgrad{a}{\mu} - \pd_\mu\cgrad{a}{\nu}) = 0.
\end{equation}

%Thus, we have two equivalent forms of the PDE for the configuration gradient which we shall 
%use interchangeably:
%\begin{equation}\label{eqn.Lie.cgrad0}
%u^\nu \pd_\nu \cgrad{\sA}{\mu}  + \cgrad{\sA}{\nu}\pd_\mu u^\nu = 0 \quad \text{ 
%and } \quad
%\pd_\mu (u^\nu\cgrad{\sA}{\nu}) + u^\nu (\pd_\nu\cgrad{\sA}{\mu} - 
%\pd_\mu\cgrad{\sA}{\nu}) = 0.
%\end{equation}

The fact that the evolution of the configuration gradient is the Lie derivative can 
be used to replace the standard 
partial derivatives $ \pd_\mu $ by the covariant derivatives $ \nabla_\mu $.
This is valid for the torsion-free 
spacetimes. Also, it is evident that for the torsion-free connection, the covariant derivatives can 
be used instead of $ \pd_\nu $ in \eqref{eqn.PDE.cgrad} as well.  Thus, evolution equations
\eqref{eqn.A.Lie}, \eqref{eqn.PDE.cgrad}
can be written in covariant forms
\begin{equation}\label{eqn.Lie.cgrad1.cov}
u^\nu \nabla_\nu \cgrad{a}{\mu}  + 
\cgrad{a}{\nu}\nabla_\mu u^\nu = 0, 
\qquad
u^\nu (\nabla_\nu\cgrad{a}{\mu} - \nabla_\mu\cgrad{a}{\nu}) = 0,
\end{equation}

Eventually, in order to accomplish the discussion at the beginning of 
Section\,\ref{sec.LagrPDE}, we remark that the equation \eqref{eqn.A.Lie} 
cannot be obtained as a consequence of the definition of the potentials $ 
\myxi^a(x^\mu) $ if the variational principle is formulated in 
the Eulerian 
settings. Indeed, it is not clear  whether $ \pd_\nu\cgrad{a}{\mu} - \pd_\mu\cgrad{a}{\nu} = 0 
$ (the direct consequence of the definition of $ \myxi^a(x^\mu) $) or $ 
u^{\nu}(\pd_\nu\cgrad{a}{\mu} - \pd_\mu\cgrad{a}{\nu}) = 0 $ has to be used as the evolution 
equation for $ \cgrad{a}{\mu} $. In contrast, if the variational principle
is formulated in the Lagrangian frame, 
equations~\eqref{eqn.A.Lie} are obtained as a rigorous consequence of 
\eqref{eqn.integrability}.

\section{Irreversible deformations}\label{sec.Irreversible}

The classical Lagrangian formalism  without any changes was used in Section~\ref{sec.Motion} to 
obtain equations of motion \eqref{eqn.Euler-Lagrange}--\eqref{eqn.integrability} in the Lagrangian 
coordinates $ \myxi^a $. These equations are valid for arbitrary curvilinear coordinates $ \myxi^a 
$ and 
$ x^\mu $. We then transformed these equations into their 
Eulerian counterparts \eqref{EMD.conserv.cov }, \eqref{eqn.Lie.cgrad1.cov} by means of the change 
of variables $ \myxi^a \rightarrow x^\mu$. Formally, these equations 
are applicable to arbitrary continua, either 
fluid or solid. However, a few important ingredients are still missing.
% if one wants to apply these 
%equations to modeling of irreversible demonstrations, that 
%is flows. 

First of all, the equations are not closed, i.e. the energy 
potential $ E(\cgrad{a}{\mu}) $, which generates the energy-momentum $ 
\EMTD{\nu}{\mu} = -(\cgrad{a}{\mu} \ED_{\cgrad{a}{\nu}} - \ED \, \delta^{\nu}_{\ \mu}) $, is 
remained
unspecified. Here, 
however, we run into two principal problems of the pure Lagrangian description 
of motion if 
one tries to apply this approach to modeling of irreversible deformations. Indeed, the field of 
labels $ \xi^a(x^\mu) $ contains the complete 
information about 
the macroscopic motion and 
geometry of the continuum including the information about the initial configuration. The 
configuration gradient $ \cgrad{a}{\mu} $ was introduced in Section~\ref{sec.4Jac} as the gradient 
of the field of labels $ \myxi^a $ and thus, represents the \textit{total} 
(\textit{observable}) \textit{deformation} encoded in the laws of 
motion~\eqref{eqn.motion.laws1}. The problem is that the fluid or plastic solid 
should not be able to ``remember'' this complete information abut the history 
of motion, i.e. the stress state should not depend on the complete flow history. Such an 
information should be removed (dissipated) from the system 
in a thermodynamically consistent way. Therefore, we still need to introduce dissipation in the so 
far reversible equations \eqref{EMD.conserv.cov }, \eqref{eqn.Lie.cgrad1.cov}.

\begin{figure}[!htbp]
	\begin{center}
		\includegraphics[draft=false,trim = 0mm 0mm 0mm 0mm, clip, 
		scale=1]{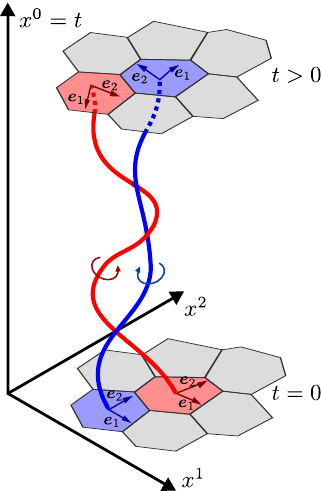}
		\caption{A pictorial illustration of the material triad evolution and 
		continuum 
		particle (finite volumes) rearrangement. The red and blue particles change their neighbors 
		which results in the incompatibility of the local deformation of particles represented by 
		the material basis triads (only $ \bm{e}_1 $ and $ \bm{e}_2 $ are depicted). Initially 
		holonomic $ \bm{e}_1 $, $ \bm{e}_2 $ become non-holonomic at later times. 
		}\label{fig.distortion}
	\end{center}
\end{figure}

Secondly, the irreversible deformations are due to the structural changes in the medium. The 
structural changes mean that the material elements 
(parcels of molecules) that were attached to each other in space may become 
disconnected after the 
irreversible process of material element rearrangements, which is, in fact,
the essence of any flow, see Fig.\,\ref{fig.distortion}. Such structural rearrangements are in 
apparent 
contradiction with the Lagrangian viewpoint relying on the existence of the single-valued mapping $ 
\myxi^a(x^\mu) $ implying that the portions of matter that were 
connected initially are remained so at all later time instants.

\subsection{Distortion field}\label{sec.dist}

In order to overcome these contradictions and, at the same time, to \textit{retain 
the variational structure} of the equations of motion  \eqref{EMD.conserv.cov }, 
\eqref{eqn.Lie.cgrad1.cov} compatible with GR, we follow Godunov and 
Romenski~\cite{GodRom1972,God1978,GodRom2003} (and our Newtonian 
papers~\cite{HPR2016,DPRZ2016,HYP2016,SHTC-GENERIC-CMAT}) and introduce two deformation fields one 
of which 
describes the observable macroscopic deformation while the other describes the internal (non 
observable for a macroscopic observer) deformation, it is called \textit{effective} deformation in 
\cite{GodRom1972,God1978,GodRom2003}. As the first field, we keep the configuration 
gradient $ \cgrad{a}{\mu} $ which is \textit{integrable} or \textit{holonomic}, i.e. $ \pd_\mu 
\cgrad{a}{\nu} - \pd_\nu \cgrad{a}{\mu} = 0 $, due to its definition. But this 
field will play no role in our theory any further. The second deformation 
field, 
denoted as $ \dist{m}{\mu} $,
is called the \textit{4-distortion} and 
defined as the solution to the following evolution equation (coupled with the energy-momentum 
conservation which will be discussed later)
\begin{equation}\label{eqn.dist.evol}
u^\nu \nabla_\nu\dist{m}{\mu} + \dist{m}{\nu} \nabla_\mu u^\nu  = S^m_{\ \mu},
\quad \text{or} \quad
\Lie{\boldsymbol{u}}\dist{m}{\mu}  = S^m_{\ \mu},
\end{equation}
which are different from the evolution of the configuration gradient \eqref{eqn.A.Lie} by the 
presence of algebraic source term $ S^m_{\ \mu}(\dist{a}{\mu}) $ which acts as the source of 
anholonomy. 

In order to emphasize that, in general,
the stress-free matter configuration 
corresponding to the distortion field $ \dist{m}{\mu} $ might be different from 
the initial stress-free 
configuration corresponding to $ \cgrad{a}{\mu} $ and labeled with $ \myxi^a $, 
from now on, we shall use middle Latin alphabet 
letters $ 
m,n,l=0,1,2,3 $ and $ \sM,\sN,\sL=1,2,3 $ to denote matter-time components, $ 
\dist{m}{\mu} $, and pure 
matter components, $ \dist{\sM}{\mu} $, of the distortion field.

Because the mapping $ x^\mu(\myxi^a) $ may not exist in general, the definition of the 
4-velocity \eqref{eqn.4vel.def} cannot be used. Nevertheless, as previously in \eqref{eqn.Jac}, we 
still can define 
the 
4-velocity 
as the 1-st column of the inverse distortion $ \idist{\mu}{a} $ ($ \idist{\mu}{a}\dist{a}{\nu} = 
\delta^\mu_{\ \nu} $, $ \idist{\mu}{a}\dist{b}{\mu} = 
\delta^b_{\ a}$), e.g. see \cite{PRD-Torsion2019},
\begin{equation}\label{4vel.irrevers}
u^\mu := \idist{\mu}{0}.
\end{equation}
In particular, 
\begin{equation}
\dist{0}{\mu} u^\mu = 1, \qquad \dist{\sM}{\mu} u^\mu = 0.
\end{equation}

Furthermore, we define the source of non-holonomy $ S^m_{\ \mu} $ satisfying the following four 
properties
\begin{subequations}\label{eqn.A.cond}
\begin{align}
\text{invertibility: } &\quad \det(\dist{m}{\mu}) 
%= \det(\cgrad{a}{\mu})  = \frac{\rho}{\rho_0}\sqrt{-g} 
> 0 
\label{eqn.A.invert},\\[2mm]
\text{orthogonality: } &\quad \dist{\sM}{\mu}
u^{\mu} = 0, \label{eqn.A.orth}\\[2mm]
\text{non-integrability: } &\quad
T^m_{\ \mu\nu}: = \pd_\mu \dist{m}{\nu} - \pd_\nu \dist{m}{\mu}
\neq 0,\label{eqn.A.nonintegr}\\[2mm]
\text{irreversibility: } &\quad \nabla_\mu(s u^\mu) \geq 0. \label{eqn.A.diss}
\end{align}
\end{subequations}
Here,  $ s $ is the entropy density, $ T^m_{\ \mu\nu} $ is the torsion tensor  
\cite{PRD-Torsion2019} and, 
in general, is not 
assumed to be equal to zero. 
In other words, the condition \eqref{eqn.A.nonintegr} is the mathematical expression of the 
microscopic structural rearrangements which imply that if a 4-potential, say $ 
A^m(x^\mu) $, would 
exist 
it would be a discontinuous (a closed loop on the material manifold becomes open during the 
evolution) because the connectivity between microscopic parts of the media may not be
preserved in the irreversible deformations. The torsion tensor $ T^m_{\ \mu\nu} 
$ thus represents a 
continuous distribution of the density of such microscopic discontinuities, e.g. 
see~\cite{PRD-Torsion2019,KleinertMultivalued} and references therein.

\subsubsection{Local relaxed reference frame}
Because a global coordinate system $ A^m(x^\mu) $ such that $ \pd_\mu A^m = \dist{m}{\mu} $ does 
not 
exist in 
general due to the non-integrability condition \eqref{eqn.A.nonintegr}, 
one 
therefore 
needs 
to define what is meant under \textit{unstressed reference frame}
%\footnote{It 
%is the 
%coordinate basis vectors $ \boldsymbol{\pd}_a $ who represent the reference frame in 
%the case when 
%it is possible to introduce the Lagrangian coordinates $ \myxi^a $.} 
in the 
context of 
irreversible deformations. In the absence of the global coordinate system, a 
reference 
frame can be defined \textit{locally} for each material element \textit{individually}. 
%This is the distortion field which actually introduces such a reference frame. 
Actually, this is the distortion field itself who introduces the local reference frame. Indeed, due 
to the invertability condition~\eqref{eqn.A.invert}, the 
inverse distortion  $ \idist{\mu}{n} $ exists and represents a basis tetrad. Such a tetrad, if 
pulled-back by the distortion $ 
\dist{m}{\mu} \dist{\mu}{n} = \delta^m_{\ \,
n}$, becomes the orthonormal tetrad $ \delta^m_{\ \,n} $. The orthonormal tetrad $ \delta^{m}_{\ 
\,n} $ 
represents the \textit{local reference frame} with respect to which one can measure the 
deformation 
of the given material element. The \textit{locality} means that in different 
material elements, the 
spatial 
directions (triad) of such  an
orthonormal tetrad may differ by an arbitrary spatial rotation, while the time 
direction is implied to be 
the same for all the material elements and be oriented in the direction of the 
proper time, see 
\eqref{eqn.A.orth}. Furthermore, the local reference frame is also identified with the new
\textit{relaxed} or \textit{stress-free} state of a given material element. In other words, if 
the 
tetrad $ \idist{\mu}{n} $ and $ \delta^m_{\ n} $ differ by only a spatial 
rotation $ R^m_{\ \mu} 
$, 
that is $ R^m_{\ \mu}A^\mu_{\ n} = \delta^m_{\ n} $, we say that the material element is locally 
relaxed.

The key 
difference between $ \dist{m}{\mu} $ and $ \cgrad{a}{\mu} $ is that the 
stress-free state
associated 
with  $ \cgrad{a}{\mu} $ can be reached 
globally for all the material elements \textit{simultaneously}, 
while the local stress-free states associated with the distortion field $ \dist{m}{\mu} $ cannot be 
reached simultaneously because 
there is no such a continuous motion $ A^m(x^\mu) $ which 
could map all the material elements from their current deformed state to the 
stress-free state.

\subsubsection{Strain measure}

Because the stresses in the irreversibly deforming matter depend not on 
the 
observable deformation 
encoded in $ \cgrad{a}{\mu} $, but on the deviation from the local stress-free 
state encoded now in $ \dist{m}{\mu} $, the 
spatial metric $ \G_{\mu\nu} = \G_{ab}\cgrad{a}{\mu}\cgrad{b}{\nu}$ 
cannot be used to measure material distances anymore. Therefore, it is implied that, in the local 
reference frame given by $ \dist{m}{\mu} $, the mater-time element is characterized with the 
axiomatically given  metric 
$ \Kmn_{mn} $ which, in the absence of phase transformations and other 
physicochemical changes at the molecular level, is assumed to be locally identical 
to $ \Kab{ab} $ from Section~\ref{sec.4Jac}, i.e. it is assumed to be locally flat $ 
\Kmn_{mn} = \text{diag}(-1,1,1,1) $. Furthermore, we introduce 
a new material metric (it is called the \textit{effective} metric tensor in 
\cite{God1972,GodRom2003} in the non-relativistic context),
\begin{equation}\label{eqn.Geff}
\Geff{\mu\nu} := \LagrProjEff_{mn} \dist{m}{\mu}\dist{n}{\nu} = 
\KMN{\sM\sN}\dist{\mathsmaller 
M}{\mu}\dist{\mathsmaller N}{\nu},
\end{equation}
where, similar to \eqref{eqn.projectorMT}, we have defined the matter projector 
in the local relaxed frame
\begin{equation}\label{eqn.projectorMT.eff}
\LagrProjEff_{mn} := \Kab{mn} + \uLag_m \uLag_n =
%\left(\begin{array}{rccc}
%-1 & 0 & 0 & 0\\ 
% 0 &   &   &  \\ 
% 0 &   & \KMN_{\sM\sN} & \\
% 0 &   &   &  
%\end{array}\right) 
%+ 
%\left(\begin{array}{rccc}
% 1 & 0 & 0 & 0 \\ 
% 0 & 0 & 0 & 0 \\ 
% 0 & 0 & 0 & 0 \\
% 0 & 0 & 0 & 0 
%\end{array}\right) 
%= 
\left(\begin{array}{rccc}
 0 & 0 & 0 & 0\\ 
 0 &   &   &  \\ 
 0 &   & \KMN{\sM\sN} & \\
 0 &   &   &  
\end{array}\right).
\end{equation}
As in \eqref{eqn.projectorMT}, $ \uLag_m = (-1,0,0,0) $ because the 
introduction of the local reference frame concerns only the spatial directions,
while it keeps the proper time direction unaffected. 
Note that if the media is an elastic solid then we 
have $ \cgrad{\sA}{\mu} 
= 
\dist{\sA}{\mu}$ and $ \G_{\mu\nu} = \Geff{\mu\nu} $. Moreover, due to the orthogonality 
property \eqref{eqn.A.orth} of the distortion field, we also have
\begin{equation}
\Geff{\mu\nu}u^\mu = 0.
\end{equation}
Also, we shall need the contravariant components of the matter projector
\begin{equation}
\GeffContr{mn} = \KabContr{mn} + \uLag^m \uLag^n, \quad \GeffContr{\mu\nu} = 
\GeffContr{mn}\idist{\mu}{m}\idist{\nu}{n}.
\end{equation}

%
%{\color{blue}
%%
%\begin{align}\label{eqn.Geff}
%\Geff{\mu\nu} &:= \LagrProj_{mn} \dist{m}{\mu}\dist{n}{\nu} = 
%\KMN_{\sM\sN}\dist{\mathsmaller 
%M}{\mu}\dist{\mathsmaller N}{\nu},\\
%\Geff{\mu\nu} &:= \LagrProjEff_{mn} \xi^{m}_{\mu}\xi^{n}_{\nu}
%\end{align}
%%
%}
%%
%%
%{\color{Green}
%%
%$$ \text{if}\qquad \kappa = k $$
%\begin{equation}\label{eqn.Geff}
%\LagrProj_{\mu\nu} := \LagrProj_{mn} \xi^{m}_\mu \xi^n_\nu, \qquad 
%\Geff{\mu\nu} := \KMN_{\sM\sN}\dist{\mathsmaller 
%M}{\mu}\dist{\mathsmaller N}{\nu} \equiv  \LagrProj_{MN} \dist{\mathsmaller 
%M}{\mu}\dist{\mathsmaller N}{\nu}  =\LagrProj_{mn} \dist{\mathsmaller 
%m}{\mu}\dist{\mathsmaller n}{\nu} \neq  \LagrProj_{\mu\nu} 
%\end{equation}
%%
%where $ \LagrProj_{mn} $ is the matter projector with the same structure as 
%$ 
%\LagrProj_{ab} $ in 
%\eqref{eqn.projectorMT}. Note that if the media is an elastic solid then we 
%have $ \cgrad{\sA}{\mu} 
%= 
%\dist{\sA}{\mu}$ and $ \G_{\mu\nu} = \Geff{\mu\nu} $.
%}
%
%{\color{Green}
%Here we should define also $\kappa_{mn}$. Does $\kappa_{mn}$ take the same 
%form 
%of $\kappa_{ab}$? i.e. from eq. (\ref{eqn.projectorMT})
%\begin{align}
%\kappa_{ab}=\left(\begin{array}{rccc}
%-1 & 0 & 0 & 0\\ 
% 0 &   &   &  \\ 
% 0 &   & \KAB_{\sA\sB} & \\
% 0 &   &   &  
%\end{array}\right) 
%\end{align}
%Then we should write it explicitly.
%I think this will tell us if $P^a_m$ operates only on the spatial coordinates 
%$A,B$. I think we should rewrite  (\ref{eqn.projectorMT}) also for $G_{mn}$ 
%explicitly. 
%}

\subsubsection{Canonical structure of the energy-momentum tensor}

We have defined two deformation fields, one of which, the 
configuration gradient $ \cgrad{a}{\mu} $, is \textit{global} and describes the 
observable deformation of the medium but it says nothing about the changes in 
the internal structure. The second deformation field, the distortion field $ 
\dist{m}{\mu} $, is \textit{local} and characterizes only deformations and 
rotations of the material elements, while the global deformation of the 
continuum cannot be recovered from $ \dist{m}{\mu} $. This, however, rises the following issue. 
The variational principle suggests that the energy-momentum tensor-density $ 
\EMT{\nu}{\mu} = -(\cgrad{a}{\mu} 
\E_{\cgrad{a}{\nu}} - \delta^{\nu}_{\ \mu}\E) $ introduced in \eqref{eqn.divEnMom1} is 
completely 
specified by 
defining the energy potential $ E(\cgrad{a}{\mu}) $, which, in turn, implies 
that $ \cgrad{a}{\mu} $ should be treated as a state variable. 
However, 
as we just discussed, $ \cgrad{a}{\mu} $ is not related to the internal 
structure and hence, to the 
stress state of the media in the case of irreversible deformations (in the elastic case $ 
\cgrad{\sA}{\mu} $ and $ \dist{\sM}{\mu} $
coincide), while the 
distortion field does characterize the real deformation of the material 
elements and should play the role of the state variable. In fact, the Euler-Lagrange
structure of the energy-momentum tensor-density allows to overcome this 
contradiction. Indeed, because any two invertible matrices can be related to 
each 
other as
\begin{equation}\label{eqn.plastdist}
\cgrad{a}{\mu} = \plast{a}{m}\dist{m}{\mu},
\end{equation}
for some non-degenerate matrix $ \plast{a}{m} $, and since we assume that the stress state can 
depend only on actual distances 
between molecules, the energy may depend on the configuration gradient only via 
its dependence on the distortion field $ \dist{m}{\mu} $ which dose describe the distances between 
molecules, see \eqref{eqn.Geff}, i.e.
\begin{equation}\label{eqn.E}
\E(\cgrad{a}{\mu}) = \hat{\E}(\plast{m}{a}\cgrad{a}{\mu}) = 
\hat{\E}(\dist{m}{\mu}),
\end{equation}
where $ \plast{m}{\, a} $ is the inverse of $ \plast{a}{m} $. Therefore, applying 
the chain rule exactly as in~\eqref{eqn.chainrule}, \eqref{eqn.chainrule2}, we 
arrive at
\begin{equation}\label{eqn.stress.dist}
-\EMT{\nu}{\mu} = \cgrad{a}{\mu} 
\E_{\cgrad{a}{\nu}} - \delta^{\nu}_{\ \mu} \E = \dist{m}{\mu} 
\hat{\E}_{\dist{m}{\nu}} - \delta^{\nu}_{\ \mu}\hat{\E},
\end{equation}
that is the canonical structure of the energy-momentum is preserved. In the rest of the paper, 
we remove the ``hat'' in \eqref{eqn.E} and write $ \E(\dist{m}{\mu}) $.

\subsection{Governing equations}\label{sec.gov.pdes}

Eventually, we can write down the system of governing equations
\begin{subequations}\label{GRGPR}
\begin{align}
&\nabla_\mu\left( \dist{m}{\nu} \E_{\dist{m}{\mu}} - \delta^\mu_{\ \nu} \E \right) = 0, 
\label{GRGPR.T}\\[2mm]
&u^\nu \nabla_\nu\dist{m}{\mu} + \dist{m}{\nu} \nabla_\mu u^\nu  = -\frac{1}{\theta(\tau)} 
\GeffContr{mn} g_{\mu\nu} \E_{\dist{n}{\nu}},\label{GRGPR.A}\\[2mm]
&\nabla_\mu (\rho u^\mu) = 0, \label{GRGPR.rho} \\[2mm]
&\nabla_\mu (s u^\mu) = \frac{1}{\E_s \theta(\tau)} \GeffContr{mn} g_{\mu\nu} 
\E_{\dist{m}{\mu}} 
\E_{\dist{n}{\nu}} \geq 0, \label{GRGPR.s} 
\end{align}
\end{subequations}
which should be closed by providing appropriate functions $ \E(\dist{\sM}{\mu},s) $ and $ 
\theta(\tau) $ where, from now on, $ \tau $ stands for the characteristic strain dissipation (or 
stress relaxation) time 
but not for the proper 
time. They will be specified in Sec.\,\ref{sec.EOS}. Also, here, $ \E_s = \pd \E/\pd s $ stands for 
the temperature.

The source of non-holonomy $ S^m_{\ \mu} = - \GeffContr{mn} g_{\mu\nu} 
\E_{\dist{n}{\nu}}/\theta(\tau)$,
as it is defined in \eqref{GRGPR.A}, satisfies the four properties \eqref{eqn.A.cond}.
In particular, the entropy source term is obviously positive. It is constructed based on the SHTC 
principle discussed in 
Sec.\,\ref{sec.SHTC}, that is the over-determined system \eqref{GRGPR} has to be compatible, 
see details in Sec.\,\ref{sec.therm.consist}.

The orthogonality condition \eqref{eqn.A.orth}, is automatically satisfied due the definition of 
the 4-velocity \eqref{4vel.irrevers}. Nevertheless, in 
Appendix~\ref{app.orthogonality}, we show that the orthogonality condition \eqref{eqn.A.orth} is 
fulfilled as long as $ \E_{\dist{\sM}{\mu}}u_\mu = 0$ (which is the case in this 
paper, see Section~\ref{sec.Specification.T}), i.e. 
the relaxation of the time components $ \dist{\sM}{0} $ is just a consequence 
of the 4-dimensional formalism.
% independent of the relaxation of the pure matter components $ \dist{\sM}{i} 
%$, $ i=1,2,3 $.

The non-integrability condition \eqref{eqn.A.nonintegr} is also automatically fulfilled for any 
non-zero source $ S^m_{\ \mu} $, and the torsion is non-zero in general, e.g. see 
\cite{PRD-Torsion2019}.

Finally, we have to show that our choice for the source $ S^m_{\ \mu} $ does not violate the 
invertibility property \eqref{eqn.A.invert}. Moreover, as was mentioned in 
Section\,\ref{sec.matter.current}, we have to show that 
for the 
non-holonomic case, i.e. when $ \pd_\nu \dist{m}{\mu} - \pd_\mu\dist{m}{\nu} \neq 0 $, the rest 
mass density conservation law \eqref{GRGPR.rho} can be derived from the distortion evolution 
equation 
\eqref{GRGPR.A}. In fact, in order to satisfy our needs, it is sufficient to initiate the 
distortion field with the condition 
\begin{equation}\label{det.A}
	\det(\dist{m}{\mu}) = \det(\cgrad{a}{\mu}) = \frac{\rho}{\rho_0}\sqrt{-g}.
\end{equation}

Indeed, in this case, by multiplying \eqref{GRGPR.A} by $ \rho_{\dist{m}{\mu}}=\pd 
\rho/\pd \dist{m}{\mu} = 
\rho \idist{\mu}{m}$, 
one obtains
\begin{equation}
\rho_{\dist{m}{\mu}} \left( u^\nu \nabla_\nu\dist{m}{\mu} + \dist{m}{\nu} \nabla_\mu u^\nu \right) 
= % \\
u^\mu\nabla_\mu\rho + \rho \idist{\nu}{m}\dist{m}{\mu}\nabla_\nu u^\mu = \nabla_\mu (\rho u^\mu)
\end{equation}
for the left hand-side of \eqref{GRGPR.A}. Therefore, it remains to show that 
\begin{equation}\label{orth.source}
	\rho_{\dist{m}{\mu}} \LagrProjEff^{mn} g_{\mu\nu} \E_{\dist{n}{\nu}} = 
	\rho_{\dist{\sM}{\mu}} \KMNContr{\sM \sN} g_{\mu\nu} \E_{\dist{\sN}{\nu}} = 0.
\end{equation}
This, however, cannot be proved in the general case but is conditioned 
by 
the choice of the energy $ \E $. Thus, we prove in Section\,\ref{sec.EOS} (see equation 
\eqref{orth.source.prove}) that for our 
particular choice of the equation of state,  condition \eqref{orth.source} is satisfied.

\subsection{Conventional structure of the energy-momentum tensor}\label{sec.Specification.T}

Because the energy potential $ \E $ is supposed to transform as a scalar-density under 
transformations of both
the spacetime and matter, it may depend on $ \dist{m}{\mu} $ only via the 
scalars that can be made from $ \Geff{\mu\nu} = \Kab{\sM\sN} 
\dist{\sM}{\mu}\dist{\sN}{\nu}$. Also, because $ \Geff{\mu\nu} $ has one zero 
eigenvalue, there are only three independent scalar invariants. For example, 
\begin{equation}\label{eqn.G.inv}
I_1 = g^{\mu\lambda}\Geff{\lambda\mu} = \GeffMix{\mu}{\mu}, 
\quad 
I_2 = \GeffMix{\mu}{\nu}\GeffMix{\nu}{\mu}, 
\quad 
I_3 = \GeffMix{\mu}{\nu}\GeffMix{\nu}{\lambda}\GeffMix{\lambda}{\mu}.
\end{equation}
However, it is convenient to replace the third invariant with the rest matter 
density 
$ \rho $. Indeed, as discussed in~\cite{Karlovini2003,Gundlach2012}, 
the spacetime tensor $ \GeffMix{\mu}{\nu} = g^{\mu\lambda} \Geff{\lambda\nu} $ has 
the 
same eigenvalues as the pure matter tensor $ \GeffMix{\sM}{\sN} := g^{\mu\nu} 
\KMN{\sN\sL}
\dist{\sM}{\mu}\dist{\sL}{\nu}$ in addition to one zero eigenvalue. 
Furthermore, in~\cite{Gundlach2012}, it is also shown that the determinant $ 
\det(\GeffMix{\sM}{\sN}) $ relates 
to the rest matter density as $ \det(\GeffMix{\sM}{\sN}) = \rho^2 $. 
Therefore, we assume that 
\begin{equation}\label{eqn.Energy.inv}
\E(\dist{m}{\mu},s) = \E(I_1,I_2,\rho,s),
\end{equation}
where we also consider a dependence on the entropy density $ s = \rho \Entropy$ 
with $ \Entropy $ 
being the specific entropy in the rest frame.
%where $ s $ is the rest frame entropy density. 
%Furthermore, we assume the following 
%decomposition
%%(see the definition of 
%%the matter density~\eqref{eqn.rho.definition} in which, without the loss of 
%%generality, one may assume that the matter metric has $ \kappa^{\rm m} = 
%%\det(g_{\sA\sB}) = 1 $)
%\begin{equation}\label{eqn.energy.rhoE}
%\tilde{E}(I_1,I_2,\rho,\sigma) = {\tilde{E}}^{\rm 
%e}(\rho,s) + {\tilde{E}}^{\rm 
%n}(I_1,I_2,\rho,s), %= \frac{1}{\sqrt{-g}}\rho(1 
%%+ \epsilon).
%\end{equation}
%where $ s = \sigma/\rho $ is the specific entropy, $ e^{\rm e} $ is the \textit{equilibrium} 
%specific 
%energy, while $ e^{\rm n} $ is the \textit{non-equilibrium} 
%contribution (for elastic media it is responsible for the elastic stresses while the viscous media 
%for the viscous stresses).
We are now ready to unveil the conventional structure of the 
energy-momentum tensor $ \EMT{\nu}{\mu} = -(\dist{m}{\mu} \E_{\dist{m}{\nu}} - \delta^\nu_{\ 
\mu}\E)$ of our theory. Thus, because of 
\eqref{eqn.Energy.inv}, one can write $ \dist{m}{\mu} \E_{\dist{m}{\nu}} = 
\dist{\sM}{\mu} \E_{\dist{\sM}{\nu}} $. Then, using that $ \rho = 
\frac{1}{\sqrt{-g}}\sqrt{-J^\mu J_\mu}$, $ u^\mu = \idist{\mu}{0}$ and the identity (see 
\cite{Kijowski1990})
\begin{equation}\label{eqn.rhoJ.identity}
\frac{\pd J^\lambda}{\pd \dist{\sM}{\nu}} \dist{\sM}{\mu} = J^\lambda 
\delta^\nu_{\ \mu} - J^\nu \delta^\lambda_{\ \mu},
\end{equation}
we may write
\begin{eqnarray}
-\EMT{\nu}{\mu} = \dist{\sM}{\mu} \left(\E_\rho \frac{\pd \rho}{\pd 
J^\lambda}\frac{\pd J^\lambda}{\pd \dist{\sM}{\nu}} + 
\E_{\dist{\sM}{\nu}}\right)  - \E \delta^\nu_{\ \mu} &=& \nonumber \\
-\frac{1}{\sqrt{-g}}(\E_\rho + \Entropy \E_s) u_\lambda (J^\lambda 
\delta^\nu_{\ 
\mu} - 
J^\nu \delta^\lambda_{\ \mu}) + \dist{\sM}{\mu} \E_{\dist{\sM}{\nu}} - 
\E \delta^\nu_{\ \mu} &=& \nonumber \\
%\frac{1}{\sqrt{-g}}
(\rho \E_\rho + s \E_s) (\delta^\nu_{\ \mu} +
 u^\nu u_\mu) + \dist{\sM}{\mu} \E_{\dist{\sM}{\nu}} - 
\E \delta^\nu_{\ \mu},
\end{eqnarray}
which now can be rewritten in the conventional form
\begin{equation}\label{eqn.energy-momentum.conv}
-\EMT{\nu}{\mu} = \E u^\nu u_\mu + p h^\nu_{\ \mu} + \sigma^\nu_{\ 
\mu},
\end{equation}
where one can identify the pressure $ p $ and the anisotropic part $ \sigma^\nu_{\ \mu} $ of the 
energy-momentum tensor as
\begin{equation}\label{eqn.sigma.p}
p := \rho \E_\rho + s \E_s - \E,
%\frac{1}{\sqrt{-g}} 
%\rho^2 e_\rho, 
\qquad 
\sigma^\nu_{\ \mu} 
:= 
\dist{\sM}{\mu} \E_{\dist{\sM}{\nu}}.
\end{equation}
%where $ e(\rho,\Entropy) = E(\rho,s)/\rho $ is the specific energy.
It is necessary to emphasize that while computing $ \sigma^\nu_{\ \mu} $, the rest mass density $ 
\rho $ and 
the distortion $ \dist{\sM}{\mu} $ should be treated as independent variables 
because their dependence has been already taken into account in the pressure. More explicitly, $ 
\sigma^\nu_{\ \mu}  = \dist{\sM}{\mu} \left (\frac{\pd \E}{\pd I_1} \frac{\pd I_1}{\pd 
\dist{\sM}{\nu}} + \frac{\pd \E}{\pd I_2} \frac{\pd I_2}{\pd 
\dist{\sM}{\nu}}\right )$.
%The scalar $ p $ should be identified with the pressure which may also include non-equilibrium 
%contribution $ \rho^2 e^{\rm n}_\rho $ due to \eqref{eqn.energy.rhoE}, while the tensor $ 
%\sigma^\nu_{\ \mu} $ is the anisotropic stress due to elastic or viscous 
%effects.

Finally, we note that due to the orthogonality condition $ \dist{\sM}{\mu} u^\mu = 0 $, the 
following orthogonality conditions are hold for the anisotropic part of the energy-momentum
\begin{equation}\label{eqn.Cauchy.orthogon}
\Cauchymix{\nu}{\mu} u^{\mu} = 0, \qquad  \Cauchymix{\nu}{\mu} u_{\nu} = 0.
\end{equation}
While \eqref{eqn.Cauchy.orthogon}$ _1 $ is obvious, \eqref{eqn.Cauchy.orthogon}$ _2 $ follows from 
\eqref{eqn.Energy.inv} and the formulas  
\begin{equation}\label{eqn.dIdA}
\frac{\pd I_1}{\pd \dist{\sM }{\mu}} = 2\,\KMN{\sM \sN}\dist{\sN}{\lambda} 
{g^{\lambda\mu}} = 2\,\KMN{\sM \sN}\dist{\sN}{\lambda} 
{h^{\lambda\mu}}, 
\quad
\frac{\pd I_2}{\pd \dist{\sM }{\mu}} = 4\,\KMN{\sM \sN}\dist{\sN}{\lambda} 
\GeffMix{\lambda}{\nu} {g^{\nu\mu}} = 
4\,\KMN{\sM \sN}\dist{\sN}{\lambda} 
\GeffMix{\lambda}{\nu} {h^{\nu\mu}}.
\end{equation}
These derivatives are obviously orthogonal to $ u_\mu $. 

%\IP{A strange factor $ \frac{1}{\sqrt{-g}} $ appears in the definition of the 
%pressure. I'll try to fix it later.... }

\subsection{Equation of state}\label{sec.EOS}

In this section, we give a particular expression to the specific energy 
\begin{equation}
1 + \varepsilon(I_1,I_2,\rho,\Entropy) := \rho^{-1}\E(I_1,I_2,\rho,\entropy).
\end{equation}

%\begin{equation}
%\GeffContr{\mu}_{\ \nu} = g^{\mu\lambda} \Geff{\lambda\nu} = g^{\mu\lambda} g^{\sM 
%\sN} 
%\dist{\sM }{\lambda}\dist{\sN}{\nu} =  h^{\mu\lambda} \Geff{\lambda\nu} = 
%h^{\mu\lambda} 
%g^{\sM \sN} 
%\dist{\sM }{\lambda}\dist{\sN}{\nu}.
%\end{equation}
%Tensor $ \GeffMix{\mu}{\nu} $ is exactly the one used 
%in~\cite{Karlovini2003,Kijowski1992}.
%
%Therefore, the energy should be the function of the three invariants (the 
%fourth invariant depends on these three because the fourth eigenvalue of $ 
%\Geff{\mu\nu}
%$ is zero 
%~\cite{Karlovini2003,Gundlach2012})

Following our papers on Newtonian continuum mechanics~\cite{DPRZ2016,DPRZ2017}, we shall decompose 
tensor $ \Geff{\mu\nu} $ on the traceless, $ \devG{\mu\nu} $, 
and spherical parts
\begin{equation}\label{eqn.devG}
\Geff{\mu\nu}  = \devG{\mu\nu} + 
\frac{\GeffMix{\lambda}{\lambda}}{3}
h_{\mu\nu}, \quad \text{ where } \quad
\devG{\mu\nu} := \Geff{\mu\nu} - 
\frac{\GeffMix{\lambda}{\lambda}}{3}
h_{\mu\nu}.
\end{equation}
Note that, in this definition, $ \devG{\mu\nu} $ refers to $ h_{\mu\nu} $ and not to the full 
spacetime metric $ g_{\mu\nu} $. We then use the norm of the traceless part (here one has to use $ 
h^{\mu}_{\ \mu} = g^{\mu\lambda} 
h_{\lambda\mu} = 3 $)
\begin{equation}\label{eqn.dev.norm}
\devGMix{\lambda}{\nu}\devGMix{\nu}{\lambda} = I_2 - 
I_1^2/3,
\end{equation}
as an indication of the presence of non-volumetric deformations, and define the 
specific energy $ \varepsilon(I_1,I_2,\rho,\Entropy) $ as
\begin{equation}\label{eqn.EOS}
\varepsilon(I_1,I_2,\rho,\Entropy) = \varepsilon^{\rm eq}(\rho,\Entropy) + 
\frac{c_s^2}{4} \devGMix{\lambda}{\nu}\devGMix{\nu}{\lambda}
\end{equation}
where $ \varepsilon^{\rm eq} $ is the internal (equilibrium) part of the total 
energy, and $ c_s $ is the sound speed for the shear perturbation propagation. This sound speed may 
depend on $ \rho $ and $ \Entropy $ but we do not consider this possibility in 
this paper. The 
non-equilibrium term (the second term on the right hand-side of \eqref{eqn.EOS}) is due to the 
distortion of the material elements. Such a simple equation of state serves only as an example 
which nevertheless allows us to demonstrate all of the important features of our unified framework 
including dynamics of solids and fluids. It is also important to emphasize that the shear momentum 
transfer (as well as other transfer processes) occurs in an intrinsically transient way in the SHTC 
framework in contrast to the steady-state classical transport theory based on the viscosity concept 
and the steady-state Newton's viscous law. The transient character of our theory is expressed in 
that the transport coefficient are the velocity and time characteristics, $ c_s $ and $ \tau $, 
accordingly. Such characteristics can be recovered only from transient experiments such as high
frequency sound wave propagation, see~\cite{DPRZ2016,HYP2016}. We 
thus stress that the classical steady-state viscosity concept is not used in our theory either 
explicitly nor implicitly.

When we shall deal with gases in the numerical examples we shall use the \textit{ideal gas} 
equation of state for the equilibrium part $ \varepsilon^{\rm eq} $  of the energy
\begin{subequations}\label{eqn.EOS.examples}
\begin{equation}\label{eqn.ideal_gas_eos}
\varepsilon^{\rm eq}(\rho,\Entropy) = \frac{c^2_0}{\gamma(\gamma-1)}, \quad
c_0^2=\gamma\rho^{\gamma-1}e^{\Entropy/c_V},
\end{equation}
or the so-called \textit{stiffened gas equation of state}
\begin{equation}\label{eqn.stiff_gas_eos}
\varepsilon^{\rm eq}(\rho,\Entropy) = \dfrac{c^2_0}{\gamma(\gamma-1)}\left ( 
\dfrac{\rho}{\rho_0}\right 
)^{\gamma-1}e^{\Entropy/c_V} + %\\
\dfrac{\rho_0 c_0^2-\gamma p_0}{\gamma\rho}, \quad c_0^2=const,
\end{equation}
\end{subequations}
if we want to deal with liquids or solids.
In both cases, $c_0$ has the meaning of the adiabatic sound speed\footnote{It is the sound speed 
in the low frequency limit as shown in \cite{DPRZ2016} via the dispersion analysis.}, $ c_V $ is 
the specific heat 
capacity at constant volume, $ \gamma $ is the ratio of the specific heats, \textit{i.e.} $ 
\gamma=c_p/c_V $, if $ c_p $ is the specific heat capacity at constant pressure. 
In~\eqref{eqn.stiff_gas_eos}, $ \rho_0 $ is the reference mass density, $ p_0 $ is the reference 
(atmospheric) pressure. 
%\IP{If one uses $ \gmunu $ instead of $ \hmunu $ in \eqref{eqn.devG} then $ 
%\G^{\langle\lambda}_{\nu\rangle}\G^{\langle\nu}_{\lambda\rangle} = I_2 + 
%\frac{2}{3} I_1^2 $\\}

%We shall also use the mixed tensor
%\begin{equation}\label{eqn.dev.mixed}
%\G^{\langle\mu}_{\nu\rangle} = g^{\mu\gamma}\devG{\G}{\gamma}{\nu} = 
%h^{\mu\gamma}\devG{\G}{\gamma}{\nu} = \G^\mu_\nu - 
%\frac{\GeffMix{\lambda}{\lambda}}{3}
%h^\mu_\nu,
%\end{equation}
%where in the second equality one has to use $ h^{\mu\gamma} h_{\gamma\nu} 
%= h^\mu_\nu $, see \eqref{eqn.h.operators}$ _2 $.

Giving an equation of state for $ \E $ closes the model formulation. Thus, we now 
may compute the anisotropic part $ \sigma^\nu_{\ \mu}  $ of the energy-momentum tensor and the 
dissipative source term in the evolution equation \eqref{GRGPR.A} for the distortion field. 
Differentiating of $ \E $ with respect to $ \dist{\sM }{\mu} $ gives the 
formula similar to its Newtonian analog (see equation (9) in \cite{DPRZ2016}) 
\begin{equation}\label{eqn.eA}
\E_{\dist{\sM }{\mu}} =
%\rho c_s^2
%\KMN^{\sM \sN}\dist{\sN}{\lambda}g^{\lambda\alpha}\left( \Geff{\alpha\beta} - 
%\frac{1}{3} I_1 
%h_{\alpha\beta} \right) h^{\beta \mu} = 
\rho\, c_s^2
\KMN{\sM \sN}\dist{\sN}{\lambda}g^{\lambda\alpha} \devG{\alpha\beta} h^{\beta 
\mu}, 
%\qquad {\color{red} = \rho c_s^2
%\KMN_{\sM\sN}\dist{\sN}{\lambda}g^{\lambda\alpha} \devG{\alpha\beta} h^{\beta 
%\mu}}
\end{equation}
where we have used formulas~\eqref{eqn.dIdA}.

Furthermore, it is now clear how to choose function $ \theta(\tau) $ in 
\eqref{GRGPR.A} to make the 
physical units on the right and left-hand side of \eqref{GRGPR.A} agree, e.g. 
$ \theta(\tau) \sim \rho \,
\tau c_s^2 $. We choose
\begin{equation}\label{eqn.f_tau}
\theta(\tau) = \rho_0\, \tau c_s^2 \GeffMix{\lambda}{\lambda}/3. 
\end{equation}
The trace $ \GeffMix{\lambda}{\lambda}/3 $ appears here for convenience only which later makes it 
easier the computation of the effective viscosity of our model.

%\begin{equation}
%\frac{\pd I_1}{\pd \AAA} = 2\, \og{\bm{g}}{}\AAA\bm{h}^{-1}, 
%\label{eqn.I1A} \qquad \frac{\pd I_2}{\pd \AAA} = 4\,
%\og{\bm{g}}{}\AAA\bm{g}^{-1}\GG\bm{h}^{-1}, \label{eqn.I2A}
%\end{equation}
%where $ \GG $ denotes the matrix with the entries $ \Gmunu $.
Eventually, formula~\eqref{eqn.sigma.p} for the covariant component of the stress tensor-density 
$\Cauchy{\mu\nu}$ 
can be written explicitly 
(recall that we treat $ \rho $ and $\dist{\sM }{\mu} $ as independent variables  in 
\eqref{eqn.sigma.p})
\begin{equation}\label{eqn.shear.stress}
\Cauchy{\mu\nu} = g_{\mu\lambda}\Cauchymix{\lambda}{\nu} = \,g_{\mu\lambda} 
\dist{\sM }{\nu}  \E_{\dist{\sM }{\lambda}} = \rho\, c_s^{2} \devG{\mu\lambda} 
\GeffMix{\lambda}{\nu}.
\end{equation}

Also, note that if one wants to incorporate the volume relaxation (volume viscosity) effect one 
just need to add an extra 
non-equilibrium term in the energy potential:
\begin{equation}\label{sum.E.bulck}
\varepsilon(\dist{\sM }{\mu},\rho,\Entropy) = \varepsilon^{\rm 
eq}(\rho,\Entropy) + 
\frac{c_s^2}{4} \devGMix{\lambda}{\nu}\devGMix{\nu}{\lambda} + \frac{c_v^2}{2}  \left ( 
\frac{w}{\sqrt{-g}} - 
\frac{\rho}{\rho_0}\right )^2,
\end{equation}
where $ c_v $ is a velocity that will contribute to the sound speed of propagation of volume 
perturbations at high frequencies, $ w = \det(\dist{m}{\mu}) $, see \eqref{det.A}. This extra 
term, of course, affects the computation of
\eqref{eqn.eA} and \eqref{eqn.shear.stress}. An effective volume viscosity then can be identified 
as $ \sim \tau (c_0^2 + c_v^2) $ in the same way as we identify the effective shear viscosity, see 
details in 
Sec.\,\ref{sec.asymptotic}. Here, $ c_0 $ is the adiabatic (low frequency) sound speed from 
\eqref{eqn.EOS.examples}.

Finally, we have to prove that our choice of the energy potential is consistent with the rest mass 
density conservation law, i.e. that the condition \eqref{orth.source} is fulfilled. Indeed, one has 
(without losing the generality, we shall omit $ \rho c_s^2 >0 $ factor in $ \E_{\dist{\sN}{\nu}} 
$):
\begin{align}\label{orth.source.prove}
\rho_{\dist{\sM}{\mu}} \KMNContr{\sM \sN} g_{\mu\nu} \E_{\dist{\sN}{\nu}} = \rho \idist{\mu}{\sM} 
\KMNContr{\sM 
\sN} g_{\mu\nu} \E_{\dist{\sN}{\nu}} &= \nonumber\\
\rho \idist{\mu}{\sM} \KMNContr{\sM \sN} g_{\mu\nu} \KMN{\sN 
\sK}\dist{\sK}{\lambda}g^{\lambda\alpha} 
\devG{\alpha\beta} h^{\beta \nu} &= \nonumber\\
\rho \devG{\alpha\beta} h^{\beta \alpha} = \rho h^{\beta\alpha}\left (G_{\alpha\beta} - 
\frac{I_1}{3} h_{\alpha\beta}\right ) &= \nonumber\\
\rho \left( I_1 - \frac{I_1}{3} h^{\beta\alpha} h_{\alpha\beta}\right) &= 0.
\end{align}

\subsection{Asymptotic analysis in the diffusive regime}\label{sec.asymptotic}

As we discussed in the previous section, our theory does not rely on the classical steady-state 
viscosity approach, but nevertheless, we may perform a formal asymptotic analysis for $ \tau 
\rightarrow 
0 $ (i.e. we are in the diffusive regime) and show that, in the leading terms, 
the viscous stress is similar to the relativistic Navier-Stokes stress obtained by Landau and 
Lifshitz~\cite{Landau-Lifshitz6} and thus we may obtain an expression for the \textit{effective} 
viscosity coefficient in terms of transient characteristics $ c_s $ and $ \tau $. However, it is 
necessary to recall that the 
relativistic Navier-Stokes stress leads to the \textit{acausal} governing PDEs, 
e.g.~\cite{MullerRuggeri1998,RezzollaZanottiBook}, which results also in the 
numerical instabilities. Because our model is hyperbolic and hence causal, it 
is thus clear that this is the high order terms in $ \tau $ of the model who damp the unstable 
modes in the relativistic Navier-Stokes equations.

In this section, it is convenient to use the evolution equation for the effective metric tensor $ 
\Geff{\mu\nu} $ since the stress $ \Cauchy{\mu\nu} $ depends on $ \dist{\sM}{\mu} $ only through $ 
\Geff{\mu\nu} $. The 
evolution of $ \Geff{\mu\nu} $ reads as\footnote{One may also recognize that the reversible part 
of 
the $ \Geff{\mu\nu} $ evolution is the Lie derivative along the 4-velocity.}
\begin{equation}\label{eqn.G.PDE}
u^\lambda \nabla_\lambda \Geff{\mu\nu} + \Geff{\mu\lambda}\nabla_\nu 
u^\lambda + 
\Geff{\lambda\nu}\nabla_\mu u^\lambda = 
-\frac{2}{\tau\,\vartheta}\Cauchy{\mu\nu},
%\dist{\sM }{\mu}g^{\sM \sN} S^B_\nu + S^\sM _\mu g^{\sM \sN}\dist{\sN}{\nu}
\end{equation}
where $ \vartheta = \theta/\tau = \rho_0 \, c_s^2 \GeffMix{\lambda}{\lambda}/3 $.
This PDE is the direct consequence of the distortion evolution \eqref{GRGPR.A}, the advection 
equation\footnote{Here, we use that the components of the material metric field 
$\KMNContr{\sM \sN}$ are transformed like scalars with respect to the change of the 
Eulerian 
coordinates $ x^\mu $.} for 
$\KMNContr{\sM \sN}$, 
$u^\lambda\nabla_\lambda \KMNContr{\sM \sN} = 0,$
and the identity
\begin{equation}\label{eqn.dG}
\rmd\Geff{\mu \nu}=\dist{\sM }{\mu}\KMNContr{\sM \sN}(\rmd\dist{\sN}{\nu})+
\dist{\sM }{\mu}(\rmd \KMNContr{\sM \sN})\dist{\sN}{\nu}+(\rmd\dist{\sM 
}{\mu})\KMNContr{\sM 
\sN}\dist{\sN}{\nu}.
\end{equation}

After taking the traceless part\footnote{Note that, in order to write the 
deviatoric part $ \devG{\mu\nu} $ under the derivative $ 
u^\lambda\nabla_\lambda \devG{\mu\nu} $ one has to use 
the fact that $ \nabla_\lambda g_{\mu\nu} = 0 $.} of the covariant PDE for $ 
\Geff{\mu\nu} $ \eqref{eqn.G.PDE}
%\begin{equation}
%u^\lambda \nabla_\lambda \Geff{\mu\nu} + \Geff{\mu\lambda}\nabla_\nu 
%u^\lambda + 
%\Geff{\lambda\nu}\nabla_\mu u^\lambda = -\frac{2}{\tau \, 
%\vartheta}\sigma_{\mu\nu}
%\end{equation}
and replacing $ \Geff{\lambda\beta} $ by $ \devG{\lambda\beta} + 
\frac{1}{3}\GeffMix{\lambda}{\lambda} h_{\lambda\beta}( = \Geff{\lambda\beta}) $,
we 
have the following PDE
\begin{equation}
u^{\lambda }\nabla _{\lambda }\devG{\mu \nu }+\devG{\mu \lambda }\nabla 
_{\nu }u^{\lambda }+\devG{\lambda \nu }\nabla _{\mu }u^{\lambda
}+2\traceG  D_{\mu \nu } - % \\
\frac{2}{3} \left( h^{\alpha \beta }G_{\lambda 
\beta }\nabla _{\alpha }u^{\lambda }\right ) h_{\mu \nu } + {\traceG 
u^\lambda \nabla_\lambda h_{\mu\nu}} = -\frac{2}{\tau \, 
\vartheta}\devsigma{\mu\nu},
\end{equation}
where
\begin{subequations}
	\begin{align} \label{eq:dsk}
		D_{\mu \nu } &:= \frac{1}{2}\left(h_{\mu \lambda }\nabla _{\nu }u^{\lambda 
		}+h_{\lambda \nu }\nabla _{\mu }u^{\lambda }\right), \\
		\devsigma{\mu\nu} &:= \sigma_{\mu\nu} - \frac{\sigma^\lambda_{\ 
		\lambda}}{3} h_{\mu\nu}, \qquad \traceG := \frac{1}{3}\GeffContr{\lambda}_{\ 
		\lambda}
	\end{align}
\end{subequations}
are the symmetric 4-velocity gradient, the trace-less part of the stress 
tensor, and one-third of the trace of $ \Geff{\mu\nu} $, accordingly.

After once more replacing $ \Geff{\lambda\beta} $ by $ \devG{\lambda\beta} + 
\traceG 
h_{\lambda\beta} $ we arrive at the final PDE for the 
trace-less 
part $ \devG{\mu\nu} $ 
\begin{equation}\label{eqn.devG.PDE}
u^{\lambda }\nabla _{\lambda }\devG{\mu \nu }+\devG{\mu \lambda 
}\nabla _{\nu }u^{\lambda }+\devG{\lambda \nu }\nabla _{\mu }u^{\lambda
} + 2\traceG  \mathring{D}_{\mu \nu } - % \\
\frac{2}{3}\left (h^{\alpha \beta 
}\devG{\lambda \beta }\nabla _{\alpha 
}u^{\lambda }\right ) h_{\mu \nu } + {\traceG 
u^\lambda \nabla_\lambda h_{\mu\nu}}= 
-\frac{2}{\tau \, \vartheta}\devsigma{\mu\nu},
\end{equation}
where $\mathring{D}_{\mu\nu} = D_{\mu\nu} - (D^\lambda_{\ 
\lambda}/3) h_{\mu\nu} $ is the trace-less part of the symmetric velocity gradient $ D_{\mu\nu} $. 
We now assume that the solution to \eqref{eqn.devG.PDE} can be written as 
\begin{subequations}
	\begin{align}\label{eqn.devG.series}
		\devG{\mu\nu} &= \devG{\mu\nu}^{(0)} + \tau \devG{\mu\nu}^{(1)} + \tau^2 
		\devG{\mu\nu}^{(2)} + \ldots \, , \\
		\rho &= \rho_0 + \tau \rho_1 + \tau^2 \rho_2 + 
		\ldots\, , \\
		\traceG &= \traceG^{(0)} + \tau \traceG^{(1)} + \tau^2 \traceG^{(2)} + \ldots\, ,
	\end{align}
\end{subequations}
and will show that the relativistic Euler equations are recovered if only zeroth order terms are 
retained, and the relativistic Navier-Stokes equations are recovered if the first-order terms in $ 
\tau $ are retained.
\paragraph{Zeroth-order approximation (Euler fluid):} After plugging 
\eqref{eqn.devG.series} into the PDE 
\eqref{eqn.devG.PDE}, and collecting leading 
order terms ($ \tau^{0} $), we have that
\begin{equation}\label{eqn.devG.0}
\devG{\mu\nu}^{(0)} = 0,
\end{equation}
and hence
\begin{equation}\label{eqn.traceG0}
\Geff{\mu\nu}^{(0)} = \traceG^{(0)} h_{\mu\nu} \quad \text{ with } \quad 
\traceG^{(0)} = \frac{1}{3}g^{\alpha\beta}\Geff{\beta\alpha}^{(0)} = 1,
\end{equation}
which results in that $ \Cauchy{\mu\nu} = 0$ and hence, the energy-momentum tensor reduces to the 
one of the relativistic Euler equations.

\paragraph{First-order approximation (Navier-Stokes):} If we now plug \eqref{eqn.devG.series} 
into the stress tensor  $ \sigma_{\mu\nu} =  \rho\, c_s^{2} 
\devG{\mu\lambda} \GeffMix{\lambda}{\nu} $ and keep only leading terms of the order $ \tau^1 $, 
we have 
\begin{equation}\label{eqn.sigma.series}
\sigma_{\mu\nu} =  \, c_s^2 \left((\rho_0 + \tau \rho_1)\devG{\mu\lambda}^{(0)}
\Geff{\alpha\nu}^{(0)} + %\right. \\ \left. 
\tau \rho_0 \left (\Geff{\alpha\nu}^{(1)}\devG{\mu\lambda}^{(0)} + 
\devG{\mu\lambda}^{(1)} \Geff{\alpha\nu}^{(0)}\right)\right)h^{\lambda\alpha} 
+ 
\ldots
\end{equation}
which, because of \eqref{eqn.devG.0}, \eqref{eqn.traceG0}, and the orthogonality 
condition, simplifies to
\begin{equation}\label{eqn.sigma.series2}
\sigma_{\mu\nu} = \rho_0 c_s^2 \tau \devG{\mu\lambda}^{(1)} 
\Geff{\alpha\nu}^{(0)}h^{\lambda\alpha} = % \\
\rho_0 c_s^2 \tau 
\devG{\mu\lambda}^{(1)} 
\traceG^{(0)} h^{\lambda}_{\ \nu} =\rho_0 \tau c_s^2 
\devG{\mu\lambda}^{(1)} h^{\lambda}_{\ \nu}.
\end{equation}
It also follows from this result that $ h^{\alpha\beta} \Cauchy{\alpha\beta} = 
0 $, 
i.e. at first order the stress $ \Cauchy{\mu\nu} $ is trace-less, $
\devsigma{\mu\nu} = \Cauchy{\mu\nu} $.

From the other hand, if we plug expansion~\eqref{eqn.devG.series} into the PDE 
\eqref{eqn.devG.PDE}, in leading terms, we have 
\begin{equation}\label{eqn.NSE}
2 \mathring{D}_{\mu\nu} + { u^\lambda \nabla_\lambda h_{\mu\nu}} 
= - \frac{2}{\rho_0 c_s^2}\rho_0 c_s^2 \devG{\mu\lambda}^{(1)} h^{\lambda}_{\ \nu}.
\end{equation}

Now, using that $ \devG{\mu\lambda}^{(1)} h^{\lambda}_{\ \nu} = 
\Cauchy{\mu\nu}/(\rho_0 \tau c_s^2) $, see \eqref{eqn.sigma.series2},
and the definition of $ 
\mathring{D}_{\mu\nu} $, the last equality transforms into
\begin{equation}\label{eqn.NSE.stress}
-\frac{1}{ \rho_0 \tau c_s^2}\Cauchy{\mu\nu} = \frac{1}{2}\left( 
h_{\mu\lambda}\nabla_\nu u^\lambda
+ h_{\lambda\nu}\nabla_\mu u^\lambda - \vphantom{\frac{2}{3}} 
% \right. \\ \left.
\frac{2}{3} (h^\alpha_{\ \lambda} \nabla_\alpha u^\lambda) h_{\mu\nu} + 
{ u^\lambda \nabla_\lambda h_{\mu\nu}}
 \right),
\end{equation}
which is equivalent to the Landau-Lifshitz version of the relativistic Navier-Stokes stress 
\cite{Landau-Lifshitz6,RezzollaZanottiBook} (to see this, it is only necessary to take into account 
that $ u_\lambda \nabla_\nu u^\lambda = 0 $ and $ \nabla_\lambda g_{\mu\nu} = 0$)
\begin{equation}\label{IS.NSstress1}
\tau^{\mathsmaller{\rm NS}}_{\mu\nu} := -\mu \left ( \nabla_\nu u_\mu + \nabla_\mu u_\nu + 
u^\lambda \nabla_\lambda (u_\mu u_\nu) - \vphantom{\frac{2}{3}}
%\right.\\  \left. 
\frac{2}{3} 
(h^{\alpha\lambda}\nabla_{\lambda}u_\alpha) h_{\mu\nu} \right ).
\end{equation}
Hence, for small $ \tau $, we may identify an \textit{effective shear viscosity} coefficient for 
our model:
\begin{equation} 
    \mu := \frac{1}{6} \rho_0 \tau c_s^2.
\end{equation}

%Also, note that the last term on the right-hand side of \eqref{eqn.NSE.stress} 
%is due to the relativistic effect and it vanishes in the Newtonian limit thanks 
%to that $h_{\mu\nu}$ becomes the identity matrix.

%Finally, it worths to mention that in such an approximation, the anisotropic part $ 
%\Cauchy{\mu\nu} 
%$ 
%of the energy-momentum tensor is identical to the on of

%One may see that in the special relativistic context, formulas 
%\eqref{eqn.NSE.stress} and \eqref{eqn.Landau.stress} differ only in the 
%definition of the symmetric velocity gradient (first two terms on the 
%right-hand 
%sides). The last terms on the right-hand sides are equivalent because  $ 
%u^\lambda\nabla_\lambda h_{\mu\nu} = u^\lambda\nabla_\lambda (u_\mu u_\nu) $ 
%thanks to $ \nabla_\lambda g_{\mu\nu} = 0 $.

\paragraph{3+1 version:}\label{eqn.asympt.3plus1}

Since, in the numerical simulation, we use the 3+1 split of the spacetime, it 
is 
also necessary to know the viscosity in this case. Thus, using that $ 
(\gamma,\gamma\hat{v}^i) = (u^0,u^i) $, $ \gamma = \alpha^{-1} W $, where $ \alpha $ is the time 
lapse and $ W $ is the Lorentz factor, we can 
rewrite the stress \eqref{eqn.NSE.stress} as
\begin{equation}\label{eqn.stress.split}
-\sigma_{\mu\nu} = \mu \left(h_{\mu 0}\pd _{\nu }\gamma +h_{\mu  
j}\pd _{\nu }(\gamma \hat{v}^j)+h_{0\mu }\pd _{\mu }\gamma
+ \vphantom{\frac{2}{3}}  % \right .\\ 
h_{j \nu }\pd _{\mu }(\gamma 
\hat{v}^j)-\frac{2}{3}(h^{\alpha }_{\ 0}\pd _{\alpha }\gamma 
+h^{\alpha }_{\ j}\pd _{\alpha }(\gamma
\hat{v}^j))h_{\mu \nu } + %\right. \\
\vphantom{\frac{2}{3}}
\gamma \pd _0 h_{\mu \nu }+\gamma 
\hat{v}^j\pd _j h_{\mu \nu }\right).
\end{equation}
Further, using that $ h_{\mu\nu} $ is the projector, we may write
\begin{equation}
h_{\mu 0}\pd _{\nu }\gamma + h_{\mu j}\pd _{\nu }(\gamma \hat{v}^j) = h_{\mu 
0}\pd _{\nu }\gamma + % \\ 
h_{\mu j}\hat{v}^j\pd_\nu \gamma  + \gamma h_{\mu j}\pd_\nu \hat{v}^j
= \gamma h_{\mu 
j}\pd_\nu \hat{v}^j,
\end{equation}
and hence
\begin{equation}
-\sigma_{\mu\nu} = \mu \gamma \left(h_{\mu  j}\partial _{\nu }\hat{v}^j+h_{j 
\nu 
}\partial _{\mu }\hat{v}^j - 
\vphantom{\frac{2}{3}} 
% \right .\\   \left . 
\frac{2}{3}(h_j^{\alpha }\partial _{\alpha 
}\hat{v}^j)h_{\mu
\nu } 
+ \partial_t h_{\mu \nu }+\hat{v}^j\partial _j h_{\mu \nu }\right).
\end{equation}

Therefore, if $ \mu $ denotes the viscosity coefficient in the covariant 
version \eqref{eqn.NSE.stress} and $ \hat{\mu} $ denotes the viscosity 
coefficient in the 3+1 split then they relate to each other as
\begin{equation}\label{eqn.2viscosities}
\hat{\mu} = \gamma \mu, \qquad \gamma = \alpha^{-1}W.
\end{equation}

\subsection{Comparison with the M\"{u}ller-Israel-Stewart model}\label{sec.IS}

In this section we compare our equations with the state of the art relativistic 
dissipative fluid dynamics model, the Israel-Stewart model (also known as 
M\"{u}ller-Israel-Stewart model) \cite{Mueller1966PhD,Israel1976,Stewart1977}, 
which is actively used in the special relativistic context, and in 
particular for relativistic heavy-ion collisions, 
e.g.~\cite{DelZanna2013,Karpenko2014,Shen2016}. However, since we ignore in this paper the 
volume 
relaxation effect and the heat conduction, we shall compare only the PDEs for 
the evolution of the shear stress tensor. The volume relaxation can be
incorporated in the current framework as it is discussed in \eqref{sum.E.bulck}. The hyperbolic 
heat conduction in the SHTC 
framework also has a variational nature, e.g. see~\cite{SHTC-GENERIC-CMAT}, 
and can be incorporated in a straightforward manner in the present model. This will 
be the subject for further publications. 

The Israel-Stewart equation for the trace-less shear stress tensor $ 
\tau_{\mu\nu} $ (without the 
temperature terms) reads as, see e.g. \cite{RezzollaZanottiBook},
\begin{equation}\label{IS.equations}
\lambda \, h^{\alpha}_{\ \mu} h^{\beta}_{\ \nu}\DT{\tau}_{\alpha\beta}
+ 
\tau_{\mu\nu} = -2 \mu \tau^{\mathsmaller{\rm 
NS}}_{\mu\nu}
%- \left[\mu T\nabla_\gamma \left(\frac{\lambda}{2 \mu T}u^\gamma\right) 
%\tau_{\mu\nu} \right]
,
\end{equation}
where 
\begin{equation*}\label{IS.NSstress}
\tau^{\mathsmaller{\rm NS}}_{\mu\nu} := \frac{1}{2}\left (L_{\nu\mu} + L_{\mu\nu} + 
u^\gamma \nabla_\gamma (u_\mu u_\nu) \right ) - \frac{1}{3} 
(h^{\alpha\gamma}\nabla_{\gamma}u_\alpha) h_{\mu\nu},
\end{equation*}
is the relativistic Navier-Stokes stress tensor, $ \lambda $ is the relaxation time associated with 
the time scale of the relaxation of the shear stress to its equilibrium value $ 
\tau^{\mathsmaller{\rm NS}}_{\mu\nu} $, the overdot symbol 
stands for 
the \textit{convective 
derivative} $ \DT{\tau}_{\mu\nu} := u^\gamma \nabla_\gamma \tau_{\mu\nu} $, 
while $ L^\mu_{\ 
\nu} = 
\nabla_\nu u^\mu $, $ L_{\mu\nu} = g_{\mu\gamma} L^\gamma_{\ \nu}$ is the 
velocity gradient tensor. Therefore, it only remains to obtain the evolution equation for the 
shear stress tensor~\eqref{eqn.shear.stress} and to compare it with \eqref{IS.equations}. This is 
however not a trivial 
task in the general case even for the simple equation of state \eqref{eqn.EOS} due to the nonlinear 
relation between the distortion 
field $ \dist{\sM}{\mu} $ and the shear stress $ \sigma_{\mu\nu} $. Nevertheless, this can be 
done relatively easily if to assume the smallness of $ \sigma_{\mu\nu} $ in $ 
\devG{\mu\nu} $, that is we shall assume the approximation
\begin{equation}\label{IS.smallness}
\sigma_{\mu\nu} = \rho \, c_s^2 \devG{\mu\lambda} G^\lambda_{\ \nu} = \rho \, c_s^2 
\devG{\mu\lambda} (\devGMix{\lambda}{\nu} + \traceG h^\lambda_{\ \nu}) \approx 
\rho \, c_s^2 \, \traceG \, \devG{\mu\nu},  \qquad % \\ 
\traceG = \frac{1}{3}\GeffContr{\lambda}_{\ \lambda}.
\end{equation}
Also, recall that in this paper it is assumed that $ c_s = const$. We note that, in general, the 
shear stress $ \sigma_{\mu\nu} $ is not necessary 
trace-less and hence already has some contribution to the bulk viscosity effect but its linear 
approximation in $ \devG{\mu\nu} $ 
\eqref{IS.smallness} has obviously zero trace.  

We then apply the convective derivative $ u^\gamma \nabla_\gamma 
$ to the  approximation~\eqref{IS.smallness} to obtain
\begin{equation}\label{IS.conv.deriv.sigma1}
\DT{\sigma}_{\mu\nu} = \DT{\rho} c_s^2 \traceG \devG{\mu\nu} + \DT{\traceG} 
\rho 
c_s^2 \devG{\mu\nu} + \DT{\devG{}}_{\mu\nu} \rho c_s^2 \traceG.
\end{equation}
Note that if one wants to obtain the PDE for the full shear stress tensor, 
\eqref{IS.conv.deriv.sigma1} should be replaced with 
\begin{equation}\label{IS.conv.deriv.sigma11}
\DT{\sigma}_{\mu\nu} = \DT{\rho} c_s^2 ( \traceG \devG{\mu\nu} + 
\devG{\mu\lambda}\devGMix{\lambda}{\nu}) + \DT{\traceG} \rho 
c_s^2 \devG{\mu\nu} + % \\
\rho c_s^2 (\traceG \DT{\devG{}}_{\mu\nu} + 
\DT{\devG{}}_{\mu\lambda}\devGMix{\lambda}{\nu}) + \rho c_s^2 
\devG{\mu\lambda}\DT{\devGMix{\lambda}{\nu}}.
\end{equation}
From the rest mass conservation $ \nabla_{\mu}(\rho u^\mu ) = 0$ we can obtain that 
$ \DT{\rho} = -\rho L^\lambda_{\ \lambda} $, while the PDE for $ \DT{\traceG} $ 
can be obtained by applying $ u^\gamma \nabla_{\gamma} $ to \eqref{eqn.G.PDE}. 
Thus, we have
\begin{equation}\label{IS.dot.kappa}
\DT{\traceG} = -\frac{2}{3}(\devGMix{\nu}{\lambda} + \traceG h^\nu_{\ \lambda}) 
L^\lambda_{\ \nu}.
\end{equation}
Eventually, the evolution equation for $ \DT{\devG{}}_{\mu\nu} $ is given by 
\eqref{eqn.devG.PDE}. 

After plugging expressions for $ \DT{\rho} $, $ \DT{\traceG} $, and  $ 
\DT{\devG{}}_{\mu\nu} $ into \eqref{IS.conv.deriv.sigma1} and after a few term
rearrangements we arrive at
\begin{subequations}
\begin{equation}\label{IS.conv.deriv.sigma2}
\frac{\tau}{2}\left( 
\DT{\sigma}_{\mu\nu} + \sigma_{\mu\nu} L^\lambda_{\ \lambda} + \frac{5}{3} 
\sigma_{\mu\lambda} L^\lambda_{\ \nu} + \sigma_{\lambda\nu} L^\lambda_{\ \mu} - 
\vphantom{\frac23} 
% \right.\\ \left.
\frac{2}{3} (\sigma^\alpha_{\ \lambda} L^\lambda_{\ \alpha})  h_{\mu\nu} 
\right) 
+
\sigma_{\mu\nu} = - 2 \mu \sigma^{\mathsmaller{\rm NS}}_{\mu\nu},
\end{equation}
where
\begin{equation}\label{IS.sigma.NS}
\sigma^{\mathsmaller{\rm NS}}_{\mu\nu} = \frac{1}{2}\left( 
L_{\mu\nu} + L_{\nu\mu} + u^\gamma \nabla_\gamma (u_\mu u_\nu) 
\right) -
\frac{1}{3}(h^\alpha_{\ \lambda} L^\lambda_{\ \alpha}) h_{\mu\nu}.
\end{equation}
\end{subequations}

Thus, we can directly compare equations \eqref{IS.equations} and 
\eqref{IS.conv.deriv.sigma2}. It is obvious that $ \tau^{\mathsmaller{\rm 
NS}}_{\mu\nu} $
and $ \sigma^{\mathsmaller{\rm NS}}_{\mu\nu} $ are identical and the main 
difference is in the first terms of \eqref{IS.equations} and 
\eqref{IS.conv.deriv.sigma2}. Thus, one may expect that the solutions to the Israel-Stewart 
theory and our theory are close for the flows not far from the equilibrium, 
i.e. when the Navier-Stokes terms dominate. The solutions will diverge as long 
as the flow becomes more and more non-equilibrium. A detailed comparison of the 
theories in the non-equilibrium settings is outside the scope of this study.
Moreover, it is necessary to underline that the PDE 
\eqref{IS.conv.deriv.sigma2} is the result of the specific and very simple choice \eqref{eqn.EOS} 
of the energy 
$ \E (\dist{\sM}{\mu},\rho,s) $. Each time we specify the energy, the stress tensor 
\eqref{eqn.shear.stress} $ \sigma_{\mu\nu} = g_{\mu\alpha} 
\dist{\sM }{\nu}  \E_{\dist{\sM }{\alpha}} $ changes, and hence, 
\eqref{IS.smallness}, \eqref{IS.conv.deriv.sigma1}, and \eqref{IS.conv.deriv.sigma2} change as 
well.  It is thus unlikely that the PDE for the stress tensor may have a certain structure. In 
contrast, the structure of the distortion evolution equation~\eqref{GRGPR.A} is canonical in the 
sense that it is invariant with respect to the choice of the closure
due to the geometrical nature of the distortion field (non-holonomic bases tetrad).
Apparently,  the use of the full shear stress tensor $ \sigma_{\mu\nu} = \rho c_s^2 
\devG{\mu\lambda} G^\lambda_{\ \nu} $ instead of its  approximation \eqref{IS.smallness}, and hence 
\eqref{IS.conv.deriv.sigma11} instead of \eqref{IS.conv.deriv.sigma1}, only increases the 
complexity of the structure of \eqref{IS.conv.deriv.sigma2} by adding new nonlinear terms to the 
existing 
ones. However, to obtain an explicit time evolution even in the case of the, perhaps, most simplest 
energy expression \eqref{eqn.EOS} which leads to the shear stress $ \sigma_{\mu\nu} = \rho c_s^2 
\devG{\mu\lambda} G^\lambda_{\ \nu} $ is hopeless.

It may also seem that the connection of the Israel-Stewart theory with the Boltzmann 
equation via Grad's moment method, e.g. \cite{HuovinenMolnar2009,Denicol2013,Molnar2012}, and 
the absence of such a connection for the SHTC equations, says in favor of the use of the 
Israel-Stewart-type models in relativistic fluid dynamics.  On the other hand, the recently 
established consistency~\cite{SHTC-GENERIC-CMAT} of the SHTC equations with the GENERIC (General 
Equation for 
Non-Equilibrium Reversible-Irreversible Coupling) provides a possibility to connect the SHTC theory 
with the fundamental equation of statistical mechanics, the Liouville equation for the $ N 
$-particle 
distribution function, which, as well as the SHTC equations, has a Hamiltonian structure and 
simultaneously applicable to gases, liquids 
and 
solids. Recall that the Boltzmann equation can 
be obtained as a 
reduction of the Liouville equation for the one-particle distribution function~\cite{Pavelka2016}. 
Moreover, the lack of a Hamiltonian formulation for the Israel-Stewart-type models destroys the 
variational structure of the Einstein field equations if one tries to couple such models with 
the gravity field. This means that the Israel-Stewart stress tensor has to be added to the matter 
energy-momentum of GR in an \textit{ad hoc} manner.

\subsection{Thermodynamical consistency of the governing equations} 
\label{sec.therm.consist}

In this section, we demonstrate that the governing equations \eqref{GRGPR} 
constitute a thermodynamically compatible system, that is the first and the 
second law of the
thermodynamics are fulfilled on the solution to \eqref{GRGPR}.

As in Section~\ref{sec.SHTC}, we may specify the conservative state variables
and the thermodynamic potential as (here, we need to use the orthogonality 
property 
\eqref{eqn.Cauchy.orthogon}$ _2 $) $ \mathscr{E}(q_\ell):= -\EMT{\mu}{0} u_\mu $ 
and $ q_\ell := 
(-\EMT{\mu}{i} u_\mu, \cgrad{\sA}{\nu}u^\mu u_\mu,\rho u^\mu u_\mu, s  u^\mu 
u_\mu) $, $ \ell = 
1,2,\ldots,17 $. Hence, we have
\begin{equation}\label{symm.E.Q}
\scE(q_\ell) = -u_0 \E, \qquad q_\ell = (-u_i \E,-\dist{\sM}{\nu} 
-\rho,-s ).
\end{equation}
Now, using the fact that $ \scE = -u_0 \E $ is not an unknown but is a 
potential of $ q_\ell $, we conclude that the PDE system \eqref{GRGPR} is 
an overdetermined system of equations because the number of equations is one 
more than the number of unknowns. This means that one of the equations, say the 
energy conservation, can be obtained as a linear combination of the others. 
In order to see this, let us introduce the new variables $ p^{\ell} $ 
and the new potential $ L(p^\ell) $ as the 
Legendre conjugates to $ q_{\ell} $ and  $ \scE(q_\ell) $
\begin{equation}\label{symm.p}
p^\ell := \mathscr{E}_{q_\ell}, \quad L := q_\ell \scE_{q_\ell} - \scE = 
q_\ell\, p^\ell - \scE, 
\quad
%= -\frac{u^i}{u^0}, \qquad p^4 := \mathscr{E}_{q_4} = -\frac{\ED_\rho}{u^0}, 
%\qquad p^5 := \mathscr{E}_{q_5} = -\frac{\ED_s }{u^0},
\end{equation}
or explicitly
\begin{subequations}
\begin{align}
p^\ell &= \left( -\frac{u^i}{u^0}, -\frac{\E_{\dist{\sM}{\lambda}}}{u^0}, 
-\frac{\E_\rho}{u^0}, -\frac{\E_s}{u^0} \right), \label{therm.p.explicit} \\
L &= \frac{1}{u^0}\left(\dist{\sM}{\lambda} \E_{\dist{\sM}{\lambda}} + \rho 
\E_\rho 
+s \E_s - \E\right). \label{therm.L.explicit} 
\end{align}
\end{subequations}
Then, it is a matter of a straightforward verification that the energy 
conservation (the zeroth equation in \eqref{GRGPR.T}) can be obtained as 
the linear combination of the other equations with the conjugate 
variables $ p^\ell $ playing the role of the coefficients
\begin{multline}\label{therm.summation}
	\nabla_\mu\left[ \E \, u^\mu u_0 + p h^\mu_{\ 0} + 
	\Cauchymix{\mu}{0} \right] = \\  
	- \dfrac{u^i}{u^0} \nabla_\mu\left[ 
	\E \, u^\mu 
	u_i + p h^\mu_{\ i} + 
	\Cauchymix{\mu}{i} \right] - % \\[1mm]
	\dfrac{\E_{\dist{A}{\mu}}}{u^0} 
	\left[ u^\nu \nabla_\nu\dist{m}{\mu} + \dist{m}{\nu} \nabla_\mu u^\nu \right] - 
	\dfrac{\E_\rho}{u^0} \nabla_\mu (\rho 
	u^\mu) - \dfrac{\E_s }{u^0}
	\nabla_\mu (s u^\mu).
\end{multline}
The dissipative sources in \eqref{GRGPR.A} and \eqref{GRGPR.s} are 
obviously annihilated in \eqref{therm.summation} and hence the way the 
dissipation is introduced in \eqref{GRGPR} does not violate the
energy conservation law (first law of thermodynamics). Moreover, because $ 
\KMNContr{\sM \sN}$ and $ g_{\mu\nu}  $ are assumed to be positive definite, the 
entropy production (the right hand-side in \eqref{GRGPR.s}) is positive and 
thus the second law of thermodynamics is 
also fulfilled. In other words, the system of governing equations 
\eqref{GRGPR} is a thermodynamically compatible system.

%This system is overdetermined because there are one more equations than the 
%unknowns. Thus, one may 
%show that one of the equations, e.g. the entropy 
%conservation, can be obtained as a linear combination of the others
%\begin{multline}\label{symm.summation1}
%u^\nu \nabla_\mu\left[ E\,u^\mu u_\nu + p h^\mu_{\ \nu} + 
%\cgrad{A}{\nu}\ED_{\cgrad{A}{\mu}} \right] + \ED_{\cgrad{A}{\mu}} \left[ 
%\nabla_\mu 
%(u^\nu 
%\cgrad{A}{\nu}) + u^\nu (\nabla_\nu \cgrad{A}{\mu} - 
%\nabla_\mu \cgrad{A}{\nu}) \right] + \ED_\rho \nabla_\mu (\rho u^\mu) = \\ - 
%\ED_s 
%\nabla_\mu (\sigma u^\mu).
%\end{multline}

\subsection{Causality and hyperbolicity}
\label{sec.SHTC.relasticity}

One of the key features of the SHTC equations is that the reversible part (left 
hand-side of \eqref{GRGPR}) and irreversible part (right hand-side of 
\eqref{GRGPR}) of the evolution are treated separately. Because the 
irreversible terms are algebraic, the type of the governing equations is 
defined only by the reversible part which constitutes the principal symbol of 
the PDE system \eqref{GRGPR}. In particular, this means that if the 
reversible part of the evolution is hyperbolic then the entire dissipative 
system is hyperbolic as well even in the diffusive regime when the relaxation 
time goes to zero, e.g. see the dispersion analysis in~\cite{DPRZ2016}. 
Moreover, all signals propagate at the speeds slower than the characteristic 
speeds (eigenvalues) of the reversible part. These characteristic speeds are 
the upper bounds for the signal speeds (sound speeds) of the complete 
system~\eqref{GRGPR} (including the relaxation terms) at high frequencies, 
see~\cite{DPRZ2016}.

It has appeared that in contrast to the 
non-relativistic case, the symmetric form of the relativistic equations written 
in the Eulerian coordinates is not easy to obtain due to their covariance. 
Moreover, due to the nonlinearity of the system ~\eqref{GRGPR}, it is also 
not clear if it is possible to compute the eigenvalues (characteristic speeds) 
and 
eigenvectors of the system analytically. Therefore, in this paper we may give 
only indirect evidences of that the system ~\eqref{GRGPR} is hyperbolic. 
First of all, one can be certain that the characteristic speeds 
of~\eqref{GRGPR} are real and finite because they can be obtained from the 
characteristic speeds of the Lagrangian system~\eqref{eqn.Lagr.motion.U} which 
is symmetric hyperbolic (if $ U $ is convex) and hence has real eigenvalues. {
Moreover, the same dispersion analysis as for the non-relativistic equations in ~\cite{DPRZ2016} 
can be performed which, in particular, gives the maximum material sound speed $ c_\infty^2 = c_0^2 
+ 4 c_s^2 /3 $ at high frequencies, i.e. $ \omega \rightarrow \infty $.}
Secondly, Eulerian system~\eqref{GRGPR} is a thermodynamically compatible system 
which is the key for the symmetrization in the framework of SHTC equations. The 
problem here is that the technique we used to symmetrize the non-relativistic 
equations can not be straightforwardly generalized to the covariant 
relativistic equations. Thirdly, the fact that we were able to compute the 
numerical solution in the wide range of the state variables (e.g. the 
distortion field) with the methods specifically designed for hyperbolic 
equations and which do not employ different stability improvement techniques 
(artificial dumping of high 
frequencies modes, etc.) says that the initial value problem for 
~\eqref{GRGPR} is likely well-posed. We plan to investigate the 
hyperbolicity, and symmetric 
hyperbolicity in particular, of~\eqref{GRGPR} in a more rigorous way in 
future publications.

\section{Numerics} \label{sec.Numerics}

\subsection{Summary of the 4D equations}\label{sec.summary.cov.eqs}

In this section, we summarize the equations of our theory. The governing PDEs are 
\begin{subequations}\label{sum.GRGPR}
\begin{align}
&\nabla_\mu\left( \E \, u^\mu u_\nu + p h^\mu_{\ \nu} + 
\dist{\sM}{\nu}\E_{\dist{\sM}{\mu}} \right) = 0, \label{sum.GRGPR.T}\\[2mm]
&u^\nu \nabla_\nu\dist{\sM}{\mu} + \dist{\sM}{\nu} \nabla_\mu u^\nu = -\frac{1}{\theta(\tau)} 
\KMNContr{\sM \sN} g_{\mu\nu} 
\E_{\dist{\sN}{\nu}}, 
\label{sum.GRGPR.A}\\[2mm]
&\nabla_\mu (\rho u^\mu) = 0, \label{sum.GRGPR.rho} \\[2mm]
&\nabla_\mu (s u^\mu) = \frac{1}{\E_s \theta(\tau)} \KMNContr{\sM \sN} g_{\mu\nu} 
\E_{\dist{\sM}{\mu}} 
\E_{\dist{\sN}{\nu}} \geq 0, \label{sum.GRGPR.s} 
\end{align}
where, we recall, $ s = \rho \Entropy $ is the entropy per unit volume of the 
rest mass, $ \Entropy $ 
is the specific entropy, $ 
\E(\dist{\sM}{\mu},\rho,\entropy) = \rho 
\left (1 + \varepsilon(\rho,\Entropy,\dist{\sM}{\mu})\right ) $, the pressure $ 
p = \rho \E_{\rho} 
+ s \E_s - 
\E = \rho^2 
\varepsilon_{\rho}$, the 
anisotropic 
stresses $ 
\dist{\sM}{\nu}\E_{\dist{\sM}{\mu}} $. In this paper, we choose
\begin{equation}\label{sum.E}
\varepsilon(\rho,\entropy,\dist{\sM }{\mu}) = \varepsilon^{\rm 
eq}(\rho,\Entropy) + 
\frac{c_s^2}{4} \devGMix{\lambda}{\nu}\devGMix{\nu}{\lambda},
\end{equation}
where $ \varepsilon^{\rm eq} $ is either the ideal gas \eqref{eqn.ideal_gas_eos} or stiffened gas 
\eqref{eqn.stiff_gas_eos} equations of state. The temperature is $ \E_s = \pd 
\E/ \pd s = \pd 
\varepsilon^{\rm eq}/\pd \Entropy$, $ c_s^2 = const $. Our choice 
for the non-equilibrium part of the energy
results in 
\begin{equation}\label{sum.eA}
	\E_{\dist{\sM }{\mu}} = \rho \, c_s^2
	\KMN{\sM \sN}g^{\lambda\alpha} \dist{\sN}{\lambda}\devG{\alpha\beta} 
	h^{\beta 
	\mu},
	\qquad
	\Cauchymix{\mu}{\nu} = 
	\dist{\sM }{\nu}  \E_{\dist{\sM }{\mu}} = \rho\, c_s^{2} 
	\devGMix{\mu}{\lambda} 
	\GeffMix{\lambda}{\nu}.
\end{equation}
Eventually, we chose $ \theta(\tau) = \rho_0 \tau c_s^2 
\GeffMix{\lambda}{\lambda}/3 $ and $ \tau = 
const 
$. 
Note, that in general $ \tau = \tau (\rho,\entropy,\dist{\sM}{\mu}) $, e.g. for 
elastoplastic solids, 
complex fluids, or mixtures, e.g. see~\cite{Hyper-Hypo2019,Jackson2018}.
\end{subequations}

\subsection{Summary of the 3+1 split of the equations}
\label{sec:finalPDE}
%
%\begin{align}
%\nabla_\mu J^\mu = 0\\
%\nabla_\nu T^{\mu\nu} = 0 \label{eq:nablaT}
%\end{align}
%\begin{align}
 %J^\mu & = \rho u^\mu \\
%T^{\mu\nu}  &= (e+p)u^\mu u^\nu + p g^{\mu\nu} + \Sigma^{\mu\nu} 
 %%T^{\mu\nu} & = S^{\mu\nu} + S^\mu n^\nu + n^\mu S^\nu + E n^\mu n^\nu \label{eq:divT}
%\end{align}
%where  
We recall here the definition of the momentum-energy stress tensor $T^{\mu\nu}$ and the pure anisotropic stress tensor component $\sigma^{\mu}_{\ \nu}$
\begin{align} \label{eq:NumT}
\EMT{\mu\nu}{}  &= \E u^\mu u^\nu + p h^{\mu\nu} + \sigma^{\mu\nu} \\
\sigma^\mu_{\ \nu} &:=   \frac{\partial \E}{\partial \dist{\sM}{\mu}}  
\dist{\sM}{\nu} = \rho c_s^2 
\devGMix{\mu}{\lambda} G^{\lambda}_{\ \nu}, \\
\E(\rho,s,\dist{\sM }{\mu}) & := \rho \left[ 
\varepsilon^{\rm eq}(\rho,\Entropy) + 
\frac{c_s^2}{4}
\devGMix{\lambda}{\nu}\devGMix{\nu}{\lambda} \right]\\ % %eq.113
\Geff{\mu\nu} & :=  \KMN{\sM\sN} \dist{\sM}{\mu} \dist{\sN}{\nu} , \qquad  
\devG{\mu\nu} :=  
\Geff{\mu\nu} 
- \frac{\GeffMix{\lambda}{\lambda}}{3} h_{\mu\nu}
. %=  h^{\mu\lambda} \G_{\lambda\nu} = 
%h^{\mu\lambda} \og{g}{AB} \dist{A}{\lambda}\dist{B}{\nu}.
\end{align}
After expanding $\EMT{\mu\nu}{}$ and $\sigma^{\mu\nu}$ with respect to the 
spatial and temporal projection operator $\gamma^{\mu}_{\ \nu}$ and $N^{\mu}_{\ 
\nu}$ as
\begin{align}
\EMT{\mu\nu}{}  = S^{\mu\nu} + S^\mu n^\nu + n^\mu S^\nu + U n^\mu n^\nu,\quad 
\quad  \quad
\sigma^{\mu\nu} =  \vartheta^{\mu\nu} + \vartheta^\mu n^\nu + n^\mu \vartheta^\nu + \vartheta n^\mu n^\nu
\end{align}
i.e.
\begin{align}
S^{\mu\nu}&:=\hphantom{+} \gamma^\mu_{\ \alpha}\gamma^\nu_{\ \beta} \EMT{\alpha 
\beta}{} ,      &  
\vartheta^{\mu\nu} := & \gamma^\mu_{\ \alpha}  \gamma^\nu_{\ \beta}  \sigma^{\alpha \beta}\\ %=   
%\gamma^i_\alpha  \gamma^j_\beta  \pi^{\alpha \beta} \\          \\ 
S^\mu&:=-\gamma^\mu_{\ \alpha} n_\beta \EMT{\alpha 
	\beta}{}, &                             
\vartheta^\mu :=  &- \gamma^\mu_{\ \alpha} n_\beta \sigma^{\alpha \beta} \\%=   v^j \pi_{ij} \equiv 
%\lapse \pi^{0}_{\oo i} \equiv \lapse \pi^{\oo 0}_{i} \\
U&:=\hphantom{+}n_\alpha n_\beta \EMT{\alpha 
	\beta}{} ,         &                            \vartheta := & n_\alpha 
	n_\beta \sigma^{\alpha\beta}
\end{align}
the GRGPR equations for static-spacetimes (Cowling approximation) read  
as (see details in Appendix~\ref{sec:Atlas}).
\begin{align}
& \partial_t \left(  \gamma^{\frac{1}{2}} D \right) +\partial_i \left[  \gamma^{\frac{1}{2}}D \left(   \lapse v^i -    \shift^i  \right) \right]  = 0 \label{eq:finalPDE} \\
 &\partial_t \left( \gamma^{\frac{1}{2}}    S_j \right) + \partial_i \left[ \gamma^{\frac{1}{2}} \left( \lapse   S^i_{\oo j}-\shift^iS_j \right)\right]   -\gamma^{\frac{1}{2}} \left(  \frac{1}{2} \lapse S^{ik}\partial_j \gamma_{ik}  + S_i \partial_j \shift^i   - U \partial_j \lapse \right) = 0   \nonumber  \\
 %&  \partial_t \left( \gamma^{\frac{1}{2}} E  \right)    + \partial_i \left[  \gamma^{\frac{1}{2}} \left( \lapse S^i  -   E \shift^i \right)\right]  -  \gamma^{\frac{1}{2}}  \left(  \lapse S^{ij}  K_{ij} - S^j    \partial_j     \lapse \right) = 0\label{eq:PDEe_b}\\
&  \partial_t \left( \gamma^{\frac{1}{2}} \mathcal{U}  \right)    + \partial_i \left[  \gamma^{\frac{1}{2}} \left( \lapse \left(S^i - D v^i\right)  -   \mathcal{U} \shift^i \right)\right] - \gamma^{\frac{1}{2}}  \left(  \lapse S^{ij}  K_{ij} - S^j    \partial_j     \lapse \right) =   0 \nonumber   \\
%\end{align}
%The equation of the configuration gradient reads as
%&\begin{equation}
&\pd_t \dist{i}{j} 
 +  \pd_j  \left( \dist{i}{k}\hat{v}^k \right) + \hat{v}^k \left( \pd_k \dist{i}{j}-   \pd_j   \dist{i}{k} \right)  = 
-\frac{1}{\theta(\tau)} \dist{i}{\mu}  \devGMix{\mu}{j}  \nonumber  \\ 
%\go{g}{iC}
%\Piolamix{\lambda}{C} g_{\lambda j},\label{eqn.A_b} \\ 
%\end{equation}
%The advection equation for the  scalar-transported (along worldlines) matter metric 
%\begin{align} 
& \partial_t \KAB_{\sA\sB}  +\hat{v}^k\partial_k \KAB_{\sA\sB}  = 0, \nonumber  \\
%& \partial_t \og{g}{AB}  +\hat{v}^k\partial_k \og{g}{AB} = 0, \label{eq:ogg} \\
%\end{align}
%%\begin{align}
%%\nabla_\mu T^\alpha_\beta &:=  \partial_\mu T^\alpha_\beta + \Gamma^\alpha_{\nu\mu} T^\nu_\beta -  \Gamma^\nu_{\beta\mu} T^\alpha_\nu \\
%%\nabla_\mu T_{\alpha\beta} &:=  \partial_\mu T_{AB} + \Gamma^\alpha_{\nu\mu} T_{\nu B} -  \Gamma^\nu_{\beta\mu} T_{A\nu} \label{eq:dcovT}
%%\end{align} 
%The equations for the metric tensor components read as
%\begin{align}
& \pd_t \left[ \begin{array}{ccc} 
\lapse  , %\\ 
\shift_j , %\\ 
\tilde{\gamma}_m 
\end{array} \right] = 0, \nonumber  %\;\;\; j=1,2,3;\;\;\; m=1,\ldots,6.
\end{align}
where 
\begin{align}
%\end{align}
%\begin{align}
S^i_{\oo j} & = \rho h W^2 v^i v_j + p \gamma^i_j   + \vartheta^i_{\oo j}     \\
S_i & = \rho h W^2 v_i  + \vartheta_i , \\
U &=  \rho h W^2 - p   +   \vartheta \\
D & = \rho W,\\
%S_i & =  \rho h W^2 v_i  {\color{red} + \sigma_i} , \\
%E &=  \rho h W^2 - p  {\color{red}+   \sigma}\\
\mathcal{U} & = U - D
\end{align} 
and in the Cowling approximation the following relation holds 
\begin{align}
\lapse S^{ij}  K_{ij} \equiv \frac{1}{2} S^{ik} \shift^j \partial_j \gamma_{ik} + S^j_{\oo i} \partial_j\shift^i.
\end{align}
%with
%\begin{align}
%%\sigma^i_{\oo j} :=  &\gamma^i_\alpha  \gamma_{j\beta}  \Sigma^{\alpha \beta}\\ %=\pi^i_{\oo j}   +  \shift^i  \pi^0_{\oo j} \\
 %\sigma^{ij} := & \gamma^i_\alpha  \gamma^j_\beta  \Sigma^{\alpha \beta}\\ %=   \gamma^i_\alpha  \gamma^j_\beta  \pi^{\alpha \beta} \\
 %\sigma_i :=  &- \gamma_{i\alpha} n_\beta \Sigma^{\alpha \beta} =   v^j \sigma_{ij}  \\%=   v^j \pi_{ij} \equiv \lapse \pi^{0}_{\oo i} \equiv \lapse \pi^{\oo 0}_{i} \\
	 %%\sigma^i  := & - \gamma^i_{\alpha} n_\beta \Sigma^{\alpha \beta} \\%= \lapse  \gamma^{ik} \pi^0_{\oo k}\\
 %\sigma := & n_\alpha n_\beta \Sigma^{\alpha\beta} = v^iv^j \sigma_{ij}  %= v^iv^j \pi_{ij}  \equiv \gamma^{ij}\pi_{ij}\\
%\end{align} 

%-------------------------------------------------------------
%\section{Numerical methods}
%\label{sec:Numerical_method}
%-------------------------------------------------------------

%=============================================================
\subsection{ADER discontinuous Galerkin schemes}
%=============================================================

With the aim of validating the physical model, in addition to the outlined theoretical results we propose a series of non-trivial numerical tests, grouped within the following benchmark-classes on flat or curvilinear coordinates: 1) the limit of viscous Newtonian fluid-dynamics, i.e. setting parameters \ldots ; 2) the limit of Newtonian linear elasticity; 3) the limit of special relativistic viscous fluid-dynamics, i.e. setting \ldots; 4) the limit of general relativistic hydrodynamics, i.e. \ldots.
Since the nonlinear PDE system (\ref{eq:finalPDE}) is provably strongly hyperbolic, we have chosen as ideal candidate numerical method the ADER discontinuous Galerkin (DG) scheme supplemented with an
\emph{a-posteriori} finite-volume subcell limiter approach, which has been presented in a well known series of papers \cite{Dumbser2014,Zanotti2015c,Zanotti2015d,ADERDGVisc} and more recently in \cite{ADERGRMHD} and it has been shown to be very robust even against complex shock-wave dominated scenarios in the context of the Euler equations of compressible
gas dynamics, ideal MHD, special relativistic RMHD, but also compressible
Navier-Stokes and viscous and resistive MHD and general relativistic MHD equations. 

In this section we summarize the main feature of the adopted ADER-DG strategy with subcell finite-volume limiter (SCL).

First, let us denote with $\boldsymbol{V}$ the array of the 30~\emph{primitive variables} 
\begin{align}
\boldsymbol{V} := \left( \rho, v_j, p, \dist{i}{j}, \tilde{k}_m , \lapse, \shift^j, \tilde{\gamma}_m \right)\,,\;\;\; i,j=1,2,3; \;\;m=1,\ldots,6\,,
\end{align}
where the rest-mass density is denoted by $\rho$, as measured by the comoving frame, $\boldsymbol{v}$ is the three-velocity vector, the fluid
pressure if $p$, $\dist{i}{j}$ are the spatial components of the so called \emph{4-distrortion 
field}, $\tilde{\boldsymbol{k}}$ is the array of the six
{independent} components of the \emph{matter} metric $\boldsymbol{k}$, \ie
\begin{align}
\tilde{\boldsymbol{k}} = \left( k_{11}, k_{12}, k_{13},
k_{22}, k_{23}, k_{33}\right)
\end{align}
$\lapse$ is the {lapse} function, $\boldsymbol{\shift}$ is the {shift} vector, and
$\tilde{\boldsymbol{\gamma}}$ the array of the six
{independent} components of the three (spatial) metric $\boldsymbol{\gamma}$, \ie
\begin{align}
\tilde{\boldsymbol{\gamma}} = \left( \gamma_{11}, \gamma_{12}, \gamma_{13},
\gamma_{22}, \gamma_{23}, \gamma_{33}\right).
\end{align}
Then, the corresponding state vector $\boldsymbol{Q}$ of {conserved}
variables with respect to the PDE system (\ref{eq:finalPDE}) can be defined as
\begin{align}
\boldsymbol{Q} := \left( \sqrt{\gamma} D, \sqrt{\gamma} S_j, \sqrt{\gamma}
\tau, \dist{i}{j}, \tilde{k}_m , \lapse, \shift^j,
\tilde{\gamma}_{m} \right)\,.
\end{align}
If the transformation from primitive to conserved variables is explicit and straightforward, the 
definition of '\emph{an}' inverse transformation, i.e. conserved to primitive, is far from being 
simple. Indeed, since the elements of the state vector $\boldsymbol{Q}$ are defined as a non-banal 
non-linear combination of the components of $\boldsymbol{V}$, in this work, the inversion of the 
primitive to conserved function is computed iteratively. The very basic and adopted strategies are 
briefly described in appendix \ref{sec:cons2prim}.

Then, it is very easy to verify that the governing PDE system (\ref{eq:finalPDE}) can be casted into a (numerically) very elegant form, i.e.
\begin{align}
\partial_t \QQ  + \nabla \cdot
  \boldsymbol{F}(\QQ)+ \boldsymbol{\mathcal{B}}(\QQ) \cdot\nabla
  \QQ  =  \boldsymbol{\mathcal{S}}(\QQ) \label{eq:PDEQ}
\end{align}
where $\FF$ is the tensor of the non-linear conservative fluxes, $\boldsymbol{\mathcal{B}}$ is the 
matrix-tensor of the non-conservative product $\boldsymbol{\mathcal{B}}(\QQ) \cdot\nabla
  \QQ $,  and $\boldsymbol{\mathcal{S}}(\QQ)$ is the prescribed non-linear source term, which, depending on the  physical regime, it may become stiff.
	In particular, we have
\begin{align}
% fluxes 
\boldsymbol{F} :=  \left(\begin{array}{c}  \gammabar{}\left(\lapse v^i D -
    \shift^i D\right) \\  \gammabar{}\left( \lapse T^i_j - \shift^i S_j\right)   \\  \gammabar{}\left( \lapse \left( S^i - v^i
    D\right) - \shift^i \tau\right) \\ 
		\dist{i}{k}\hat{v}^k \\ 
		0\\ 0\\ 0\\ 0
  \end{array}\right), &&
	\boldsymbol{S} := \left(\begin{array}{c} 
	0 \\  0 \\ 0 \\ 
-\frac{1}{\theta(\tau)} \dist{i}{\mu}  \devGMix{\mu}{j}  \\ 0\\ 
	 0\\ 0\\ 0
  \end{array}\right)
\,,\label{eq:fluxes}  
\end{align}
\begin{align} 
\boldsymbol{\mathcal{B}}(\boldsymbol{Q})\cdot \nabla \boldsymbol{Q} :=
\left( \begin{array}{c} 0 \\ 
\gammabar{}\left(U\partial_j\lapse -\frac{1}{2}\lapse T^{ik}\partial_j \gamma_{ik} - S_i \partial_j  \shift^i\right)  \\ 
\gammabar{}\left(S^j\partial_j \lapse - \frac{1}{2} T^{ik} \shift^j \partial_j \gamma_{ik} - T_i^j \partial_j \shift^i \right) \\
	\hat{v}^k \left( \pd_k \dist{i}{j}-   \pd_j   \dist{i}{k} \right)   \\ 
		\hat{v}^k\partial_k k_{AB}    \\
		0 \\ 0 \\ 0 \end{array}\right)\,,
 \label{eq:NCP} 
\end{align}
Equation (\ref{eq:PDEQ}) can be alternatively expressed in the quasi-linear form
\begin{align}
\partial_t \QQ  +  \boldsymbol{\mathcal{A}}(\QQ) \cdot\nabla
  \QQ  =  \boldsymbol{\mathcal{S}}(\QQ) \label{eq:PDEA}
\end{align}
after defining $\boldsymbol{\mathcal{A}}(\QQ) := \partial \boldsymbol{F}/\partial \boldsymbol{Q} + 
\boldsymbol{\mathcal{B}}(\QQ)$. System (\ref{eq:PDEA}) is said to be \emph{hyperbolic} if  the 
matrix $\boldsymbol{\mathcal{A}} \cdot \boldsymbol{n}$ is diagonalizable for all normal vectors 
$n\neq 0$ with only real eigenvalues
and a complete set of bounded linearly independent eigenvectors, see \cite{toro-book}.
Here, the differential operator $\nabla$ is intended to be $\nabla = (\partial_x, \partial_y, \partial_z)$.
Even if the form of (\ref{eq:PDEA}) cover a very large class of complex non-linear hyperbolic PDE systems, it becomes numerically friendly if approached by the mentioned ADER-DG techniques, mainly because of the use of three numerical tools: i) the so called path-conservative integration, which allows to give sense to the quasi-linear product $\boldsymbol{\mathcal{A}}(\QQ)\cdot \QQ$ even in the presence of discontinuities in the state variables,  within ii) the arbitrary high-order accurate \emph{explicit} and \emph{local} ADER-DG predictor, which allows to solve stiff and non-stiff source terms, with the support of a very robust \emph{a-posteriori} and resolution-preserving finite volume limiter. Details and references will be given in the following.

Given a Cartesian mesh partition $\Omega_h = \{ \Omega_i \}$, such that
\begin{align}
\Omega = \bigcup \limits_{i=1,\ldots N_E} \Omega_i, \hspace{1cm} \bigcup
\limits_{i\neq j; \;\;i,j=1,\ldots N_E} \Omega^{\circ}_i \cap
\Omega^{\circ}_j = \varnothing \label{eq:mesh}
\end{align}
where $\Omega\subset\mathbb{R}^d$ is the computational domain in $d$ space-dimensions, which is discretized within $N_E$ total number of spatial elements and symbol ``$\left.\right.^{\!\!\circ}$'' denotes the interior
operator, the following weak formulation of the governing equations (\ref{eq:PDEA}) is obtained after multiplication by a \emph{test function} with compact support integration along a space-time control-volume $\Omega_i \times [t^n,t^{n+1}]$, where $[t^n, t^{n+1}]$ is the future time interval where the solution is still unknown, and $t=t^n$ is the time slice where the solution is known or from the initial condition, or from the previously computed time-step, i.e.
\begin{align}
& \int \limits_{t^n}^{t^{n+1}} \int\limits_{\Omega_i } \phi_k \left(
  \partial_t \QQ  + \boldsymbol{\mathcal{A}}(\QQ) \cdot\nabla
  \QQ - \boldsymbol{\mathcal{S}}(\QQ)\right) \,d\boldsymbol{x}\,dt = 0\,. \label{eq:weakPDE}
\end{align}
In the DG framework, the test function $\phi_k$ is a basis element for the
vector space $\mathcal{U}_h^N$ of {piecewise} polynomials of maximum
degree $N\geq 0$ over $\Omega$. Notice that, since the chosen  basis functions are piecewise polynomials, they are allowed to be
discontinuous across the element interfaces $\partial \Omega_i$.
%After choosing a local basis
%$\mathcal{B}^i_{N,d}=\{\phi_k\}_{k=1,2,\ldots \ndof= (N+1)^d}$, \ie
In this work, the set of Lagrange interpolation polynomials of degree
$N$ over $\Omega_i$ with the property
\begin{align}
\phi_k(\x_{\text{GL},i}^m) = \left\{ \begin{array}{rl} 1 & \text{if}\;\;
  k=m; \\0 & \text{otherwise}; \end{array}\right.\hspace{0.7cm}
k,m=1,\ldots,(N+1)^{\text{d}}
\end{align}
 %through the Gauss-Legendre quadrature points of the element $\Omega_i$.
has been chosen as \emph{nodal} polynomial basis, with $\{\x_{\text{GL},i}^m\}$ being the set of the Gauss-Legendre (GL)
quadrature points in $\Omega_i \in \Omega_h$.

In the present formulation, the grid is locally Cartesian and, thanks to the polynomial expansion and quadrature rules for numerical integration, the multi-dimensional spatial integrals of Eq. (\ref{eq:weakPDE}) %{eq:ADER-DG})
 can be factorized as the multiplication of one-dimensional independent integrals in $x$, $y$ and $z$ direction. Moreover, the domain of integration $\Omega_i$ is first rescaled to the unit element $[0,1]^d$, and therefore, basis function are defined accordingly to the only tensor product of the GL quadrature points in the unit interval $[0,1]$, denoted by
$\{\xi_{\text{GP}}^m\}_{m=1,\ldots,N+1}$.

\noindent
%
 %Note that the total number of GL quadrature points $\{\x_{\text{GP}}^m\}$ in
%$\Omega_i$, as well as the total number of basis elements $\{ \phi_k\}$,
%is $(N+1)^d$.
%
%After integration by parts of the flux-divergence term,
%Eq.~(\ref{eq:weakPDE}), can be rewritten as
%%
%\begin{align}
%&\int \limits_{t^n}^{t^{n+1}} \int\limits_{\Omega_i } \phi_k\, \partial_t
  %\QQ \,d\boldsymbol{x}\,dt + \int
  %\limits_{t^n}^{t^{n+1}} \int\limits_{\Omega_i } \phi_k \,
  %\boldsymbol{\mathcal{B}}(\QQ) \cdot \nabla \QQ
  %\,d\boldsymbol{x}\,dt= 0\,. \label{eq:DGPDE}
%\end{align}
%%
%
Then, schematically: 
\begin{enumerate}[i)]
\item after introducing a {space-time} polynomial $\q_h(\x,t)$ as an only-locally implicit predictor solution for $t\in[t^n,t^{n+1}]$, which is in general discontinuous at the element edges $\partial \Omega_i$ and whose details will be outlined in the next section; %for   the time interval $t\in[t^n,t^{n+1}]$ as higher order accurate, but only element-local and boundary independent, solution, \ie
%\begin{align}
%\QQ (\boldsymbol{x},t) \approx  \boldsymbol{q}_h(\boldsymbol{x},t) %= \phi_k (\boldsymbol{x})\;
%%\hat{\boldsymbol{u}}^n_k, \;\;\;k=1,\ldots,(N+1)^d, \quad \boldsymbol{x} \in
%%\Omega_i, \nonumber
%\end{align}
 %and
\item  after choosing the set of piecewise and purely spatial
polynomials $\mathcal{U}_h^N$  over $\Omega_h$ as the space of solutions for the problem 
(\ref{eq:weakPDE}),  such that for every time slice $t=t^n$ the solution can be approximated as 
%$\boldsymbol{u}_h$ and expanded along the polynomial basis as 
%% $u_h \in \U_h^N$ over $\Omega$, \ie $u_h(\mathbf{x}_i,t) = \phi_k
%% (\mathbf{x}_i ) u_k (t)$ with $\mathbf{x}_i \in \Omega_i$,
\begin{align}
\QQ (\boldsymbol{x},t^n) \approx  \boldsymbol{u}_h(\boldsymbol{x},t^n) = \phi_k (\boldsymbol{x})\;
\hat{\boldsymbol{u}}^n_k, \;\;\;k=1,\ldots,(N+1)^d, \quad \boldsymbol{x} \in
\Omega_h; \nonumber
\end{align}
\item then, a higher order accurate and \emph{path-conservative} ADER-DG
scheme for the time-evolution of the expansion coefficients 
$\hat{\boldsymbol{u}}_k^n$, named also degrees of freedom, can be written in the following form
\begin{align}
\left( \, \int\limits_{\Omega_i} \phi_k \phi_l \, d\boldsymbol{x}\right)
  \left( \hat{\boldsymbol{u}}_l^{n+1} - \hat{\boldsymbol{u}}_l^{n} \, \right) + \int \limits_{t^n}^{t^{n+1}} \!\!
  \int\limits_{\partial \Omega_i } \!\! \phi_k \mathcal{D}\left(
  \q_h^-,\q_h^+ \right) \cdot \boldsymbol{n} \, dS \, dt   + & \nonumber \\
	+ \int
  \limits_{t^n}^{t^{n+1}} \!\! \int\limits_{\Omega_i } \phi_k
  \boldsymbol{\mathcal{A}}(\q_h) \cdot \nabla \q_h
  \,d\boldsymbol{x}\,dt & = \int
  \limits_{t^n}^{t^{n+1}} \!\! \int\limits_{\Omega_i} \phi_k
  \boldsymbol{\mathcal{S}}(\q_h)
  \,d\boldsymbol{x}\,dt  \,, 
\label{eq:ADER-DG}
\end{align}
for any spatial element $\Omega_i\in\Omega_h$, and where the extent of the time-interval is dependent on the local CFL stability condition, see Eq. (\ref{eq:CFL}) in the next. Notice here, the purely spatial integral of $\phi_k \phi_l$ on the left is can be regarded as the Gram matrix of the basis functions, also called 'mass-matrix', it is positive definite, but also purely-diagonal because we have chosen an orthogonal basis set.
\end{enumerate}

Dredging up the fact that the predictor solution $\q_h$ is allowed to be discontinuous at the element
edges $\partial \Omega_i$, %the surface integral of the fluxes is
%computed by means of an approximate Riemann solver $\mathcal{G}$
%depending on the boundary extrapolated data $\q_h^-$ and $\q_h^+$
%evaluated at the left and right of an element interfaces,
%respectively. In this paper we use the simple Rusanov flux
%\cite{Rusanov1961a}
%%
%\begin{equation}
  %\mathcal{G}\left(\q_h^-, \q_h^+ \right) \cdot \boldsymbol{n} = \frac{1}{2}
  %\left( \boldsymbol{F}(\q_h^+) + \boldsymbol{F}(\q_h^-) \right) \cdot \boldsymbol{n}
  %- \frac{1}{2} s_{\max} \left( \q_h^+ - \q_h^- \right)\,.
	%\label{eq.rusanov} 
%\end{equation} 
%%
%where $s_{\max}$ denotes the maximum signal speed computed in $\q_h^-$
%and $\q_h^+$. On the other hand, 
%the flux term of 
the non-conservative product has been approximated by means of a
{path-conservative} scheme (see \cite{pares2006,Castro2006,CASTRO2017_ch6,PARES2010_ch2,ParesMunoz2009} for a detailed discussion about the topic) of the
form
\begin{align}
\mathcal{D}\left( \q_h^-,\q_h^+ \right) \cdot \boldsymbol{n} = & \frac{1}{2}
\left(\int \limits_{0}^{1} \boldsymbol{\mathcal{A}} \left(
\boldsymbol{\psi}(\q_h^-,\q_h^+,s) \right)\cdot\boldsymbol{n} \, ds
\right)\cdot\left(\q_h^+ - \q_h^-\right)- \frac{1}{2} s_{\max} \left( \q_h^+ - \q_h^- \right)\,, \label{eq:PC}
\end{align}
depending on the boundary extrapolated data $\q_h^-$ and $\q_h^+$,
and which mathematical definition is based on the theory of \cite{DLM1995} on hyperbolic partial
differential equations with nonconservative products. In particular, the numerical flux, or better, 
numerical jump (\ref{eq:PC}) takes the form of a line-integral along a path $\boldsymbol{\psi}$ in 
the configuration space, with a consistency condition represented by
\begin{align}
  &
  \mathcal{D}\left( \q_h^-,\q_h^+ \right) \cdot \boldsymbol{n} -
  \mathcal{D}\left( \q_h^+,\q_h^- \right) \cdot \boldsymbol{n} =  \int \limits_{0}^{1} 
  \boldsymbol{\mathcal{A}} \left( \boldsymbol{\psi}(\q_h^-,\q_h^+,s)
  \right)\cdot\boldsymbol{n} \, \partial_s \boldsymbol{\psi} \, ds\,,
\end{align}
named also as generalized Rankine-Hugoniot condition.

In principle, any choice of the path $\boldsymbol{\psi}$ that admits a parametrization in the form
\begin{align}
 \boldsymbol{\psi} =\boldsymbol{\psi}(\q_h^-,\q_h^+, s), &&
 \boldsymbol{\psi}(\q_h^-,\q_h^+, 0 ) = q_h^-,   &&
 \boldsymbol{\psi}(\q_h^-,\q_h^+, 1 )  = q_h^+, 
\end{align}
with $\boldsymbol{\psi}$ being a Lipschitz continuous function in the variable $s$, is possible. In 
practice, we adopted the straight-line path connecting the states $\q_h^-$ and $\q_h^+$, i.e.
\begin{align}
\boldsymbol{\psi} = \boldsymbol{\psi}(\q_h^-, \q_h^+, s) = \q_h^- + s \left( \q_h^+ -
\q_h^- \right)\,, && s \in [0,1]\,, 
\end{align}
and approximate the line integral in (\ref{eq:PC}) with a sufficiently accurate numerical quadrature rules [see also \cite{ADERNC,OsherNC} for details].
	
Whenever the matrix-tensor $\boldsymbol{\mathcal{A}}$ is a pure Jacobian, i.e. the pure 
conservative case $\boldsymbol{\mathcal{A}}= \partial \bf{F} /\partial \q$, then it is easy to 
verify that the numerical flux (\ref{eq:PC})
is a generalization  to
the non-conservative case of the widely used, single-wave, Rusanov (or local Lax-Friedrichs) approximate Riemann solver, i.e.
\cite{Rusanov1961a}
\begin{equation}
 \mathcal{D}\left( \q_h^-,\q_h^+ \right) \cdot \boldsymbol{n}  
 \;\;\xrightarrow{\boldsymbol{\mathcal{A}}= \partial \bf{F} /\partial \q}\;\; 
 \mathcal{G}\left(\q_h^-, \q_h^+ \right) \cdot \boldsymbol{n} = \frac{1}{2}
  \left( \boldsymbol{F}(\q_h^+) + \boldsymbol{F}(\q_h^-) \right) \cdot \boldsymbol{n}
  - \frac{1}{2} s_{\max} \left( \q_h^+ - \q_h^- \right)\,.
	\label{eq.rusanov} 
\end{equation} 
In principle, more sophisticated and little dissipative schemes, based on a wider eigen-spectrum of 
the matrix-tensor $\boldsymbol{\mathcal{A}}$ may also be used [see \eg an
{HLLEM}-type version in \cite{NCP_HLLEM}, or the path-conservative Osher schemes in \cite{OsherNC}].

Note also that in this work we follow \cite{ADERGRMHD} in rewriting the gravity terms of the fluid-equations as non-conservative products.

Finally, we stress the fact that the proposed space-time ADER-DG scheme (\ref{eq:ADER-DG}) is an explicit DG scheme, and it is $(N+1)$-th order accurate both in space and time, for smooth solutions. Then, the standard CFL-type stability condition of DG schemes constrains the time-step to 
\begin{align}
\Delta t_{\text{DG}} < \text{CFL}\frac{h_{\text{min}} }{d
  \left(2N+1\right)} \frac{1}{|\lambda_{\text{max}}|}, \label{eq:CFL}
\end{align}
 $h_{\text{min}}$ being the minimum characteristic mesh-size, $d$ 
the number of spatial dimensions, $\lambda_{\text{max}}$ the maximum
signal velocity of the PDE, and CFL is a real number within  $0 < \text{CFL} < 1$. In our tests, if not
stated otherwise, we chose $\text{CFL}=0.9$. 

\noindent
In the following sections we describe the space-time DG predictor and the implementation of the subcell finite-volume limiter, see also \cite{Dumbser2014,Zanotti2015c,Zanotti2015d,ADERDGVisc,ADERGRMHD}.

%=============================================================
\subsection{Spacetime discontinuous Galerkin predictor}
\label{sec:STDG}
%=============================================================
In this section we give a brief description of the  the spacetime {predictor}
$\q_h$, appearing in Eq. (\ref{eq:ADER-DG}), and how we compute it. 

\noindent
First, one introduces a the new (nodal) basis set $\{\theta_k=\theta_k(t,\boldsymbol{x})\}$ 
spanning the vector space $\mathcal{Q}_h^N$  of all piecewise {spacetime} polynomials of
maximum degree $N$ over $\Omega_h$. According to this new basis,  any discrete solution $\q_h(\x,t)\in \mathcal{Q}_h^N$ can be expanded as
\begin{equation}
  \q_h(\boldsymbol{x},t) = \theta_k(\boldsymbol{x},t) \, {\hat \q}_k\,,
\label{eq.stdof}
\end{equation}
where ${\hat \q}_k$ are real-valued expansion coefficients, named also spacetime degrees of freedom of $\q_h$.
\noindent
Then, based on the following weak formulation of (\ref{eq:weakPDE}) 
(\ref{eq:weakPDE}) in space and time
\begin{align}
 \int \limits_{t^n}^{t^{n+1}} \!\! \int\limits_{\Omega_i^{\circ} }
 \theta_k \, \partial_t \q_h \,d\boldsymbol{x} \, dt + \int
 \limits_{t^n}^{t^{n+1}} \!\! \int\limits_{\Omega_i^{\circ} } \theta_k \,
 \nabla \cdot \boldsymbol{F}(\q_h) \,d\boldsymbol{x}\,dt  + \int
 \limits_{t^n}^{t^{n+1}} \!\! \int\limits_{\Omega_i^{\circ} } \theta_k
 \boldsymbol{\mathcal{B}}(\q_h ) \cdot \nabla \q_h \,d\boldsymbol{x}\,dt=
  \int
  \limits_{t^n}^{t^{n+1}} \!\! \int\limits_{\Omega_i^{\circ}} \theta_k
  \boldsymbol{\mathcal{S}}(\q_h)
  \,d\boldsymbol{x}\,dt   \,, \label{eq:predictor}
\end{align}
the spacetime {predictor} 
$\q_h$ can be regarded as an \emph{``interior''} solution of the partial differential
equations within each space-time element. Indeed, the spatial domain of integration in (\ref{eq:predictor}) has chosen to be the
{interior} of the space elements $\Omega^{\circ}_i$, that means approximating boundary contributions. Thanks to this first approximation, a system
of $N_E$ independent  and element-local equation systems of the type (\ref{eq:predictor})  is obtained.
Then, similarly to the procedure of above, one invokes the the Gauss-Legendre quadrature rules and, after integration by parts of the time-derivative term, equation (\ref{eq:predictor}) reduces to the following (element-local) system of $(N+1)^{(d+1)}$ nonlinear equations in the spacetime degrees of freedom
$\hat{\q}_k$
\begin{align}
   \int\limits_{\Omega_i^{\circ}} \theta_k(\boldsymbol{x},t^{n+1})
   \q_h(\boldsymbol{x},t^{n+1}) \, d\boldsymbol{x} -
   \int\limits_{\Omega_i^{\circ}} \theta_k(\boldsymbol{x},t^{n})
   \boldsymbol{u}_h(\boldsymbol{x},t^{n}) \, d\boldsymbol{x} %&\nonumber\\ 
	- \int
   \limits_{t^n}^{t^{n+1}} \!\!  \int\limits_{\Omega_i^{\circ} } \!\!\!
   \partial_t \theta_k \q_h(\boldsymbol{x},t) \,d\boldsymbol{x}
   \, dt +&\nonumber\\ + \int \limits_{t^n}^{t^{n+1}} \!\!
   \int\limits_{\Omega_i^{\circ} } \!\!\! \theta_k \nabla \cdot
   \boldsymbol{F}(\q_h) \,d\boldsymbol{x}\,dt  + \int
   \limits_{t^n}^{t^{n+1}} \!\!  \int\limits_{\Omega_i^{\circ} } \theta_k
   \boldsymbol{\mathcal{B}}(\q_h) \cdot \nabla\q_h \,d\boldsymbol{x}\,dt &=  \int
  \limits_{t^n}^{t^{n+1}} \!\! \int\limits_{\Omega_i^{\circ}} \theta_k
  \boldsymbol{\mathcal{S}}(\q_h)
  \,d\boldsymbol{x}\,dt   \,, \nonumber
   \\ i=1,2,\ldots,N_E;\quad
   k=1,2,\ldots,(N+1)^{(d+1)}. \quad\quad\quad& \label{eq:DOFpredictor}
\end{align}
that holds for all the space-elements of the partition $\Omega_h$.
Notice, here we also make use the known solution $\boldsymbol{u}_h$ at the time slice $t=t^n$,
using up-winding in time for integrating the flux $\theta_k \boldsymbol{u}_h$ at the time-slice 
$t=t^n$. This choice is justified after appealing to the \emph{causality principle}.
In order to circumvent the non-linearity of system (\ref{eq:DOFpredictor}), a very simple discrete and local Picard iteration can be used \cite{Dumbser2008}. In this way, the discrete system (\ref{eq:DOFpredictor}) is solved \emph{independently} for all the $N_E$ space-elements of the partition $\Omega_h$, without needing any MPI communication.

In principle, without caring about HPC performances, a more classical alternative is represented by Runge-Kutta time-stepping schemes. On the other hand, the here-presented family of one-step ADER schemes seems to be particularly well suited for simulations on HPC systems, because  (i) the resolution of (\ref{eq:DOFpredictor}) can be performed  locally without requiring any information about the status of the neighbor cells %and (ii) the update of (\ref{eq:ADER-DG}) needs the information about the status of the only direct-neighbor spatial elements 
and, then, a consistently lower number of MPI communications is needed, see \cite{AMR3DCL,AMR3DNC} for details.

%=============================================================
\subsection{\textit{A-posteriori} subcell finite-volume limiter}
\label{sec:limiter}
%=============================================================

The numerical scheme described so far, is still incomplete. Indeed, even if formally $(N+1)$-th  order accurate,  a direct application of the purely ADER-DG scheme (\ref{eq:ADER-DG})  may generate unphysical oscillations that are potentially damaging to the stability of the simulation,  i.e.,  compromising the positivity of the solution. This is actually an unavoidable result in signal analysis, known as 'Gibbs phenomenon', that applies whenever attempting the finite-order (polynomial) approximation of discontinuities or steep gradients. Moreover, the ADER-DG scheme (\ref{eq:ADER-DG})  is also linear in the sense of the Godunov theorem \cite{Godunov59}, and then a special treatment is needed to circumvent  this problem.

The general idea is the following: whenever at a given future time-slice $t=t^{n+1}$ any unphysical solutions is \textit{locally} generated by (\ref{eq:ADER-DG}), or any 'suspicious' behavior of the physical variables is \textit{locally} detected, then, only \textit{a-posteriori} and locally, the candidate solution  is labeled as problematic and directly rejected. Then, only in the troubled zone, the initial state at the previous time-slice $t=t^n$ is evolved again in time by means of a more robust scheme. Generally, this limiting procedure can be seen as an a-posteriori and non-linear dosage of healthy numerical diffusion. This procedure is known as the 'MOOD paradigm', after \cite{CDL1,CDL2,CDL3} in the finite-volume context, applied to ADER schemes in \cite{ADER_MOOD_14}, and applied for the first time as a-posteriori limiting-technique of DG methods by \cite{Dumbser2014} on a subgrid level. The present formulation has been tested on AMR Cartesian grids against a wide class of hyperbolic systems, e.g. ideal MHD equation \cite{Zanotti2015c},  the ideal special relativistic MHD equations \cite{Zanotti2015d}, the viscous Navier-Stokes and viscous-resistive MHD equations \cite{ADERDGVisc}, and very recently the general-relativistic MHD equations on stationary space-times \cite{ADERGRMHD}. See also \cite{ADERDGALE2017} for the implementation on general moving unstructured and conforming meshes.

If for a detailed description, the reader is encouraged to refer to the previously cited papers by Dumbser and collaborators, in the following we summarize the main points. The main ingredients for the limiting-solver are:
\begin{enumerate}[i)]
\item an over-sensitive troubled cells indicator, which activate or de-activate the limiter accordingly to the chosen phyisical and mathematical admissibility criteria;
\item the limiter: i.e. a more robust shock-capturing scheme  than the high-order ADER-DG scheme (e.g. ADER-TVD or ADER-WENO);
\end{enumerate}

First, one needs to choose the mathematical and physical admissibility criteria that will drive the activation or deactivation of the limiter. In our implementation, the pure-DG 
candidate solution $\u_h^*=\u_h^*(\x,t^{n+1})$ is computed through  the ADER-DG scheme (\ref{eq:ADER-DG}) and  then, it is a-posteriori checked against the main physical \textit{admissibility} conditions, i.e. the pressure and density positivity, subluminal 
velocities, the successful  %invertability of the 
primitive-to-conserved variables conversion $\boldsymbol{V}=\boldsymbol{V}(\QQ)$, but also the absence of floating point errors (NaNs). 
On the other hand, this check is still not sufficient for the detection of possible and latent numerical instabilities. Then, as mathematical \textit{detection} criterion, a relaxed version of the discrete maximum principle (DMP) has been chosen in the following form
\begin{align} 
\min \limits_{\y\in {\cal{V}}_i} (\v_h(\boldsymbol{y},t^n))-\delta \leq
\v_h^*(\boldsymbol{x},t^{n+1})\leq \max \limits_{\boldsymbol{y}\in
  {\cal{V}}_i}(\v_h(\y,t^n))+\delta
\,, 
\label{eq:DMP}
\end{align}
where $\v_h=\v_h(\boldsymbol{x},t)$ is the piecewise-\textit{constant} representation of the piecewise polynomial solution $\u_h$, derived by means of the standard average-projection $\v_h(\x,t^n) = \mathcal{P}\left( \u_h(\x,t^n) \right)$, i.e.
\begin{align}
   \bar{\v}_{i,s}^{n} := \frac{1 }{| \Omega_{i,s} |} \int
   \limits_{\Omega_{i,s}} \u_h(\x,t^{n}) d\x \,, && 
   \bar{\v}_{i,s}^{*} := \frac{1 }{| \Omega_{i,s} |} \int
   \limits_{\Omega_{i,s}} \u_h^*(\x,t^{n+1}) d\x \,.
	\label{eq.subcellaverage} 
\end{align}
over a suitable uniform sub-grid of $N_s^d$ sub-cells $\Omega_{i,s}\subset\Omega_i$ with $\bigcup \Omega_{i,s} = \Omega_i$; $ {\cal{V}}_i$ is the set containing $\Omega_i$ and the so called Voronoi neighbor elements of $\Omega_i$, that are the spatial elements $\Omega_j\in\Omega_h$ that share at least one node with $\Omega_i$.
Then, parameter $\delta$ is just a relaxing tolerance in order to limit the number false-positive activation of the limiter. Remember, indeed, that (i) the DMP condition (\ref{eq:DMP}) can be seen just as a warning indicator and not as an admissibility condition for the computed solution; (ii) the $L_2$ projection $\v_h(\x,t^n) = \mathcal{P}\left( \u_h(\x,t^n) \right)$ already clips the original extrema of the initial state $\u_h(\x,t^n)$, and this fact makes condition (\ref{eq:DMP}) more severe. In particular, similarly to  \cite{Dumbser2014,Zanotti2015c,Zanotti2015d,ADERDGVisc} we have adopted a solution-dependent relaxation tolerance in the following form
\begin{equation}
	\delta =\max \left( \delta_0\,, \epsilon \times \left( \max
        \limits_{y\in {\cal{V}}_i}(\boldsymbol{u}_h(\boldsymbol{y},t^n))- \min
        \limits_{y\in {\cal{V}}_i}(\boldsymbol{u}_h(\boldsymbol{y},t^n))\right)\,
        \right)\,,
\end{equation}
and, following \cite{ADERGRMHD}, we have chosen a rather restrictive condition by fixing $\delta_0=10^{-8}$ and $\epsilon=10^{-7}$.

\noindent
Then, whenever new extrema are generated and detected by (\ref{eq:DMP}), the limiter will be activated even if the new extrema are compatible with the physics of the equations. For this reason, a high order limiter that does not clip extrema will be fundamental in order not to lose the original high order resolution of the ADER-DG scheme, e.g. a good candidate is actually a subcell limiter (SCL) based on the ADER-WENO finite-volume method, see \cite{Dumbser2014,AMR3DCL}. On the counter part, any essentially-non-oscillatory (ENO) scheme would possibly generate negative pressure and densities in low density flows. In such cases, a second-order accurate MUSCL-Hancock TVD finite-volume scheme, with a MinMod slope limiter (see \cite{toro-book}), would be preferred. One should mention the fact that the development of high-order and positivity preserving numerical schemes is an open topic of research in many areas, e.g. in high-energy astrophysics for the simulation of  compact objects inserted within low density atmospheres, but also in the fluvial engineering or oceanography for managing correctly the wetting-and-drying processes.

In this work, we use an a posteriori finite-volume subcell limiter but, in principle, one should also apply any favorite robust scheme, e.g. a proper shock-capturing finite-difference numerical scheme. In particular, for simplicity, we adopted the same uniform sub-grid of $N_s$ sub-cell per space-dimension introduced for for the DMP check (\ref{eq:DMP}). 
Choosing either the high order ADER-WENO or the second-order MUSCL-Hancock scheme (alias ADER-TVD), the discrete PDE system reads as
\begin{align}
&\bar{\boldsymbol{v}}_{i,s}^{n+1} - \bar{\boldsymbol{v}}_{i,s}^{n} \, + \int \limits_{t^n}^{t^{n+1}} \!\!
  \int\limits_{\partial \Omega_{i,s} } \!\!  \mathcal{D}\left(
  \q_h^-,\q_h^+ \right) \cdot \boldsymbol{n} \, dS \, dt   + \int
  \limits_{t^n}^{t^{n+1}} \!\! \int\limits_{\Omega_{i,s} }
  \boldsymbol{\mathcal{A}}(\q_h) \cdot \nabla \q_h
  \,d\boldsymbol{x}\,dt = \int
  \limits_{t^n}^{t^{n+1}} \!\! \int\limits_{\Omega_{i,s}} 
  \boldsymbol{\mathcal{S}}(\q_h)
  \,d\boldsymbol{x}\,dt  \,,  \nonumber \\
\label{eq:ADER-FV}
\end{align}
which can be regarded as the piece-wise constant version ($\phi_k=\text{const.}$) of the ADER-DG scheme (\ref{eq:ADER-DG}).
Similarly to (\ref{eq:ADER-DG}), this is a one-step  scheme with high-order of accuracy in space and time. Once the cell-averages $\bar{\boldsymbol{v}}_{i,s}^n$ are evaluated, the piecewise polynomials, denoted as  $\boldsymbol{w}_h(\x,t^n)$,
 are computed by means of a non-linear reconstruction (TVD or WENO). Once we know $\boldsymbol{w}_h(\x,t^n)$, then, for ADER-WENO the space-time predictor $\boldsymbol{q}_h(\x,t)$ is derived accordingly to Eq. (\ref{eq:DOFpredictor}) after substituting the domain of integration with $\Omega_{i,s} \times
[t^n, t^{n+1}]$ and using $\boldsymbol{w}_h(\x,t^n)$ instead of the original DG solution $\boldsymbol{u}_h(\x,t^n)$. As an alternative, if the ADER-TVD limiter is chosen, the predictor can be computed  through the MUSCL-Hancock method with a half time-step evolution $\boldsymbol{q}^{\text{TVD}}_h=\boldsymbol{q}^{\text{TVD}}_h(\x,t^{n+1/2})$, see \cite{toro-book}

\begin{figure*} 
\includegraphics[draft=false,width=0.55\textwidth]{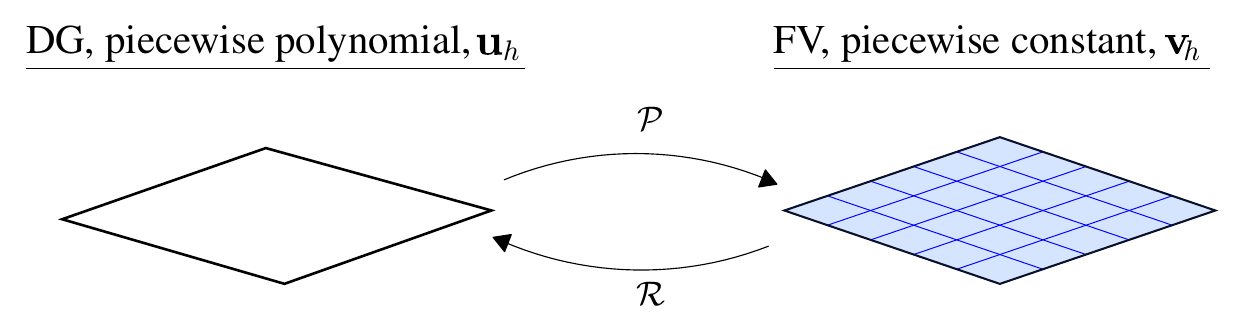}   
\caption{Mapping of the numerical solution between the piecewise
  polynomials $\u_h$ of the DG scheme and the piecewise constant data
  $\v_h$ of the finite-volume scheme.}
\label{fig:AMRmaps}
\end{figure*}

Finally, from the piecewise constant solution $\boldsymbol{v}_h(\x,t^{n+1})$ obtained after Eq. (\ref{eq:ADER-FV}), which is still high-order accurate but also (essentially) non-oscillatory, one reconstructs the so-called \textit{limited}-DG polynomial by means of a \textit{reconstruction} $\mathcal{R}$  operator associated to the projector $\mathcal{P}$, built in order to fulfill the constrain $\mathcal{R} \circ \mathcal{P} = \mathcal{I}$, where $\mathcal{I}$ is the identity operator, see  \cite{Dumbser2014}.   
Fig. \ref{fig:AMRmaps} shows  the mapping between the chosen solution spaces, piecewise polynomial (unlimited) or piecewise constant (limited).
 
Since a finite-volume scheme is used in the limited cells, then the respective CFL stability condition reads as
\begin{align}
\Delta t_{\text{FV}} < \text{CFL}\frac{h_{\text{min}}}{d \, N_s}
\frac{1}{|\lambda_{\text{max}}|}, \label{eq:CFLweno}
\end{align}
where now the number of sub-cells per space dimension $N_s$ appears at the denominator, instead of $2N+1$. A natural condition that allows to preserve the number of degrees of freedom is choosing $N_s \geq N+1$. Moreover, this choice condition allows to reconstruct the limited-DG polynomials from the respective cell-averages.  With the aim of maximizing the CFL number of the finite-volume scheme, i.e. $\Delta t_{\text{FV}} = \Delta t_{\text{DG}}$, as well as increasing the corresponding resolution properties, we have chosen $N_s = 2 N+1$ accordingly to \cite{Dumbser2014}.
Further details illustrating the main
stages of the final algorithm are outlined in \cite{Dumbser2014}.

%NUMERICAL PART:

\section{Numerical tests} \label{sec.numerical.examples}
Numerical test cases will be added in the next version of the manuscript.

\section{Conclusion and perspectives}\label{sec.Conclusion}

We have presented a new causal general relativistic continuum model for dissipative flows which may 
include flows of viscous fluids as well as elastic and inelastic deformations of solids. The 
governing PDEs belongs to the class of so-called Symmetric Hyperbolic Thermodynamically Compatible 
(SHTC) equations and
consist of two parts. The non-dissipative part of the PDEs, or the Hamiltonian part, is represented 
by the 
hyperbolic time evolution which is derived from the Hamilton principle. The second part, the 
dissipative one, is represented by algebraic source terms (low order terms) of relaxation type. The 
resulting system is consistent with the first and second laws of thermodynamics.  Thanks to the 
Hamiltonian nature of the governing equations, the theory is compatible with the Hamiltonian 
structure of the canonical energy-momentum tensor that appears as the source term in the Einstein 
field equations.

The main field of the theory is the geometric object, four-distortion field, which 
is an anholonomic basis tetrad field in differential geometry language. 
It provides the geometric settings for unified  description of flows of fluids 
and deformation of solids.

Via formal asymptomatic analysis, we demonstrated that the relativistic Navier-Stokes stress tensor 
is recovered in the leading terms. We also compared our model with the state of the art dissipative 
relativistic model, the M\"{u}ller-Israel-Stewart theory, where essential differences were observed 
in the definition of the stress rate.

The model was discretized using an advanced family of high-order ADER Discontinuous Galerkin 
(ADER-DG) and 
ADER Finite Volume (ADER-FO) methods. An extensive 
range of numerical examples was presented demonstrating the applicability of our 
theory to relativistic flows of viscous fluids and deformation of solids in 
Minkowski and curved spacetimes.

Because the SHTC formulation for other transfer process such as mass, heat and electric charge 
transfer, also admits the Hamiltonian formulation~\cite{DPRZ2017,SHTC-GENERIC-CMAT}, the extension 
of our Newtonian works is rather a 
straightforward task, and will be the subject of a near future research. The ultimate goal is to 
couple this SHTC multi-physic formulation of continuum physics with the recently proposed strongly 
hyperbolic first-order formulation of the Einstein field equations based on the Conformal
and Covariant Z4 system (CCZ4) with constraint-violation damping, which is refereed to as FO-CCZ4
\cite{FO-CCZ4}, and is implemented in the same ADER-DG framework as the presented model.

\subsection*{Acknowledgment}

The research contained in this paper has been financed by the European 
Research Council (ERC) under the
European Union's Seventh Framework Programme (FP7/2007-2013) with the research 
project \textit{STiMulUs}, 
ERC Grant agreement no. 278267. 
M.D. and F.F. have further received funding from the European Union's Horizon 2020 
Research and 
Innovation Programme under the project \textit{ExaHyPE}, grant agreement 
number  671698 (call
FETHPC-1-2014). 
E.R. acknowledges a partial support by the Program N15 of the Presidium of 
RAS, project 121 and the Russian foundation for Basic Research (grant number 16-29-15131). 
I.P. gratefully acknowledges the support of Agence Nationale de la Recherche 
(FR) 
(grant ANR-11-LABX-0040-CIMI) under program ANR-11-IDEX-0002-02.
The authors are grateful to the Leibniz Rechenzentrum (LRZ) for awarding 
access to the SuperMUC supercomputer based in Munich, Germany; they also
acknowledge the support of the HLRS computing center for providing access 
to the Hazel Hen supercomputer based in Stuttgart, Germany. M.D. has 
received further funding from the Italian Ministry of Education, 
University and Research (MIUR) in the frame of the Departments of Excellence 
Initiative 2018--2022 attributed to DICAM of the University of Trento  
and has been supported by the University of Trento in the frame of 
the Strategic Initiative \textit{Modeling and Simulation}.

\bibliographystyle{apsrev4-1}

\bibliography{library}

%\printbibliography

%%%%%%%%%%%%%%%%%%%%%%%%%%%%%%%%%%%%%%%%%%%%%%%%%%

%%%%%%%%%%%%%%%%% APPENDICES %%%%%%%%%%%%%%%%%%%%%
\appendix

%%%%%%%%%%%%%%%%%%%%%%%%%%%%%%%%%%%%%%%%%%%%%%%%%%%%%%%%
\section{Gravity effect in the Lagrangian description}\label{app.Lagr.gravity}
%%%%%%%%%%%%%%%%%%%%%%%%%%%%%%%%%%%%%%%%%%%%%%%%%%%%%%%%

In this appendix, we explain how the gravity is taken into account in the Lagrangian field 
equations~\eqref{eqn.Euler-Lagrange}. Because the material coordinates $ \myxi^a $ and metric $ 
\Kab{ab} $ can be assigned in a completely independent manner from their 
Eulerian counterparts $ 
x^\mu $ and $ g_{\mu\nu} $, the gravity cannot be associated with the curvature of the Lagrangian 
manifold $ \mattertime $. In particular, the material metric $ \Kab{ab} $ can 
be taken flat even 
though the spacetime has a non-vanishing curvature. So, how the gravity effect is accounted for in 
the Lagrangian description? In fact, the gravity field $ g_{\mu\nu} 
$ is taken into account in the Lagrangian field equations in the fluxes $ 
\Lambda_{\F{\mu}{a}} $.
Indeed, it is necessary to understand that since our theory is essentially a deformation theory, a 
single-frame, say pure Lagrangian, description of the matter motion is impossible because  
two frames are necessary in order to make comparisons of the lengths. Thus, the Lagrangian observer 
needs to 
receive some information from a non-comoving observer. The proper Lagrangian 
strain tensor therefore is (cf. \eqref{eqn.G}) 
\begin{equation}\label{app.Lagr.strain}
h_{ab} = h_{\mu\nu} \F{\mu}{a}\F{\nu}{b}
\end{equation}
which is the Eulerian projector $ h_{\mu\nu} = g_{\mu\nu} + u_\mu u_\nu$ as seen by the Lagrangian 
observer. It is then 
implied that $ \Lambda(\F{\mu}{a}) = \Lambda(h_{ab}) $ (more precisely, it is a function of the 
invariants of $ h_{ab} $)
%Because the 
%Lagrangian $ \Lambda $ is supposed to be a scalar, it may depends on $ h_{ab} $ only via its 
%invariants which are formed with the help of $ \Kab{ab} $. 
Therefore, the gravity field $ 
g_{\mu\nu} $ emerges in the Lagrangian description not by means of the covariant differentiation 
but it emerges in the fluxes $ \Lambda_{\F{\mu}{a}} $ every time when one computes the 
derivatives 
\begin{equation}\label{app.flux}
\Lambda_{\F{\mu}{a}} = \frac{\pd \Lambda}{\pd h_{bc}}\frac{\pd h_{bc}}{\pd 
\F{\mu}{a}}.
\end{equation}

%%%%%%%%%%%%%%%%%%%%%%%%%%%%%%%%%%%%%%%%%%%%%%%%%%%%%%%%
\section{Orthogonality condition as the consequences of the PDE for 
\EqTitle{$\dist{\sM}{\mu}$}{A}}\label{app.orthogonality}
%%%%%%%%%%%%%%%%%%%%%%%%%%%%%%%%%%%%%%%%%%%%%%%%%%%%%%%%

In this appendix we prove that the orthogonality condition
\eqref{eqn.A.orth} follows from the way how we define the dissipative 
source terms in the evolution equation for the distortion field~\eqref{GRGPR.A}. More precisely, 
we demonstrate that the relaxation of the time components $ \dist{\sM}{0} $ is rather 
\textit{artificial} and is completely defined by the relaxation of the pure matter components $ 
\dist{\sM}{i} $, $ i=1,2,3 $.  We denote the entire source term (the right hand-side) 
in~\eqref{GRGPR.A} as $ S^{\sM}_{\ \, \mu} $ and prove that the distortion field defined as 
the 
solution to \eqref{GRGPR.A} fulfills the orthogonality condition 
\eqref{eqn.A.orth} if and only if the source term $ S^{\sM}_{\ \mu} $ also satisfies
the orthogonality condition
\begin{equation}\label{eqn.orth.condition}
S^{\sM}_{\ \mu} u^{\mu} = 0, \quad \text{or} \quad S^{\sM} _{\ \, 0} = -S^{\sM }_{\ j} \hat{v}^j,
\end{equation}
where $ \hat{v}^j $ is the pseudo-Newtonian velocity, i.e. 
$(u^0,u^i) = \gamma (1,\hat{v}^i) := \gamma(1,\alpha v^i - \beta^i)$, 
$\gamma = 
\alpha^{-1} W
$,
and $ v^i $ is the Newtonian velocity, $ \beta^i $ is the shift vector, while $ 
\alpha $ and $ W $ are the laps 
and the Lorentz factor. We note that in the SHTC theory, $ S^{\sM}_{\ \mu} \sim 
\KMN{}^{\sM\sN} 
g_{\mu\nu}\E_{\dist{\sN}{\ \nu}} $ and therefore $ S^{\sM}_{\ \mu} $ is indeed orthogonal to $ 
u^{\mu} $ because $ 
\E_{\dist{\sM}{\mu}} u_\mu = 0 $, see \eqref{eqn.dIdA}.

In the 3+1 notations, orthogonality condition \eqref{eqn.A.orth} reads as 
\begin{equation}\label{eqn.A.orth.3D}
\dist{\sM }{0} = -\dist{\sM }{j}\hat{v}^j.
\end{equation}

We first prove that if \eqref{eqn.A.orth.3D} holds then 
\eqref{eqn.orth.condition} holds as well. 
Equations \eqref{GRGPR.A} for $ \mu=0 $ and $ \mu=k $ read as 
\begin{subequations}
\begin{equation}\label{GRGPR.A.3D.0}
\pd_t \dist{\sM }{0} + \hat{v}^j \pd_j \dist{\sM }{0} + \dist{\sM }{j} \pd_t 
\hat{v}^j = 
\gamma^{-1}S^{\sM}_{\ \, 0},
\end{equation}
\begin{equation}\label{GRGPR.A.3D.k}
\pd_t \dist{\sM }{k} + \hat{v}^j \pd_j \dist{\sM }{k} + \dist{\sM }{j} \pd_k 
\hat{v}^j = 
\gamma^{-1}S^{\sM}_{\ k}.
\end{equation}
\end{subequations}
After substituting $ 
\dist{\sM }{0} $ in \eqref{GRGPR.A.3D.0} by its expression 
\eqref{eqn.A.orth.3D}, we arrive at
\begin{equation}
-\pd_t (\dist{\sM }{k}\hat{v}^k) - \hat{v}^j\pd_j (\dist{\sM }{k}\hat{v}^k) + 
\dist{\sM }{j}\pd_t\hat{v}^j = \gamma^{-1}S_{\ \, 0}^\sM .
\end{equation}
After some term rearrangements, we have
\begin{equation}
-\hat{v}^k(\pd_t \dist{\sM }{k} + \hat{v}^j\pd_j \dist{\sM }{k}) - \dist{\sM 
}{k}\hat{v}^j\pd 
_j\hat{v}^k = \gamma ^{-1}S_{\ \, 0}^\sM .
\end{equation}
Now, substituting the terms in the brackets by \eqref{GRGPR.A.3D.k}, we 
obtain $ S^{\sM} _{\ \, 0} = 
-S^{\sM} _{\ k} \hat{v}^k$, i.e. \eqref{eqn.orth.condition} is fulfilled.

We now prove that if \eqref{eqn.orth.condition} holds then 
\eqref{eqn.A.orth.3D} is fulfilled. Indeed, assuming \eqref{eqn.A.orth.3D} is 
true, equation \eqref{GRGPR.A.3D.0} can be rewritten as
\begin{equation}
\pd_t \dist{\sM }{0} + \hat{v}^j \pd_j \dist{\sM }{0} + \dist{\sM }{j} \pd_t 
\hat{v}^j = 
-(\pd_t \dist{\sM }{k} + \hat{v}^j \pd_j\dist{\sM }{k} + 
\dist{\sM }{j}\pd_k\hat{v}^j)\hat{v}^k.
\end{equation}
After arranging the terms in a proper way, we arrive at
\begin{equation}
\pd_t (\dist{\sM }{0} + \dist{\sM }{k}\hat{v}^k) + \hat{v}^k \pd_j (\dist{\sM 
}{0} + 
\dist{\sM }{k}\hat{v}^k) = 0,
\end{equation}
which means that if \eqref{eqn.A.orth.3D} holds at the initial moment of time 
then it holds so at the all later times.

%%________________________________________________________
%%%
 %%%			FOLIATION AND COORDINATES
%%%-------------------------------------------------------

%\end{comment}

\section{Foliation of spacetime \EqTitle{$\spacetime$}{} and coordinate system}
\label{sec:Atlas}

In this section we briefly summarize what means choosing a so called $3+1$ 
foliation of the spacetime, and choosing a coordinate system on it. Everything 
in this section is referred mainly to the books by Rezzolla \& Zanotti 
\cite{RezzollaZanottiBook} and Gourgoulhon \cite{Gourgoulhon2012}.

\subsection{3+1 foliation of \EqTitle{$\spacetime$}{spacetime}}
\label{sec:3plus1} 
First, the 4D space-time is \textit{foliated} or sliced into a one-parameter family of (non-intersecting) \textit{space-like} hypersurfaces  $\Sigma$ for which a natural parameter is represented by any 'regular' scalar field $t:\spacetime \longrightarrow \reaSet$.
 The scalar field is enrolled as the \textit{time-coordinate} since its isosurfaces $\Sigma_t$ represent the set of the \textit{local} events that are simultaneous with the \textit{local}  Eulerian observers. 

Given two adjacent leaves $\Sigma_t$ and $\Sigma_{t+\delta t}$, there are infinite ways for dragging $\Sigma_t$ to $\Sigma_{t+\delta t}$.
One way is to select a 4-vector field $n_\mu=n_\mu(p)$ normal to the hypersurface  $\Sigma_t$, which is parallel to the gradient of the time-coordinate $\nabla t$, at every event $p\in\Sigma_t$.
 Due to the regularity of the foliation, the time-like vector field $n_\mu$ changes smoothly in $t$ and it can be regarded as the tangent vector with respect to a trajectory  $\psi \in \spacetime$. 
In order to let the trajectory $\psi$ to be compatible with the worldline of a \textit{local} Eulerian observer, then one choose $n_\mu=n_\mu(p)$ to be unitary in the sense $n_\mu n^\mu = -1$. Notice that, this choice allows to interpret $n_\mu$ as the 4-velocity of the local Eulerian observer. $n_\mu$ is the main building block for the $3+1$ formalism. 
\subsubsection{Lapse, shift and coordinate system}
Then, one defines the normalized 4-vector field at every point $p\in\spacetime$ as
\begin{align} \label{eq:EulV}
&n_\mu = - \lapse \nabla_\mu t = ( -\lapse, 0_i ); \quad n^\mu n_\mu = -1\\
& n^\mu = \frac{1}{\lapse}(1, - \shift^i)
\end{align}
where the last identity for the contravariant $n^\mu$ is a result of the next definitions.
Here we have introduced the so called lapse \textit{scalar} function $\lapse=\lapse(p)$ and shift \textit{spatial-vector} $\shift^\mu=\shift^\mu(p)$.
In particular, the lapse function is defined as the inverse of the norm of $\nabla t$, i.e.
\begin{align}
\lapse := \left\| \nabla t \right\|^{-1}. \label{eq:lapse}
\end{align}
The lapse represents the first arbitrariness of the chosen coordinate system.
Then we can define a 4-vector $m_\mu := \lapse n_\mu$ that is a non-unitary vector $\left\|m\right\| = -\lapse^2$, normal to the hypersurface $\Sigma_t$; 
in particular, vector $\delta t\, m_\mu$ drags every event $p\in\Sigma_t$ to a corresponding event $p'\in\Sigma_{t+\delta t}$; the 4-distance is exactly the proper time \
\begin{align} \delta \tau_\Eul = \lapse \delta t \label{eq:tau} ,\end{align} the time measured by the local Eulerian observer;
After assigning a purely-spatial coordinate system $\{x^i\}$ on each slice $\Sigma_t$ that varies 
'smoothly' between any neighbor slices $\Sigma_{t\pm \delta t}$, then the coordinate system 
$\{t,x^i\}$ can be regarded as a well-behaved coordinate system for $\spacetime$. In this work, the 
system of coordinates  $\{t,x^i\}$ is named as '\textit{Eulerian}' system of coordinates. In 
particular, there exists a natural basis $\{ \boldsymbol{\partial}_\mu \}$ for the tangent space 
$\mathcal{T}_p$ at every event $p\in\spacetime$, that is associated to the chosen coordinates 
$\{x_\mu\}$. In this notation, the so-called \textit{time-vector} is defined as $ \boldsymbol{t} 
:=\boldsymbol{\partial}_t$, or $t_\mu$ in terms of its coordinates, and it is tangent to the lines 
of constant spatial coordinates. Similarly to $m_\mu$, also $t_\mu$ drags the slice $\Sigma_t$ to 
the neighbor one $\Sigma_{t+\delta t}$\footnote{Notice, indeed, that $t^\mu \nabla_\mu t \equiv 
m^\mu \nabla_\mu t \equiv 1$.}.
It is important to notice that $t_\mu$ is not necessarily a timelike vector, and this is the second arbitrariness of the chosen coordinate system. Then, it becomes useful to define the so called \textit{shift vector} as 
\begin{align}
\shift_\mu := t_\mu - m_\mu. \label{eq:shift}
\end{align}
 By construction, it is spacelike and it lies on the hypersurface $\Sigma_t$. 

Notice that, any choice of (i) the Eulerian-velocity field $n_\mu$, (ii)  lapse $\lapse$ and  (iii) shift vector $\shift^i$ \textit{univocally defines the coordinate system} $\{t,x^i\}$, or atlas, on $\spacetime$. Vice versa, the specification of a proper atlas on $\spacetime$ univocally defines the 4-velocities of the Eulerian observers $n_\mu$, that are associated to the $t$ isosurfaces, but also the lapse $\lapse$ and shift vector $\shift^i$.

The so called \textit{Lagrangian coordinate system} is obtained after choosing a foliation so that coordinate-lines and the worldlines of the fluid particles and local observers coincide, i.e. 
\begin{align}
\boldsymbol{t}(p) \equiv \boldsymbol{n}(p) \equiv \boldsymbol{u}(p)
 \end{align} 
for every point in the continuum media $p\in\spacetime$. In this frame, the time of the coordinates, the proper time as measured by the local observer and the proper time observed by the local fluid particle actually coincide, i.e. \begin{align}t\equiv\tau_\Eul\equiv \tau_\Lag   . \end{align}
\subsubsection{Temporal and spatial projectors}
The \textit{time projection operator} and the corresponding (complementary) \textit{spatial projection operator} are defined as
\begin{align} \label{eq:NGamma}
&N^\mu_\nu := - n^\mu n_\nu, \\
&\gamma^\mu_\nu  : = \delta^\mu_\nu - N^\mu_\nu = \delta^\mu_\nu + n^\mu n_\nu.
\end{align}
These operators allows to split any vector or tensor in its corresponding spatial and time components,
\begin{align}
U^\mu = \gamma^\mu_\nu U^\nu + N^\mu_\nu U^\nu
\end{align}
where the \textit{purely spatial} vector $V^\mu=\gamma^\mu_\nu U^\nu$ is a four-vector with \textit{vanishing contravariant time component} $V^0=0$.
Moreover, one can introduce the so called \textit{spatial metric} $\gamma$
\begin{align}
\gamma_{\mu\nu} = g_{\mu\nu} - N_{\mu\nu}, \quad \gamma^{\mu\nu} = g^{\mu\nu} - N^{\mu\nu}. \label{eq:gammaproj}
\end{align}
allowing to evaluate distances and norms on $\Sigma$, being a purely spatial tensor, i.e. $\gamma^{0\mu} = 0$ and $\gamma_{ij} = g_{ij}$ ($\gamma^{ij} \neq g^{ij}$).
The explicit form of the covariant and contravariant components of the metric tensor
\begin{align} \label{eq:gammagamma}
g_{\mu\nu} = \left( \begin{array}{cc} -\lapse^2 + \shift_i\shift^i & \shift_i \\ \shift_i & \gamma_{ij} \end{array}\right);
g^{\mu\nu} = \left( \begin{array}{cc} -1/\lapse^2   & \shift^i / \lapse^2  \\   \shift^i  /  \lapse^2   & \gamma^{ij} - \shift^i\shift^j / \lapse^2 \end{array}\right) .
\end{align}
\begin{align}
\gamma_{\mu\nu} = \left( \begin{array}{cc}  \shift_i\shift^i & \shift_i \\ \shift_i & \gamma_{ij} \end{array}\right); &&
\gamma^{\mu\nu} = \left( \begin{array}{cc} 0   & 0^i \\   0^i  & \gamma^{ij} \end{array}\right) .
\end{align}
One can show further that\footnote{Cramer's rule, see \cite{Gourgoulhon2012}} 
\begin{align}
(-g)^{\frac{1}{2}} = \lapse \gamma^{\frac{1}{2}}.
\end{align}
%The four-vector $n_\mu$ is associated to the four-velocity of the so called (\textit{normal} or) \textit{Eulerian observer}, the four-vector tangent to the world-line of the Eulerian observer.
\subsubsection{4-velocity \EqTitle{$u_\mu$}{}, spatial-velocity \EqTitle{$v_\mu$}{} and fluid coordinate velocity \EqTitle{$\hat{v}_\mu$}{}}
There are three different velocities that became useful in simplifying the equations in the text. These velocities are (i) the \textit{four-velocity} $u_\mu$ of a fluid particle,  (ii) the \textit{spatial} four velocity $\mbf{v}$ and  (iii) the fluid coordinate velocity (or transport velocity), defined as
\begin{align}
(i) \;\; \boldsymbol{u} := \frac{d \boldsymbol{p}}{d \tau_\Lag} &&(ii) \;\; \boldsymbol{v} := 
\frac{ d \boldsymbol{\ell}}{d \tau_\Eul}   &&(iii) \;\; \hat{\boldsymbol{v}} := \frac{ d 
\boldsymbol{x}}{d t} 
\end{align}

Then, the spatial four-velocity $\boldsymbol{v}$ as measured by the Eulerian 
observer of a material 
particle flowing with four-velocity $\boldsymbol{u}$ is 
\begin{align}
v^\mu = \frac{\gamma^\mu_\nu u^\nu }{-n_\alpha u^\alpha} = \frac{\text{proj. of \textbf{u} along $\Sigma$}}{\text{Lorentz factor of \textbf{u} as measured by \textbf{n}}}
\end{align}
\begin{align}
n_\alpha u^\alpha = -\lapse u^0 
\end{align}
and then
\begin{align}
v^0 &= 0, \qquad   
v^i = \frac{\gamma^i_\nu u^\nu}{ \lapse u^0} = \frac{u^i - n^i n_\nu u^\nu}{ 
\lapse u^0} =\frac{u^i +\lapse n^i u^0}{ \lapse u^0} = \frac{u^i +\shift^i 
u^0}{ \lapse u^0}
= \frac{1}{\lapse} \left(\frac{u^i}{u^0} +\shift^i \right) \\
v_0 &= g_{0\nu} v^\nu = \shift_i v^i, \qquad
v_i = g_{i\nu} v^\nu  =  \shift_i v^0 + \gamma_{ij} v^j  = \gamma_{ij} v^j 
\end{align}
Then, from the normalization condition and the definition of Lorentz factor $W$ 
\begin{align}
u_\mu u^\mu = -1, \quad  W = - n_\mu u^\mu = \lapse u^t = 1/ (1- v^2)
\end{align}
\begin{align}
u^t  = \frac{W}{\lapse}, \quad  u_t = W(-\lapse +\shift_i v^i)
\end{align}
one has 
\begin{align}
&u^i =  \frac{W}{\lapse} \left( \lapse v^i -  \shift^i   \right) =: \frac{W}{\lapse} \hat{v}^i \quad &  u_i =W v_i \\
%\end{align}
%\begin{align}
&v^i = \frac{u^i}{W} + \frac{\shift^i}{\lapse} ,  \quad &v_i = \frac{u_i}{W}. \label{eq:EulVel}
\end{align}
where  $\hat{v}$ is named as the \textit{fluid coordinate velocity} (or transport velocity).
 Notice moreover
\begin{align}
&u^\mu = (\gamma^\mu_\nu + N^\mu_\nu) u^\mu =  W v^\mu - (n_\nu u^\nu) n^\mu = W(n^\mu + v^\mu) \label{eq:uEul}
\end{align}  

\section{Conservative to primitive transformation}
\label{sec:cons2prim}
Inspired by the third option of \cite{DelZanna2007} (simplified since here magnetic fields are absent, actually) we build our strategy for deriving the primitive variables from the conservative set
\begin{align}
[D,S_i,U] \longrightarrow [\rho,v_i,p]
\end{align}
\begin{align}
D & = \rho W,\\
S_i & = \rho h W^2 v_i   +  \lapse \Sigma^0_{\oo i} = \rho h W^2 v_i  +\sigma_i , \\
U &= \rho h W^2 - p  + \lapse^2 \Sigma^{00} = \rho h W^2 - p  + \sigma
\end{align}
Then we guess the \textit{initial} value of
\begin{align}
\sigma_i = v^j \Sigma_{ij}, \quad \text{and}\quad \sigma = v^iv^j\Sigma_{ij}
\end{align}
and  look for the roots of two auxiliary functions
\begin{align}
 x:=v^2, &&F_1(x,y) &:= y^2\,x-\tilde{S}^2, \label{eq:F1} \\
 y:=\rho h W^2,&&F_2(x,y) &:= y-p-\tilde{U}, \label{eq:F2}
\end{align}
after defining the four vector and the scalar
\begin{align}
\tilde{S}_\mu := S_\mu - \sigma_\mu, \quad \tilde{U}:=U- \sigma
\end{align}
Notice that guess of the couple $(\sigma_i,\sigma)$ can be substituted by an initial guess $(\rho, v)$ or $(\rho, \hat{v})$\footnote{Another alternative is: i) guess an initial value for the Lorentz factor $W$, ii) then assume the equations of perfect fluids and derive $(\rho, v_i)$ directly from the value of $(D,S_i)$.}, that are used for evaluating a corresponding initial state of  $(\sigma_i,\sigma)$.
Introducing the definition of enthalpy
\begin{align}
h = 1+e_0 +e_1 + \frac{p}{\rho}
\end{align}
where, after assuming {
\begin{align}
p(\rho, e_0) = (\gamma -1) \rho e_0 \quad \Rightarrow \quad 
h = 1+e_1 +\frac{\gamma}{\gamma-1} \frac{p}{\rho}
\end{align} 
}
and, since $\rho h = y/ W^2 = y (1-x)$, then
\begin{align}
p = \frac{\gamma-1}{\gamma}\left[ \rho h - \rho(1+e_1)\right] =
\frac{\gamma-1}{\gamma}\left[ y (1-x) - D(1+e_1)(1-x)^{\frac{1}{2}}\right].
\end{align}
Using \ref{eq:F2}, we have the roots
\begin{align}
y &= p+\tilde{U} = \frac{\gamma-1}{\gamma}y (1-x) - \frac{\gamma-1}{\gamma}D(1+e_1)(1-x)^{\frac{1}{2}}+\tilde{U} \\
&= - \left[1 -\frac{\gamma-1}{\gamma}(1-x) \right]^{-1}\left[ \frac{\gamma-1}{\gamma}D(1+e_1)(1-x)^{\frac{1}{2}}-\tilde{U}\right],\\
x &= \tilde{S}^2/y^2
\end{align}
In practice, if we use $D(1+e_1)$ instead of $D$ we may recycle the same subroutine of GRMHD. 
\\
\noindent
An alternative could be the following:
re-define the four vector and the scalar
\begin{align}
\tilde{S}_\mu &:= S_\mu - \sigma_\mu - \rho e_1 W^2 v_i, \\
 \tilde{U}&:=U- \sigma-\rho e_1 W^2\\
x&:= v^2,\\
y&:= \rho (h-e_1)W^2
\end{align}
and one obtains\begin{align}
y &= - \left[1 -\frac{\gamma-1}{\gamma}(1-x) \right]^{-1}\left[ \frac{\gamma-1}{\gamma}D(1-x)^{\frac{1}{2}}-\tilde{U}\right],\\
x &= \tilde{S}^2/y^2
\end{align}

%\newpage

%\newpage

%\input{AuxSections}

\label{lastpage}
\end{document}